\documentclass{article}
\usepackage[utf8]{inputenc}

\usepackage{amsmath} 
\usepackage{graphicx}
\usepackage{listings} 
\usepackage{xcolor} 
\usepackage{acronym}

\usepackage{checkend}	
\usepackage{graphicx}	
\usepackage{hyperref}

\usepackage{caption}
\usepackage{pdflscape} 
\usepackage{afterpage}
\usepackage{multirow}  

\usepackage{comment}

\usepackage{natbib}

\usepackage{array}  
\usepackage{tabularx}

\usepackage{acronym}
\usepackage{glossaries}

\usepackage{appendix}
\usepackage{todonotes}

\newcommand{\ts}{\textsuperscript}

\DeclareUrlCommand\DOI{}
\newcommand{\doi}[1]{\href{http://dx.doi.org/#1}{\DOI{doi:#1}}}



\definecolor{codegreen}{rgb}{0,0.6,0}
\definecolor{codegray}{rgb}{0.5,0.5,0.5}
\definecolor{codepurple}{rgb}{0.58,0,0.82}
\definecolor{backcolour}{rgb}{0.95,0.95,0.92}

\lstdefinestyle{mystyle}{
    backgroundcolor=\color{backcolour},   
    commentstyle=\color{codegreen},
    keywordstyle=\color{magenta},
    numberstyle=\tiny\color{codegray},
    stringstyle=\color{codepurple},
    basicstyle=\ttfamily\footnotesize,
    breakatwhitespace=false,         
    breaklines=true,                 
    captionpos=b,                    
    keepspaces=true,                 
    numbers=left,                    
    numbersep=5pt,                  
    showspaces=false,                
    showstringspaces=false,
    showtabs=false,                  
    tabsize=2
}

\begin{document}

\title{Preferential monitoring site location in the Southern California Air Quality Basin
\thanks{
The research reported in this paper was supported by  grants from the Natural Science and Engineering Research Council of Canada.} 
}

\date{February 12, 2023}

\author{Adrian Jones$^1$,James V Zidek$^2$ and Joe Watson$^3$\\
\small
 $^1$ Statistics Canada, Ottawa, Ontario, Canada \\\small
 adriaanjones@gmail.com\\\small
 $^2$ Corresponding author:\\\small
Department of Statistics\\\small
University of British Columbia \\\small
2207 Main Mall\\\small
Vancouver, BC \\\small
Canada V6T 1Z4\\\small
 jim@stat.ubc.ca \\\small
 $^3$ Glaxo Kline Smith, UK\\\small
 joe.x.watson@gsk.com
 }

\maketitle

\maketitle

\include{00B_Bbstract}
\cleardoublepage
\section*{Acronyms and Glossary}


\newglossaryentry{air_basin}
{
    name = {air basin},
    description = {A region of mostly similar air}
}

\newglossaryentry{crit_pol}
{
    name = {criteria pollutant},
    description = {One of 6 pollutants found throughout the US that have NAAQS.  They are: Particulates, ground level ozone, carbon monoxide, sulfur dioxide, nitrogen dioxide, and lead.}
}

\newglossaryentry{NAAQS}
{
    name = {National Ambient Air Quality Standard},
    description = {Regulation for a single pollutant considered "harmful to the public." Will have a primary and secondary standard with fixed statistic, concentration and observation duration. }
}

\newglossaryentry{PC_prior}
{
    name = {penalised complexity prior},
    description = {Prior that is designed to favour less complicated models in the absence of data supporting additional complexity. \cite{simpson2017penalising} }
}

\newglossaryentry{PM}
{
    name = {particulate matter},
    description = {}
}

\newglossaryentry{PS}
{
    name = {preferential sampling},
    description = {}
}

\newglossaryentry{PC_priors}{
    name = {Penalised Complexity Prior},
    description = {Class of priors for Baysian statistics that are designed to shrink the model's complexity to a minimum, restricted by both the data and the researcher's prior knowledge.}
}


\newglossaryentry{S}{
    name = {$S$},
    description = {Spatial domain of interest.  In this application it is two dimensional, Latitude and Longitude or their Albers transformation}
}
\newglossaryentry{T}{
    name = {$T$},
    description = {Temporal domain of interest}
}
\newglossaryentry{St}{
    name = {$S_t$},
    description = {Set of locations at which observations are made at time $t$.  A realisation of the process $P$}
}
\newglossaryentry{s,t}{
    name = {$(s,t)$},
    description = {A point in the spatio-temporal domain.  $(s,t) \in S X T$}
}
\newglossaryentry{Y}{
    name = {$Y$},
    description = {The ``true" Gaussian spatio-temporal process of interest with mean $\mu(s,t) = E[Y(s,t)]$ and covariance $\gamma(s,s') = cov{Y(s), Y(s')}$.  The covariance between two locations with $t \neq t'$ is assumed to be 0}
}
\newglossaryentry{mu}{
    name = {$\mu_{s,t}$},
    description = {The mean of the Gaussian spatio-tempoeral process $Y$.}
}
\newglossaryentry{gamma}{
    name = {$\gamma$},
    description = {The covariance function of $Y$}
}
\newglossaryentry{Z}{
    name = {$Z_{s,t}$},
    description = {Set of noisy observations taken of spatio-temporal process $Y$ at location $s$ and time $t$}
}
\newglossaryentry{X}{
    name = {$X_{s,t}$},
    description = {Set of covariates that can help predict either or both of $\mu$ and $S_t$}
}

\newglossaryentry{u}{
    name = {$u$},
    description = {The scalar distance between two points $u = ||s - s'||$.}
}
\newglossaryentry{tau}{
    name = {$\tau$},
    description = {The precision of a process, the inverse of its standard deviation $\sigma$.}
}
\newglossaryentry{range}{
    name = {$r$},
    description = {A distance at which the covariance between two sites becomes relatively unimportant compared to the overall variance.}
}

\newglossaryentry{sigma}{
    name = {$\sigma$},
    description = {The overall variance of the spatial field.}
}
\newglossaryentry{phi}{
    name = {$\phi$},
    description = {The scaling parameter of the Mat\'{e}rn function.}
}
\newglossaryentry{kappa}{
    name = {$\kappa$},
    description = {The Order of the Mat\'{e}rn function.}
}

\newglossaryentry{P}{
    name = {$P$},
    description = {Point process that describes where the point locations $S_t$ are}
}

\newglossaryentry{k}{
    name = {$k$},
    description = {In preferential sampling, the number sites included when taking the mean before correlating mean with distance between means}
}

\begin{acronym}
    \acro{ACF}[ACF]{autocorrelation function}
    \acro{ANOVA}{Analysis of Variance}
    \acro{AR}{Auto Regressive}
    \acro{AQMD}{Air Quality Manegement District}
    \acro{AQMP}[AQMP]{Air Quality Management Plan}
    \acro{CAA}{Clean Air Act}
    \acro{CAAQS}[CAAQS]{California Ambient Air Quality Standards}
    \acro{CFR}{Code of Federal Regulations}
    \acro{EPA}[EPA]{Environmental Protection Agency}
    \acro{FRM}[FRM]{Federal Reference Method}
    \acro{FEM}[FEM]{Federal Equivalent Method}
    \acro{GRF}[GRF]{Gaussian Random Field}
    \acro{IID}{Independent and Identically Distributed}
    \acro{INLA}[INLA]{Integrated Nested Laplace Approximation}
    \acro{MA}{Moving Average}
    \acro{MCMC}{Markov Chain Monte Carlo}
    \acro{MSA}{Metropolitan Statistical Area}
    \acro{NAAQS}[NAAQS]{National Ambient Air Quality Standards}
    \acro{PACF}[PACF]{Partial Autocorrelation Function}
    \acro{PC}{penalised complexity}
    \acro{PM10}[$PM_{10}$]{particulate matter with a mean diameter $<10 \mu m$}
    \acro{PM25}[$PM_{2.5}$]{particulate matter with a mean diameter $<2.5 \mu m$}
    \acro{POC}[POC]{Parameter Occurrence Code}
    \acro{PS}{preferential sampling}
    \acro{RW}{random walk}
    \acro{SOCAB}[SOCAB]{South Coast Air Basin}
    \acro{SCAQMD}[SCAQMD]{South Coast Air Quality Management District}
    \acro{SPDE}[SPDE]{Stochastic Partial Differential Equation}
    \acro{USGS}{United States Geological Survey}
    
\end{acronym}

\section{Introduction.}\label{sec:introduction}

Air pollution is a continuous three-dimensional field.  It exists on many spatial scales depending upon the pollutant, from a city block to the globe.  This report focuses on ground level \ac{PM10}.  This focus simplifies the field. It becomes a two-dimensional surface, with changes being on the scale of kilometers \citep{CFR:Title40-58} instead.   

The field can only be monitored by taking point measurements and extrapolating these over the entire region of interest. The collection of monitoring sites is called a monitoring network.  That network fulfills one or more specific purposes: overall field estimation; monitoring for pollutant compliance; assessing concentrations over population centers; forecasting.  These goals do not necessarily encompass the capture of the field's mean level, in which case the network may generate a biased assessment of the overall concentration field. This bias may not matter; if the network were meant to detect noncompliance, the sites should be located in regions most likely to be out of compliance. 

However, the data from the network may well be used for unintended applications. Since most common statistical procedures assume that sampling is not preferential, i.e. unbiased, 
applying these techniques to data can yield result in erroneous conclusions.  For example, there may 
be an inverse impact on health impact parameters: if the bias were towards high observations, the effect of pollution will be underestimated \citep{Zidek:2012}.

That leads to the study reported in the paper, which presents a way of detecting bias, if any,
in multi-level governmental networks for monitoring air quality in the United States in general and the region surrounding Los Angeles in particular.  Because the US government makes data  freely available, the data are used for many purposes, some of which are unintended. An example would be epidemiological studies that attempt to link disease frequency to pollution levels.

The paper reports evidence of bias in the sampling of $PM_{10}$ in the \ac{SOCAB}.  That bias has been acknowledged as intentional by the governmental body in charge, the \ac{SCAQMD}.  Thus, it should be considered in any work that uses those networks. Furthermore, because of the bias's possible origin in policy, caution should extend to any data from these types of compliance monitoring networks.

\subsection{Motivation for the Paper}
\label{subsec:motivation}
 This study set out to explore ways of detecting monitoring site selection bias, with a focus on the South California Air Basin SOCAB monitoring region after its several decades of data monitoring.  Southern California has a long history of recognizing air pollution as a problem, dating back to 1945 \citep{CASCAQMD:2015}.
 
The models used to describe spatial fields generally assume a random placement of monitoring sites or at least independent conditional on its latent underlying latent field.  However, it seems the placement of monitoring sites is often not random. They are often chosen to fulfill a range of constraints. Even if the monitoring network were well-designed, sites might be chosen for termination because their local air pollution fields are consistently in compliance.  In short, selection bias, referred to as \ac{PS}, can lead to models that don't reflect the actual pollution field experienced by the population. 

Concern about \ac{PS} has a decade's long history. \citet{isaaks1988spatial} discuss how clustered data make variograms poor at estimating covariance parameters. \cite{diggle:07} define \ac{PS} in their book as the stochastic dependence of site locations upon the property being measured. \cite{shaddick2012preferential} discovered  \ac{PS} in the United Kingdom's black smoke monitoring network. Numerous other papers have examined \ac{PS} in different cases, showing how the \ac{PS} results in models incorrectly attributing magnitude of pollution with impact on health or other model parameters.  An extensive list of references can be found in \citep{Zidek:2012}.  

The data gathered in the US is freely available to the public and so gets put to many different uses.  Government agencies use this data to make real time air quality warnings, to monitor general compliance of regions to meet predefined standards and to monitor point sources.  Healthcare specialists use the data from the monitoring in correlational studies to predict health impacts of pollutant levels on the general population and subsets of interest.  \cite{wong2004comparison} brought up various concerns about using different interpolation techniques with \ac{EPA} data for epidemiological studies.  This combination of circumstances led us to be curious about whether the monitoring networks of the US exhibit \ac{PS}.

\section{Air Pollution}\label{sec:introdairpollution}
Air pollutants are particulates or gases in the air that have a negative impact upon human health or the economy and are present in concentrations that are unusual compared to background levels.  They can be created as a direct result of human actions (e.g. coarse particulates from construction or wood burning), as a secondary result created by chemical or physical processes in the atmosphere (e.g. ozone or nitrous oxides), or as a result of a natural process (e.g. forest fires or dust storms).  The \ac{EPA} has defined six \glspl{crit_pol} to be monitored that provide a good overview of air quality.  One of these is \ac{PM10}, the focus of this report,  with criteria set out in the US EPA NAAQS table.

\subsection{Air Pollution Monitoring: a History}
\label{sec:apmonitoring}
Air pollution has been monitored by national government agencies since the 1950s.  The most common motivation is the regulation of polluting industries and the preservation of population health, but other concerns include damage to buildings and infrastructure, reduced crop yields, and reduced air visibility.  
        
In the United Kingdom, the 1956 Clean Air Act was passed in response to high concentrations of Black Smoke (mostly particulates from coal burning) that, in 1952, was associated with 4000 excess deaths \cite{shaddick2014case}.

In the USA, the \ac{CAA} of 1963 created a regulatory system requiring states to work towards target goals for a range of air pollutants.  The Air Quality Act of 1967 created federal powers to monitor and enforce standards of air pollution, and in 1970 the creation of the \ac{EPA} consolidated these powers in a single agency 
.  The act was amended in 1977 and again in 1990 to reflect changing understanding of pollutant creation and impact.  
 
 The Los Angeles basin in particular has a long history of poor air quality.  Efforts to regulate and monitor air quality started in 1947 with the founding of the Los Angeles County Air Pollution Control District in response to widespread smog in 1943 
 .

\section{Monitoring Air Pollution fields} \label{sec:monitoring}
We now describe our study in more detail. It set out to  detect the bias, described above, that might be present in the South California Air Basin monitoring region after several decades of data monitoring.  Southern California has a long history of air pollution dating to 1945 \citep{CASCAQMD:2015}.
 
Models that describe spatial fields generally assume a random placement of monitoring sites, or an independence of` the latent field underlying the pollution field.  However, the placement of monitoring sites is often not random. They are chosen to fulfill a range of constraints and even if the initial selection is well-designed, sites might be chosen for termination because they are consistently in compliance.  This selection bias, referred to as \ac{PS}, can result in models that don't reflect the pollution field experienced by the population. 
 
Concern about \ac{PS} has a decade's long history. \citet{isaaks1988spatial} discuss how clustered data make variograms poor at estimating covariance parameters. \cite{diggle:07} define \ac{PS} in their book as the stochastic dependence of site locations upon the property being measured. \cite{shaddick2012preferential} discovered  \ac{PS} in the United Kingdom's black smoke monitoring network. Numerous other papers have examined \ac{PS} in different cases, showing how the \ac{PS} results in models incorrectly attributing magnitude of pollution with impact on health or other model parameters.  An extensive list of references can be found in \citep{Zidek:2012}.  

The data gathered in the USA is freely available to the public and so gets put to many different uses.  Government agencies use this data to make real time air quality warnings, to monitor general compliance of regions to meet predefined standards and to monitor point sources.  Healthcare specialists use the data from the monitoring in correlational studies to predict health impacts of pollutant levels on the general population and subsets of interest.  \cite{wong2004comparison} brought up various concerns about using different interpolation techniques with \ac{EPA} data for epidemiological studies.  This combination of circumstances led us to be curious about whether the monitoring networks of the US exhibit \ac{PS}.

subsection{PM10 Pollution}
\label{sec:pm10}
Because the \ac{EPA} warehouses data on all the monitored pollutants, there is a choice of hundreds of pollutants, several of which could be reasonably chosen for analysis.  \ac{PM25} is currently considered more relevant for human health.  Ozone is a primary concern in LA because it is out of compliance.   Our study focussed on \ac{PM10} as  follow-through on black smoke work in England by \cite{zidek2010monitoring}.  As well, \ac{PM10} has a longer history of monitoring in the \ac{SOCAB} than does \ac{PM25}.
    
\subsubsection*{What is PM10?}
\label{subsubsec:pm10nature}
\ac{PM10} are particulates with a diameter less than 10 $\mu m$ in diameter.  It is reported as a mass of solids per volume of air.  The coarser ($> 2.5 \mu m$ diameter) particles are generally a product of physical wear and tear, while the finer ($< 2.5 \mu m$ diameter) particles are usually aggregates from chemical reactions producing nitrogen and sulfur oxides
.

\subsubsection*{Effects of PM10}
\label{subsubsec:pm10effects}
These particulates have various deleterious effects on human health and infrastructure.  Health effects include both short and long-term concerns.  Short-term, high concentrations of \ac{PM10} can result in acute respiratory problems.  Long-term exposure to lower pollution levels can result in a chronic reduction in functionality of the lungs and cardiovascular system 
.  Economically \ac{PM10} damages property, crops, and reduces visibility
.

\subsubsection*{Measurement of PM10}
\label{subsubsec:pm10measurement}
Measuring a field that is continuous in time and space can be challenging.  Typically, discrete measurements are taken, and a model interpolates these to estimate a field.  As a result, these measurements are interpreted as averaging over both a spatial and temporal range.

Spatial averaging at each site represents a mass of upwind air, its volume depending upon the geography of each site and the pollutant being monitored.  This is acknowledged in the spatial scale provided in the reports.   

In Appendix D Section 4.6 (b) of CFR 40-58 

Temporal averaging of the observations is a property of the measurement technique.  The US's gold standard for \ac{PM10}, the \ac{FRM}, is to pull air through filters and weigh the accumulated particles after 24 hours.  This gives an average particulate presence in the air over 24 hours.  The frequency at which these 24-hour samples are taken depends upon the pollutant levels, with levels closer to the standards requiring more frequent measurements (see table \ref{tab:EPA_monitoring_freq}) \citep{CASCAQMD:2015}.

Other measurement techniques are called \ac{FEM}.  Laser back-scatter measurements provide instantaneous readings of particulate size and concentration, generally taken every few minutes.  These are useful for delivering time-sensitive warnings.  In the yearly reports, these \ac {FEM} are averaged over 24hrs to be temporally equivalent to the \ac{FRM} instrumentation.

Individual sites often have multiple instruments monitoring the same pollutant. For example, there could be a continuous \ac{FEM} monitor for forecasting as well as air quality advisories and one \ac{FRM} monitor to fulfill statutory requirements.  Other reasons to have multiple monitors include research or sensor calibration.

\subsection{Government Administration}
\label{subsec:govtadmin}
The process of interest-choice of site location and the possible subsequent preferential sampling-is a product of governmental decisions to set and meet regulatory standards.  For the United States, these regulations are described in detail on the EPA website
but outlined here.

\subsubsection*{Regulatory Framework in the US}
\label{subsubsec:regulation}
In the United States, air quality monitoring and enforcement requires cooperation and coordination between governmental agencies at the regional, state and federal level.

At the Federal level, the \ac{EPA} defines standards for air quality levels, monitoring and reporting.   These standards define:
\begin{enumerate}
    \item the levels of pollution that must not be exceeded;
    \item how to monitor each pollutant (number and location of sites, frequency of monitoring, and what methods count as \ac{FRM});
    \item when and how to make reports to the \ac{EPA}.
\end{enumerate}

The states divide themselves into regional districts responsible for choosing site placement and report preparation.  States can set their own regulations, but must still meet the \ac{EPA}'s regulations. 

\subsubsection*{Air Quality Standards}\label{subsubsec:aqs}
The \ac{CAA} established six important pollutants, called \gls{crit_pol}, 
including particulate matter, and gave the \ac{EPA} power to define \acp{NAAQS} for each.

The \ac{EPA} sets two standards to meet health and economic goals, known respectively as the primary and secondary standards.  Primary standards:
\begin{quote}
    ``Provide public health protection, including protecting the health of `sensitive' populations such as asthmatics, children, and the elderly.'' 
\end{quote}
Secondary standards:
\begin{quote}
    ``Provide public welfare protection, including protection against decreased visibility and damage to animals, crops, vegetation, and buildings.''  
\end{quote}
These criteria can be 
seen in the US EPA NAAQS table
at the URL\\
www.epa.gov/pm-pollution \\/timeline-particulate-matter-pm-national-ambient-air-quality-standards-naaqs\#Superscript1

In 2006, the EPA revoked the primary annual \ac{NAAQS} for \ac{PM10} meaning that \ac{PM10} is no longer seen as problematic long term.  Short-term \ac{PM10} pollution remains a concern and is monitored for 24-hour primary exceedance in the US 
.  Table \ref{tab:EPA_PM10_standards} shows how the averaging time and core statistic have changed historically, and how the acceptable concentration has decreased since the initial creation of the \ac{PM10} \ac{NAAQS} in 1971 
.
\begin{table}[ht]
    \centering
    \begin{tabular}{p{0.06\textwidth}|p{0.18\textwidth}|p{0.15\textwidth}|p{0.14
    \textwidth}|p{0.30
    \textwidth}}
    Year & Final Rule / Date & Averaging Time & Level & Form \\
    \hline
    1997 & 62 FR 38652 Jul 18, 1997 & 24 hour & 150 $\mu g/m^3$ &	Initially promulgated 99th percentile, averaged over 3 years; when 1997 standards for PM10 were vacated, the form of 1987 standards remained in place (not to be exceeded more than once per year on average over a 3-year period) \\
    1997 & 62 FR 38652 Jul 18, 1997 & Annual & 50 $\mu g/m^3$ & Annual arithmetic mean, averaged over 3 years \\
    2006 & 71 FR 61144 Oct 17, 2006 & 24 hours & 150 $\mu g/m^3$ & Not to be exceeded more than once per year on average over a 3-year period \\
    2012 & 78 FR 3085 Jan 15, 2013 & 24 hour & 150 $\mu g/m^3$ & Not to be exceeded more than once per year on average over a 3-year period \\
    \end{tabular}
    \caption{Abridged history of Primary and Secondary standards for \ac{PM10}.  The \ac{EPA} revoked the annual \ac{PM10} \ac{NAAQS} in 2006, but maintains the acute 24hr standard 
    }
    \label{tab:EPA_PM10_standards}
\end{table}

In addition to these federal standards, California has its own set of standards that are more stringent (24-hour: $50 \mu g/m^3$, Annual: $20 \mu g/m^3$).  However, these are not considered for this report since it focuses on \ac{EPA} standards and monitoring.

\subsubsection*{Monitoring Requirements}\label{subsubsec:monreqs}
The requirements to design and set up a monitoring network are proscribed in the \ac{CFR}, Title 40, Subsection 58 appendix D \citep{CFR:Title40-58}.

The minimum number of sensors for a given region is based upon both the population and the concentration of \ac{PM10} relative to its \ac{NAAQS} as described in table \ref{tab:NAAQS_site_count}.
\begin{table}[ht]
    \centering
    \begin{tabular}{p{0.25\textwidth}|p{0.18\textwidth}|p{0.18\textwidth}|p{0.18\textwidth}}
         Population category of \ac{MSA} & High ($\ac{PM10} > \ac{NAAQS} *1.2$) & Medium ($\ac{PM10} > \ac{NAAQS} *0.8$) & Low ($\ac{PM10} < \ac{NAAQS}*0.8$) \\
         \hline
         $>$ 1,000,000 & 6-10 & 4-8 & 2-4 \\
         500,000 - 1,000,00 & 4-8 & 2-4 & 1-2 \\
         250,000 - 500,000 & 3-4 & 1-2 & 0-1 \\
         100,000 - 250,000 & 1-2 & 0-1 & 0 \\
    \end{tabular}
    \caption{How the required number of sites for monitoring \ac{PM10} increases with both the population of the Metropolitan Statistical Area and the severity of the ambient pollution \citep{CFR:Title40-58}. }
    \label{tab:NAAQS_site_count}
\end{table}

The monitoring frequency is dependent upon the site's pollution concentration relative to that pollutant's standard, as described in table \ref{tab:EPA_monitoring_freq}  \citep{AQMNP:2019}.

\begin{table}[ht]
    \centering
    \begin{tabular}{p{0.25\textwidth}|p{0.10\textwidth}|p{0.13\textwidth}|p{0.13\textwidth}|p{0.13\textwidth}|p{0.08\textwidth}}
         Ratio of previous Pollutant level to standard & $<0.8$ & $(0.8 < 0.9)$ & $(0.9 < 1.2)$ & $(1.2 < 1.4)$ & $1.4<$  \\
         \hline
         Monitoring frequency & Every 6th day & Every other day & Every Day & Every other day & Every 6th day \\ 
    \end{tabular}
    \caption{How the monitoring frequency of a given site depends upon how close it is to the \ac{NAAQS} exceedance threshold \citep{AQMNP:2019}.}
\label{tab:EPA_monitoring_freq}
\end{table}

\subsubsection*{Spatial Scale}
\label{subsubsec:spatscale}
Each site has a spatial scale defined in Appendix D Section 1.2 of \ac{CFR} 40. The purpose of defining the spatial scale:
\begin{quote}
   ``Is to correctly match the spatial scale represented by the sample of monitored air with the spatial scale most appropriate for the monitoring site type, air pollutant to be measured, and the monitoring objective.''  \cite{CFR:Title40-58}
\end{quote}
Sites monitoring \ac{PM10} generally have two relevant scales, the Middle scale ($100~m - 500~m$) and the Neighborhood scale ($0.5~km - 4.0~km$).  

\subsubsection*{Monitoring Purpose}
\label{subsubsec:purpose}
Each site has at least one monitoring purpose, which: 
\begin{quote}
    ``Is the reason why a certain pollutant is being measured at a certain site.'' \citep{AQMNP:2019} 
\end{quote} The full list of all purposes 
is quite lengthy
but there are only two purposes relevant for sites monitoring \ac{PM10} in the \ac{SOCAB}.  
\begin{itemize}
    \item \textbf{High concentration} monitoring is conducted at sites to determine the highest concentration of an air pollutant in an area within the monitoring network. A monitoring network may have multiple high-concentration sites (i.e., due to varying meteorology year to year).
   \item \textbf{Population exposure} monitoring is conducted to represent the air pollutant concentrations to which that populated area exposed 
   \citep{AQMNP:2019}.
 \end{itemize}

\subsubsection*{Reporting Requirements}
\label{subsubsec:ReportingRequirements}
Reporting air quality data is mandated by the \ac{EPA}.  Each year, regional monitoring districts compile the past year's data and submit it to the \ac{EPA} for entry into a publicly accessible database.  After data cleaning, the \ac{EPA} makes the reported data available as recompiled data files on its website at \url{https://aqs.epa.gov/aqsweb/airdata/download_files.html}.  

The \ac{AQMP} is a report written every 3–4 years by the local \ac{AQMD} that summarizes whether the region is in or out of attainment of the Federal levels for all \glspl{crit_pol}.
Every 5 years, a full site visit and Network Assessment is done to produce a report on the state of the network. These reports for the \ac{SCAQMD} were done in 2010 and 2015.  Each report describes the sites in the network, including the site's Monitoring Purpose and Spatial Scale.  While most regions in California prepare their reports through the California Air Resources Board, the \ac{SCAQMD} prepares and submits its report separately.

\subsection{The LA Basin} \label{subsec:labasin}
\begin{landscape}
   \begin{figure}[ht]
   \captionsetup{justification=centering}
        \caption{This map made by the \ac{AQMD} shows how the jurisdiction and the air basin overlap.\\  The main figure shows the \ac{SCAQMD} and the insert in the top right corner demonstrates \\where the \ac{SOCAB} is in relation to the \ac{SCAQMD}. All the \ac{SOCAB} lie in the \ac{SCAQMD}.\\ But the \ac{SCAQMD} stretches over parts of several airsheds. 
        }
        \centering
       \includegraphics[width = \textwidth]{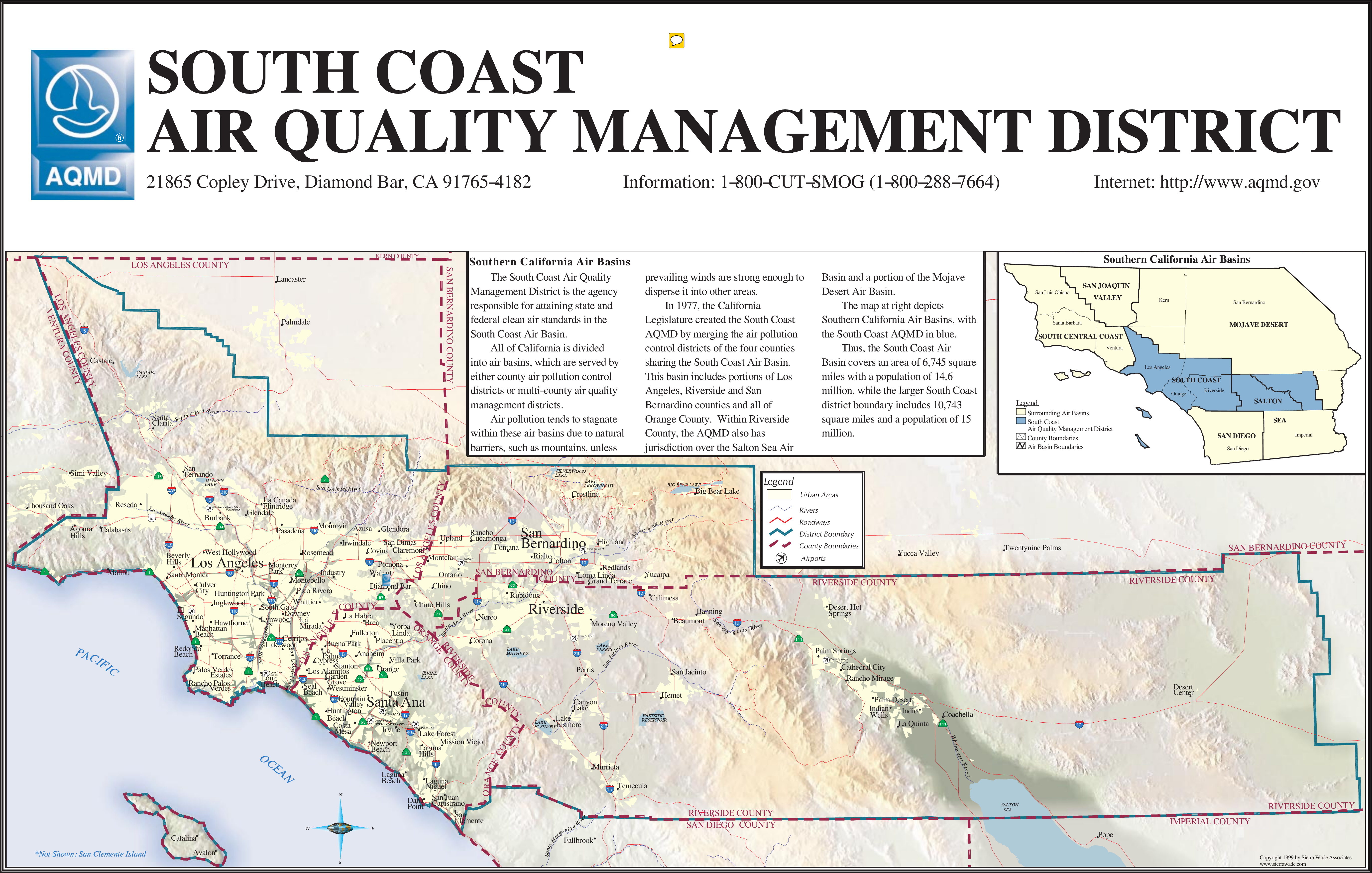}
       \label{fig:SCAQMD-jurisdiction}
   \end{figure}
\end{landscape}

A \gls{air_basin} is a geographic region of roughly similar air conditions, typically a topographic depression.  The \ac{SOCAB} is approximately 17 to 100 square kilometers \gls{air_basin} surrounding Los Angeles.  It can be seen depicted in Figure \ref{fig:SOCAB_counties}.  Its boundary is different from the jurisdiction of the body that manages the \gls{air_basin}, the \ac{SCAQMD}. The \ac{SCAQMD}, in Figure \ref{fig:SCAQMD-jurisdiction} consists of four counties and exists over several \glspl{air_basin}.

The \ac{SOCAB} is defined in the California Code of Regulations Title 17  Subchapter 1.5  Article 1.  § 60104.   
\begin{quote}
``South Coast Air Basin means the non-desert portions of Los Angeles, Riverside, and San Bernardino counties and all of Orange County as defined in California Code of Regulations, Title 17, Section 60104. The area is bounded: on the west by the Pacific Ocean; on the northwest by the Santa Susana Mountains and Simi Hills, on the north by the San Gabriel Mountains, San Bernardino Mountains, and on the east by the San Jacinto Mountains and Santa Rosa Mountains; and on the south by the San Diego County line.''
    
\end{quote}

\section{Theoretical Concepts}
\label{sec:theory}
This section outlines theoretical concepts that are used in the rest of the report. This includes a discussion of working with data on the surface of the Earth and an overview of the spatio-temporal statistical methods that will be used to analyze that data.

\subsection{Map Projection} \label{subsec:MapProjection}
The spatial statistics to be used assume a 2-dimensional surface. However, the Earth's surface is curved.  Map projections overlay this curved surface onto a flat plane, resulting in a loss of some spatial relationships.  For this reason, projections should be done with an awareness of what is lost.  

Different projections focus on maintaining the fidelity of different characteristics, typically at least one of area, shape, relative scale, or direction \citep{USGS:MapProjections}.  The projection is often named after what is preserved.  Equal Area projections maintain the ratio of surface area between the map and the surface, but can result in distorted shapes, angles and scales \citep{USGS:MapProjections}.  Consistent Shape,  aka ``Conformal'' maps, keep the local angle correct so that, for example, lines of latitude are always perpendicular to lines of longitude.  Maps cannot be both Equal Area and Conformal  \citep{USGS:MapProjections}.

At the scale of the \ac{SOCAB}, it seems reasonable to approximate the Earth as a flat plane.  However, it is still useful to have a known projection because they transform the units of latitude and longitude into flat kilometers, making the interpretation of parameters such as the range easier.  The Albers Equal Area Conic is used by \ac{USGS} for sectional maps of all 50 states in the 1970 atlas.

\subsection{Conventional Geostatistics}
\label{subsec:convgeostats}

Geostatistical processes can be divided into two components: 
\citep{diggle:07} pg 13: 
\begin{enumerate}
    \item the stationary Gaussian spatial process \gls{Y};
    \item a statistical description of data gathering conditional on the surface.
\end{enumerate}

\gls{Y} is jointly multivariate Gaussian distributed and so completely defined by its mean function \gls{mu} = $E[Y(s)]$, and covariance function, \gls{gamma} = 
Cov\{Y(s), Y(s')\}.

The observed values 
 \gls{Z} at location and time \gls{s,t} are the \gls{Y} after, including measurement error.  This makes the expectation of the observed values conditional on the surface:
\gls{mu} = 
E[\gls{Z} | \gls{Y}] 
\citep{diggle:07}.

\subsubsection*{Covariance Functions}
\label{subsubsec:covariances}
The covariance function describes how the pollution field at two separate locations relate to each other. It does this by describing their correlation as a function of the distance, \gls{u}, between those sites.  A common assumption in both temporal and spatial statistics is that the closer points are more similar than points further away from each other, and so covariance functions are typically monotonically decreasing with \gls{u}.

\subsubsection*{Mat\'{e}rn Function} \label{subsubsec:MaternIntro}
The Mat\'{e}rn function is the most commonly used covariance function for spatial statistics because of its flexibility, \citep{diggle:07} and is the one used in this report.  Its function is described in equation \ref{formula:Matern}.

\begin{equation}  \label{formula:Matern}
    \gamma(u) = {2^{\kappa -1}\Gamma(\kappa)^{-1}(u/\phi)^{\kappa}K_{\kappa}(u/\phi)}
\end{equation}

The components of equation \ref{formula:Matern} are described in \cite{diggle:07} as follows:
\begin{itemize}
    \item \gls{gamma}:  The covariance between two sites $s$ and $s'$.
    \item \gls{u}: The Euclidean distance between the two sites,  $||s_i - s_j||, i \neq j$
    \item \gls{kappa}: The order of the function, also called the shape or smoothness parameter.  \gls{kappa} controls the differentiability of the surface.  The Mat\'{e}rn function is $\kappa -1$ mean square differentiable.  A Mat\'{e}rn with $\kappa = 0.5$ is the exponential of order 1.  As $\kappa -> \infty$  the Mat\'{e}rn approaches the Gaussian Correlation function  \cite{diggle:07}.  An important note is that INLA can only compute Mat\'{e}rn functions with $0.5 \leq \kappa \leq 2$.
    \item \gls{phi}: The scaling parameter, controls the rate at which the correlation decays as the distance $u$ increases.
    \item $K_{kappa}(\cdot)$: A modified Bessel function of order $\kappa$.
\end{itemize} 

There are some challenges when implementing the Mat\'{e}rn covariance function in a modelling setting.  The parameters $\phi$ and $\kappa$ can not be estimated independently, and $\kappa$ is usually parameterized to the slightly more orthogonal $\alpha = 2\phi \sqrt{\kappa}$  \citep{diggle:07}.  In addition, it is typical to fix the smoothness, \gls{kappa} to make different models comparable. 

\subsubsection*{Anisotropy}
\label{subsubsec:anisotropy}
In the definition of the Mat\'{e}rn function (equation \ref{formula:Matern}) the distance $u$ is a scalar.  An anisotropic covariance function is dependent on the direction of $u$.  One context where this could happen is when the wind blows consistently in one direction.  In this case, there could be a faster change in conditions when moving perpendicular to the wind and so a larger variance for the same distance travelled.  

\subsubsection*{Non Stationary Trend}\label{subsubsec:nonstatiomary}
Calculation of the covariance function requires stationarity, a trend over the whole study region must be accounted for in modelling before calculating the covariance function.

\subsubsection*{Semi Variograms}
\label{subsubsec:semivariograms}
Plots of the variance (\gls{sigma}) as a function of the distance between sites (\gls{u}) are often used to examine the covariance function's goodness of fit. The semi-variance is usually plotted after binning distance measures. These plots are called semi-variograms and put three parameters with a physical interpretation on one plot:

\begin{itemize}
    \item \textbf{The Nugget:} The value of the semi variogram at \gls{u}$=0$.  The nugget is often interpreted as the variance that is inherent to each individual measurement.  This could come from the variance that exists at a spatial scale smaller than that resolved by a site, or the variance of individual monitors.
    \item  \textbf{The Sill:} The overall variance of the estimated surface.  The sill is the sum of the nugget and the variance of the spatial process.
    \item \textbf{Range ($r$):} The distance at which the covariance function between two sites is equal to the sill.  When the function is asymptotic, the range is often defined as the point where 10\% of the sill is reached.  
\end{itemize}

Examples of variograms for individual years of \ac{SOCAB} data can be seen in the next chapter's Figures \ref{fig:Variogram_1986}, \ref{fig:Variogram2013}, and \ref{fig:Variogram2019}.

\subsection{INLA}
\label{subsec:inla}
\ac{INLA} is a method to calculate posterior distributions of Gaussian fields without the computational burden of full \ac{MCMC} sampling. It approximates the Gaussian surface by projecting the observations to points on a mesh and then interpolating the whole surface using the basis functions of that mesh.

\subsubsection*{Mesh} \label{subsec:IntroMesh}
The mesh that is used to create the interpolations is an important part of \ac{INLA} modelling.  Its construction has a large impact on the resulting model.  It is made up of triangles that connect nodes and covers the study's domain.
The mesh has two regions, an inner and an outer portion.  The inner mesh covers the domain of interest, and the outer mesh is a coarser rim that reduces boundary effects.  

The triangles control the resolution of the model, with smaller triangles being more precise, but at the expense of increased computation time.  The computation time is proportional to the number of nodes in the mesh: $\propto n^{3/2}$.  The mesh construction has several tuning parameters that trade-off computational time and model fidelity.
    
\begin{itemize}
    \item \textbf{Minimum Edge Length}.  The minimum distance between two connected nodes. Larger triangle pixels reduce computational effort, but also reduce fidelity.  However, every edge should be shorter than the covariance's range.
    \item \textbf{Maximum edge length}.  The maximum distance between two nodes, it can take on one value within the study region and another value in the boundary region. 
    \item \textbf{Surplus Boundary distance}.  The \ac{INLA} algorithm has boundary effects.  Creating a buffer space between the boundary of the modelling to the region of statistical interest is a way to keep that from affecting the result.
    \item \textbf{Initial Vertices}. Permits using observation points as seeds for initial node location.
\end{itemize}

A simulation study by \citet{Righetto2020} provides the following guidelines for the creation of the mesh.  The shortest distance between points (cutoff value) has the highest impact.  Conditional on the cutoff, the maximum edge length of the inside domain has some impact.  The edge length in the outer domain is irrelevant.  They conclude by advising to keep the maximum edge length shorter than the spatial range and the cutoff value smaller than that.  Other guidelines are:
\begin{itemize}
    \item Avoid having multiple data points within the same triangle because they are part of the same basis function and therefore provide less information.
    \item Have a triangle or two between the boundary and any data point because \ac{INLA}'s algorithm has boundary effects.
\end{itemize}

\subsection{Modelling PM10 Field}\label{subsec:modelling PM10}
Following \cite{cameletti2011spatio} here is the description of the models used to describe the \ac{PM10} field from the observations made at all the sites in the network:
\begin{equation} \label{eq:obs_y}
    z(s_i,t) = x(s_i,t)\beta + y(s_i, t) + \epsilon(s_i,t).
\end{equation}

Equation \ref{eq:obs_y} describes the observed data $Z$ at location and time ${s,t}$.  It contains any covariates that explain gross trends in $\beta$, an autoregressive Gaussian field $Y$, and the (white noise) measurement error $\epsilon$ whose variance is the nugget.  The latent process is described by formula \ref{eq:xi_formula}, which shows how it is a series of Mat\'{e}rn correlation structures ($w(\cdot)$) linked by an \ac{AR}(1) process \citep{gomezGitBook, cameletti2011spatio}.

\begin{align} \label{eq:xi_formula}
    y(s_i, t) &= ay(s_i, t-1) + w(s_i,t) , &t>1, |a| < 1 \\
    y (s_i, 1) &\sim N(0, \frac{\sigma_w^{2}}{1-a^2}) , &|a| < 1 \nonumber
\end{align}

The Mat\'{e}rn covariance function is described in equation \ref{eq:cov_structure} and set to 0 when comparing different times.  This explicitly assumes that the time and space components of the model are separable. 

\begin{align} \label{eq:cov_structure}
    cov(w(s_i,t), w(s_j,t^{'})) &=
        \begin{cases} 
            0, & \text{if } t \neq t^{'} \\
            \sigma^2_w \gamma(u), & \text{otherwise}    
        \end{cases} \\
    & \text{Where } \gamma(u) \sim \text{Mat\'{e}rn, see \ref{formula:Matern}} \nonumber
\end{align} 

\subsubsection*{$\beta$ options}
\label{subsubsec:betaopts}
Several predictor effects of $\beta$ were considered, including site metadata and temporal trend.

Two general approaches to modelling the latent time effect were examined.  One with a fixed linear effect and one with a random walk.

In the first iteration of the model, the $\beta$ is an intercept and a linear slope due to time.  In this case, the $\beta$ follows equation \ref{eq:linear_beta}.
\begin{equation} \label{eq:linear_beta} 
    x(s,t_1)\beta_0 + x(s,t)\beta_1
\end{equation}  

In the second iteration, the model abandons the linear trend in favor of a random walk model, which is equivalent to a constrained spline with equidistant knots.  The $z(\cdot)\beta$ is therefore Equation \ref{eq:RW_beta}.
\begin{equation} \label{eq:RW_beta}
    z(s,t_1)\beta_0
\end{equation}

The random walk is implemented in \ac{INLA} as follows.  A prior is placed upon the difference between two years depending upon whether it is a random walk 1 (formula \ref{eq:RW1_formula}) or a random walk 2 (Formula \ref{eq:RW2_formula}).
\begin{equation} \label{eq:RW1_formula}
    f(k_{i + 1}) - f(k_i) \sim N(0,\tau), \quad i = 1, ..., K-1
\end{equation}
\begin{equation} \label{eq:RW2_formula}
    f(k_{i+1}) - 2f(k_i) + f(k_{i-1} \sim N(0, \tau), \quad i = 2, ..., K
\end{equation}

\subsubsection*{Priors} \label{subsubsec:Priors}
As a Bayesian process, \ac{INLA} requires a choice of priors for each parameter.  This includes at a minimum the  Mat\'{e}rn covariance structure and Gaussian noise.  Additional parameters could come from the time series, represented as an $AR(\cdot)$, or categorical covariates.

\Gls{PC_priors} are useful because they permit the integration of interpretable knowledge while also keeping complexity down.  They are weakly informative  \citep{fuglstad2017constructing, simpson2017penalising}.
The general idea behind the PC prior is to define a simpler version of the model that can be pushed towards a more complicated version with information.

\subsubsection*{PC Prior on the Mat\'{e}rn}
\label{subsubsec:pcprioronmatern}
The joint PC prior density for spatial range $r$ and marginal standard deviation $\sigma$ of the Mat\'{e}rn is as described in equation \ref{eq:Matern_PC_prior}.
\begin{equation} \label{eq:Matern_PC_prior}
    P(r, \sigma) = \frac{d(R)}{2 r^{-1-d/2}} e^{(-R r  ^{2 -d/2})} S e^{(-S \sigma)}
\end{equation}
$R$ and $S$ are user-defined hyperparameters that define extreme values on the distributions of the range and standard deviation respectively.

The prior is constructed to shrink the spatial effect to zero, as measured by Kullback Leibler divergence.  A model with no spatial effect (i.e. $\sigma = 0$) is the simplest model, and a model with constant spatial variance (i.e. $r = \infty$)  is simpler than a model with a spatial field \citep{fuglstad2017constructing}.

The R \ac{INLA} function \verb|inla.spde2.pcMatern()| makes the Mat\'{e}rn \ac{SPDE} model.  It uses the parameterized spatial scale parameter $\phi = \sqrt{8\kappa}/r$

The shape is defined through the user input $\alpha$ as follows:  $\kappa = \alpha -d/2$ with $\alpha$.  Where $d$ is the number of dimensions.  On the 2-dimensional surface, the differentiability $\kappa = \alpha -1$.


\subsubsection*{PC Prior on Random Walk}
\label{subsubsec:pconranwalk}
The random walk is used to detrend the time series by smoothing out changes between years, modelling the step between each observation as a Gaussian process with mean 0 and precision $\tau$.  It is equivalent to a spline.

A random walk of order one is made out of a Gaussian vector, 
$y = (y_1, ..., y_n)$,  where each step from observation $y_i$ to the next observation $y_{i+1}$ made by $\Delta y_i = y_i - y_{i-1} \sim N(0, \tau^{-1})$.  

The density of $y$ from its increments is \ref{eq:RW1_Gaussian}.
\begin{equation} \label{eq:RW1_Gaussian}
    \pi (Y|\tau) \propto \tau^{(n-1)/2} e^{-\tau/2 \Sigma (\Delta y_i)^2} 
\end{equation}

Then the \ac{PC} prior for the precision $\tau$ is defined in \ac{INLA} on $\theta = log(\tau)$ using $P(\theta > u) = \alpha$.  Where $u$ is a user-defined value and $\alpha$ is a user-defined probability.  For a Gaussian likelihood, a recommended setting for $u$ would be the empirical standard deviation of your data and $\alpha =0.01$  \citep{gomezGitBook}.


A random walk of order 2 is handled in the same way as RW1 except for the equation defining the steps in the random walk, which is different as seen in Equation \ref{eq:RW2_Gaussian}

\begin{equation} \label{eq:RW2_Gaussian}
    \Delta^2 y_i = y_1, - 2y_{i+1} + y_{i+2} \sim N(0,\tau^{-1})
\end{equation}
See 
https://inla.r-inla-download.org/r-inla.org/doc/latent/rw2.pdf
.

In both cases, we used the empirical standard deviation of the data as the informative component of the PC Prior on the precision of the random walk process.

\subsection{Preferential Sampling} \label{subsec:PreferentialSampling}
This section describes a way of modelling the sampling process and how to detect preferential sampling. 
According to 
\citet{diggle:07}, it is the result of using a joint probability distribution for a spatial field \gls{Y} that is not the same as the product of their marginal distributions, i.e. when $[Y, S] \neq [Y][S]$.  

Standard geostatistical methods assume that locations are not sampled preferentially \citep{diggle2010geostatistical}.  Using these methods when the assumptions fail, i.e. when sampling is done preferentially, may result in incorrect conclusions \citep{isaaks1988spatial}.  This issue is of concern, since numerous studies have used the \ac{SOCAB} network data to determine the impact of particulates on the region's inhabitants.

\subsubsection*{Modelling Sampling Procedures}
\'label{subsubsec:modellingsampling}
A common statistical model for the random location of sites is the log Gaussian Cox process.  This model models the probability distribution for the random number of sites in an area by using a Poisson process with intensity function, $\lambda(x)$.  
The resulting intensity function can then have various linear predictors, allowing for its adjustment in space and time.

Since site selection is an interplay of goals, budget, and site availability, and since the \ac{EPA} and \ac{SCAQMD} have criteria for site selection such as distance to road, vegetative cover, land availability, power sources and accessibility.  It is theoretically possible to define all the possible sites in the \ac{SOCAB}.  
 a comprehensive map of potential site locations.  
\citet{watson2019} suggests using either all sites in the network or a regular grid covering the study area as the population of possible sites.

\subsection{Detecting Preferential Sampling}
\label{subsec:prefsampdetection}
Several techniques have been proposed and these will now be reviewed.

\subsubsection*{Various proposals}
\label{subsubsec:various}
Schlather et al. (2004) tried two different MCMC tests.  The observed value of each test statistics were compared with values calculated from simulations using a conventional geostatistical model fitted to the data,  assuming that sampling is non-preferential \citep{schlather2004detecting}.  Guan and Afshartous (2007) partitioned the observations into non-overlapping clusters in subregions. They were then assumed to provide approximately independent replicates of the test statistics. This analysis required a large data set, so  their application used a sample size of 
$ n = 4358 $.  
\cite{diggle10} models  joint physical and sampling processes with shared spatio-temporal latent effects.

\subsubsection*{The Watson Method} \label{subsubsec:WatsonPrefSample}
\cite{watson2020} proposed a method, based on a simple premise, for detecting the preferential sampling of  sites for membership in a monitoring network. That premise states that the locations of monitoring sites within a preferentially sampled network will appear more clustered in regions recording above-average (or below-average) values of the measured response, than a network whose sites were situated for reasons independent of the response (e.g. by purely random sampling).
To be more explicit, suppose sites are picked from the population of all possible sites because they are expected to have high concentrations of an air pollutant. The result will be higher densities of sites in regions with high pollution concentrations.  This clustering effect suggests that a selected site in proximity to another site in the network will likely record a higher concentration of the pollutant than a site located far away from another site.
In other words, the  nearest neighbor distances will be negatively correlated with the observed concentrations at each site. 
This observation leads to Watson's test. It computes the non-parametric Spearman's Rho correlation between ranked nearest neighbor distances and the ranked pollutant levels of the sites. An unusual score, compared to that of simulated purely random networks, would then be an indication of preferential sampling.

Watson's test \cite{watson2019} is very general. First, it can be adjusted for real-world covariates believed to have been involved in the selection of sites to the network, and these may be correlated with the response (e.g. population density in a pollution network). Furthermore, additional realistic network restrictions (e.g. a maximum of monitoring sites allowed per jurisdiction) can be accounted for when simulating networks. Finally, an additional tuning parameter $k$ can greatly increase the power to detect PS. This tuning step proceeds as follows. At each site location, compute the average of the first \gls{k} nearest neighbor distances for $ k \geq 1$. Then, the rank correlation is computed between the ranked average distance and the response. The power of the test for a given \gls{k}, depends on how well it matches the cluster size of the actual network \citep{watson2020}. The test can be computed across a range of $k$ values, with care taken to account for the multiple comparisons. \citet{watson2020} showed that the test is highly conservative.   


The formal steps involved in  Watson's approach \cite{watson2020} to detect \ac{PS} can be summarized as follows:
\begin{enumerate}
    \item fit a point process model to the observed locations under the null hypothesis of no \ac{PS};
    \item simulate many sample networks of sites using that fitted point process;
    \item for each sampled network, estimate the value of the response at the simulated locations using a model that assumes no PS (e.g. kriging); 
    \item for each sampled network, compute the average of the \gls{k} nearest neighbor distances from the simulated locations; 
    \item compute the rank correlation test statistic for each sampled network;
    \item compare the observed vs. sampled test statistics.
\end{enumerate}



\subsubsection*{Assumptions Underlying the Method}
\label{subsubsec:underlyingassumps}
Here are the assumptions made for the test described in \cite{watson2020}.
\begin{itemize}
    \item The \ac{PS} is driven by some or all of the spatio-temporal latent effects $Y_{s,t}$.
    \item  All latent effects $Y_{s,t}$ driving the \ac{PS} are spatially ``smooth enough'' relative to both the size of the study region, $|S|$, and the number of locations chosen to sample the process
    \item The density of points within $S_t$ at space-time point $(s,t) \in (S \times T)$ depends monotonically on the values of the components of $Y_{s,t}$ driving the \ac{PS}.
\end{itemize}

Because of the monotonicity assumption, a negative correlation implies \ac{PS} for high-concentration sites.  Conversely, preferential sampling for low pollution will result in a positive correlation.

\subsection{Data Exploration}
\label{subsec:EDA}
This section describes:
\begin{itemize}
    \item the source of our case study's data;
    \item an inventory of the data;
    \item the scope of the analysis and how it was chosen;
    \item the preliminary statistics needed in preparation for more detailed modelling.
\end{itemize}

\subsubsection*{Data Sources}\label{subsec:datasources}
The data used for this report were obtained from several governmental sources. As mentioned in Section \ref{subsubsec:ReportingRequirements}, the \ac{EPA} makes all air quality monitoring data publicly available in summary files at \url{https://aqs.epa.gov/aqsweb/airdata/download_files.html}.  This data is provided in two formats.  First, as annual summaries of all pollutants and second, as daily summaries of individual pollutants.  The annual summaries contain statistics such as the mean, median, standard deviation, and various percentiles for all pollutants monitored in that year.  The daily summaries provide the observed values for a single pollutant for each day of the year.  Both time frames are .csv files.

The \ac{EPA} also publishes metadata for each monitoring site, giving information about the conditions at each site.  This includes the land use and the land urbanization, when the site started operation, and, if applicable, when it was terminated.

Metadata about the purpose of each site was also obtained from five-year reviews published in 2010 and 2015 by the \ac{SCAQMD}. In these yearly reviews, the \ac{SCAQMD} declare the scientific purpose of the sites and their expected pollutant level. This metadata is available in .pdf files, so we copied it into a .xlsx file by hand from several tables contained within the documents.

Finally, a shapefile describing the boundaries of the \ac{SOCAB} was obtained from the open data of the Southern California Association of Governments' GIS database.

\subsection{Data Choice}\label{subsec:datachoice}
With the many forms available of the data described above, one consistent domain had to be chosen for use for further analysis. 

\subsubsection*{Spatial Domain} \label{subsubsec:SpatialDomain}
An initial decision was made to constrain the study to a compact and relatively homogeneous region, to avoid possible confounding factors. It has the additional benefit of matching the pollutant process scale to the site location process scale.  Choices of site location are made by a single regulatory body, the \ac{SCAQMD}.  Restricting the spatial scale to the jurisdiction of one agency  ensures that any 
preferential sampling originates from one decision-making unit, instead of muddying the water with multiple agencies.   \cite{cressie2011statistics} describes how a change of support can result in Simpson's Paradox and recommends matching the scale of measurement to the scale of the question being investigated to avoid this risk.

The \ac{SOCAB} was chosen as the single jurisdiction because there is a long history of air pollution monitoring in the Los Angeles, LA area.  As discussed in Section \ref{subsec:labasin} the geographic and jurisdictional boundaries do not perfectly match.  So it was decided to constrain the study to the geographic extent of the \ac{SOCAB} instead of the jurisdictional extent of the \ac{SCAQMD}.  Crossing to another airshed results in a discontinuity in the covariance function.  While this discontinuity could have been modelled, the added complexity was deemed to outweigh the benefits of having the added information. The difference between the airshed and jurisdictional extent can be seen in \ref{fig:SCAQMD-jurisdiction}

Another choice that must be made is the map projection, as discussed in Section \ref{subsec:MapProjection}.  California recommends using the California (Teale) Albers projection in the CDFW Projection and Datum Guidelines 2018-02-24.  We used the Albers projection of the shape file describing the boundary of the \ac{SOCAB} for all future analyses.


\subsubsection*{Temporal Domain}
\label{subsubsec:tempdomain}
Section \ref{sec:introdairpollution} discussed how the longer monitoring time frame of \ac{PM10} is one of the main reasons for choosing \ac{PM10}.  In the \ac{SOCAB}, \ac{PM10} has been monitored from 1986 to the present day.  Every year was included in the analysis, although not all sites were  present in all years.   The times when sites provide data can be seen in fig \ref{fig:site_dotplot}.

The annual summary data was chosen over daily data for
computational efficiency, making the Bayesian estimation much faster by using a summary dataset 365 times smaller than the Daily data.  A second justification for using annual summary data lies in the nature of the process of interest, preferential site selection, which is based on annual summaries.

\subsubsection*{Other Decisions}
\label{subsubsec:otherdecisions}
Since exceptional events are generally quite rare (less than 4 per year), we saw little point in investigating differences between exceptional and unexceptional events. Thus, extreme events were excluded.

Many pollution monitors in the \ac{SOCAB} are not included in the \ac{EPA} data because they are not under its regulatory umbrella.  These could help 
produce a better model of the field, but they are not part of the sampling decision of the \ac{SCAQMD} and so were kept out of the study. That eliminated the additional effort required to find and include their data.

\subsection{Data Structure}
\label{subsec:datastructure}
Our focus thus turns to the files that give annual summaries.
The \ac{EPA} provides prepared annual summaries for each year in an individual .csv file describing all pollutants monitored at each reporting site.  Reports that cover the measurement timescale instead of the annual summary include data for only one pollutant.  

\subsubsection*{Data Rows} \label{subsubsec:DataRows}
\begin{figure}[ht]
    \centering
    \includegraphics[width = \textwidth]{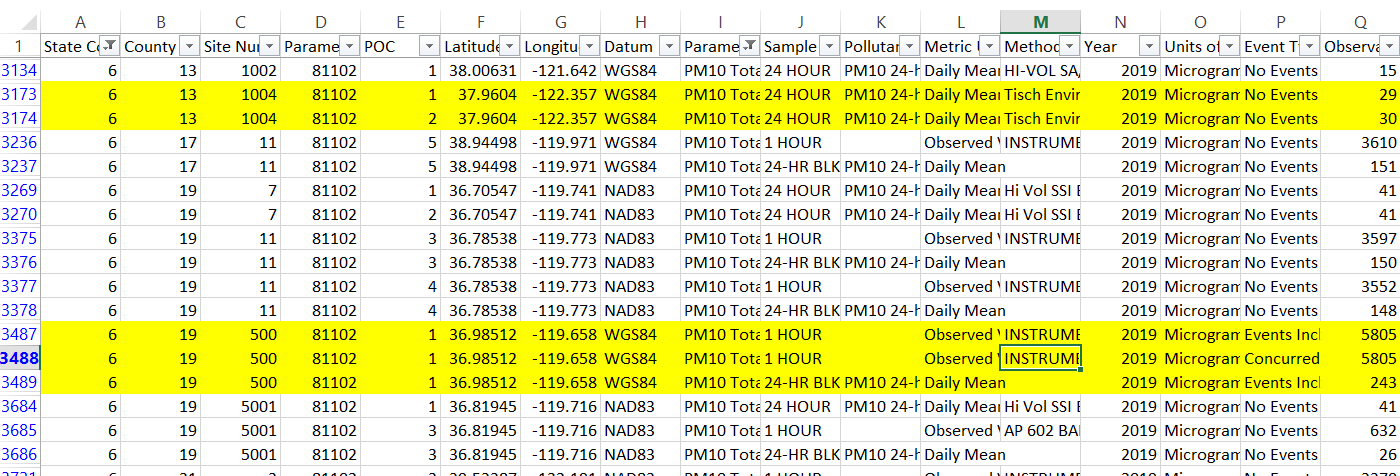}
    \caption{A selected screen capture showing part of the .csv data file from the EPA for the year 2019.  The first yellow highlight shows a site with two \ac{FRM} \ac{POC} and no extreme events.  The lower yellow highlight shows a site with a single \ac{FEM} sensor that has extreme events and has, in the third highlighted row, been smoothed from hourly averages to daily averages.  Columns A-C defines a unique site, Column E shows the \ac{POC}, Column J the sample averaging period, Column P whether extreme events are included}
\label{fig:Example_Raw_EPA_Data}
\end{figure}

In the annual summary files, each row represents a year's worth of data from a single source. There are several reasons for one site to have multiple rows for a single pollutant.  These include multiple instruments monitoring the same pollutant, and different data filters applied to the summarized data.  See Figure  \ref{fig:Example_Raw_EPA_Data} for an example of these complexities.

Multiple instruments measuring a pollutant are signified in the \ac{POC} column, with an integer value signifying each unique instrument.  Reasons to have multiple \ac{POC} include instruments being used for validation, to test new instrumentation, or for different monitoring purposes.  For example, an instrument monitoring \ac{EPA} compliance could be co-located with an instrument that is providing continuous monitoring.

Different rows for a single \ac{POC} occur when there are ``extreme events'' in the recording period.  These events are unusually high levels of pollution caused by processes outside the reporting agency's control, for example, forest fires.  When an extreme event occurs, one row will include the ``Exceptional Events'' and a second will exclude those events.  In the case of disagreement between the \ac{EPA} and local authorities (in our case the \ac{SCAQMD}), a third row, will present data including events considered to be extreme by the local authorities but not by the \ac{EPA}.

As a final reason for instruments having multiple rows, the sensor records data more frequently than the \ac{FRM}.  In this case, one row will have the raw recorded data and a second row will have the data after being averaged to the timescale of the \ac{FRM}.  

\subsubsection*{Data Columns}
\label{subsubsec:datacolumns}
Each of the \ac{EPA}'s annual summary files is a .csv with 55 columns, as described in table \ref{table:data_column_headers}.  The data columns include the Arithmetic Mean as well as the 99\ts{th}, 98\ts{th}, 95\ts{th}, 90\ts{th}, 75\ts{th}, 50\ts{th}, and 10\ts{th} Percentiles of all the observations made at that site by that instrument and for that pollutant.  For this report, the arithmetic mean 
was used as the statistic.  

In addition to the main data file, a .csv file containing metadata for each site was used. Its columns are outlined in table \ref{table:metadata_column_headers}.

\begin{landscape}
\begin{table}[ht]
    \centering
    \begin{tabular}{l | l | l | l | l }
     \hline
     Site Identification & Pollutant Metadata & Observation Metadata & Data & Other Metadata  \\
     \hline
     State Number & Parameter Name & Year & Arithmetic Mean & Local Site Name \\
     County Number & Sample Duration & Units & Arithmetic Standard Deviation & Address \\
     Site Number & Pollutant Standard & Event Type & 1\ts{st} Max Value & State Name \\
     Parameter Code & Metric Used & Observation Count & 1\ts{st} Max Date Time & City Name \\
     \ac{POC} & Method Name & Observation Percent & ... & CBSA Name \\
     Latitude & & Completeness Indicator & ... & Date of Last Change \\
     Longitude & & Valid Day Count & 4\ts{th} Max Value & \\
     Datum & & Required Day Count & 4\ts{ts} Max Date Time & \\
     & & Exceptional Data Count & 1\ts{st} Max Non Overlapping Value & \\
     & & Null Data Count & 1\ts{st} Max Non Overlapping Date Time & \\
     & & Primary Exceedance Count & 99\ts{th} Percentile & \\
     & & Secondary Exceedance Count & 98\ts{th} Percentile & \\
     & & Certification Indicator & 95\ts{th} Percentile & \\
     & & Number of Observations below MDL & 90\ts{th} Percentile & \\
     & & & 75\ts{th} Percentile & \\
     & & & 50\ts{th} Percentile & \\
     & & & 10\ts{th} Percentile & \\
     \hline
    \end{tabular}
    \caption{Names of all the column headers in a raw data file containing annual air pollution data from the \ac{EPA}. Column headers are organized by general category and then listed in order of appearance.  So State Number is the 1\ts{st} column, County Number is the 2\ts{nd} and Parameter Name is the 9\ts{th}.  The exact definitions of each column can be found at the \ac{EPA} website: \url{https://aqs.epa.gov/aqsweb/airdata/FileFormats.html\#_content_3}}
    \label{table:data_column_headers}
\end{table}
\end{landscape}

\begin{table}[ht]
    \centering
    \begin{tabular}{l|l|l|l}
    \hline
    State Code & Latitude & First Year of Data &  Networks\\
    County Code & Longitude & Last Sample Date &  Reporting Agency\\
    Site Number & Datum & Monitor Type & PQAO\\
    Parameter Code & & & Collecting Agency\\
    Parameter Name & & & Exclusions\\
    POC & & & Monitoring Objective\\
     & & & Last Method Code \\
    Local Site Name & & & Last Method \\
    Address & & & NAAQS Primary Monitor \\
    State Name & & & QA Primary Monitor \\
    County Name & & & \\
    City Name & & & \\
    CBSA Name & & & \\
    Tribe Name & & Extraction Date & \\
    \end{tabular}
    \caption{Names of column headers in metadata file describing the particulars of each site.  Full details on the \ac{EPA} website \url{https://aqs.epa.gov/aqsweb/airdata/FileFormats.html\#_format_2}}
 \label{table:metadata_column_headers}
\end{table}
   
\subsection{Data Description}
\label{subsec:datadescripion}
\begin{figure}[ht]
    \centering
    \includegraphics[width = \textwidth]{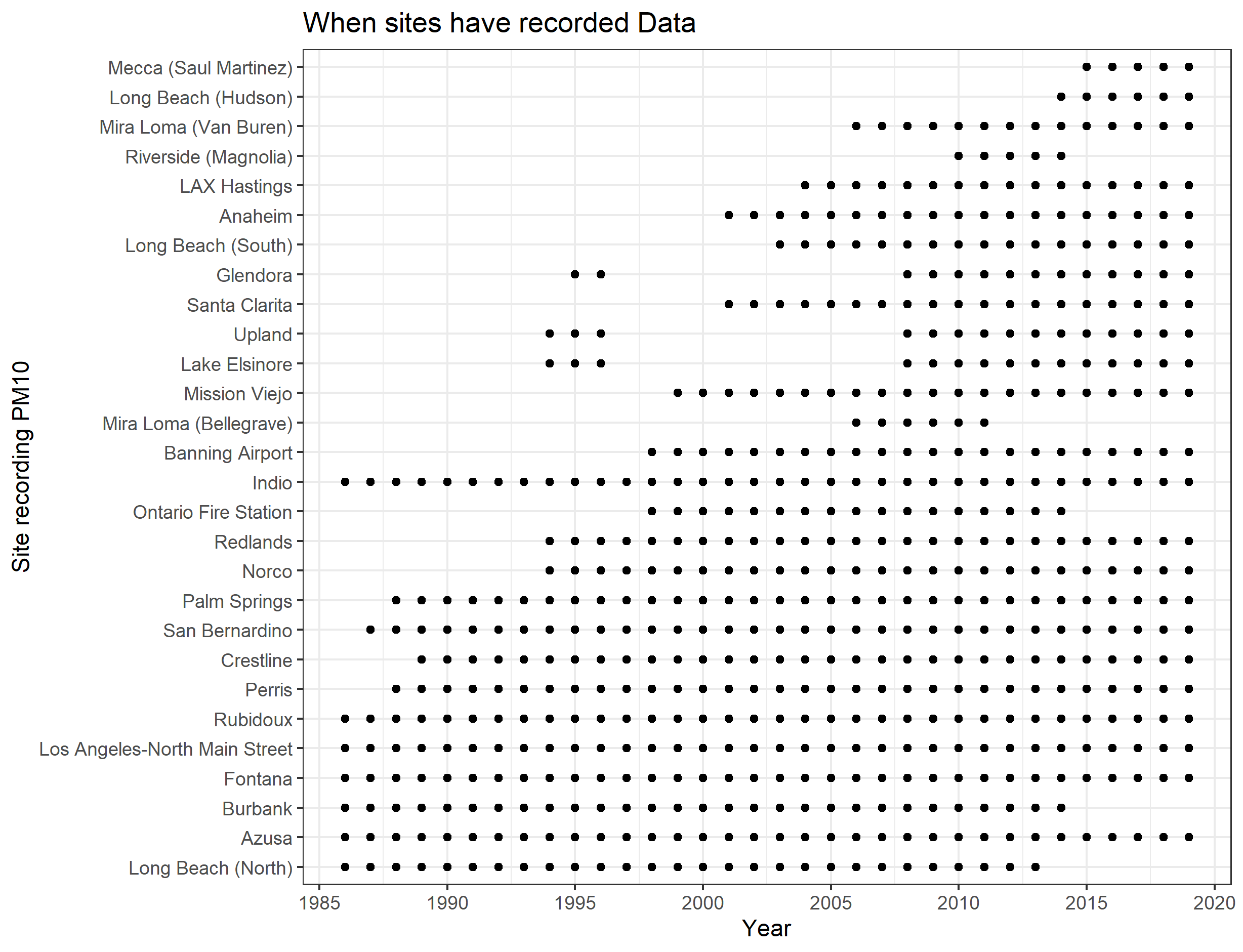}
    \caption{This figure shows how the network developed over time.  We can see that sites are generally added to the network, that 5 sites have been removed, and that 5 sites started in 1986.  A handful of sites has the unusual behavior of being taken offline and then removed.  These are sites that only had \ac{FRM} monitoring, no \ac{FEM}.}
    \label{fig:site_dotplot}
\end{figure}

\begin{figure}[ht]
    \centering
    \includegraphics[width = \textwidth]{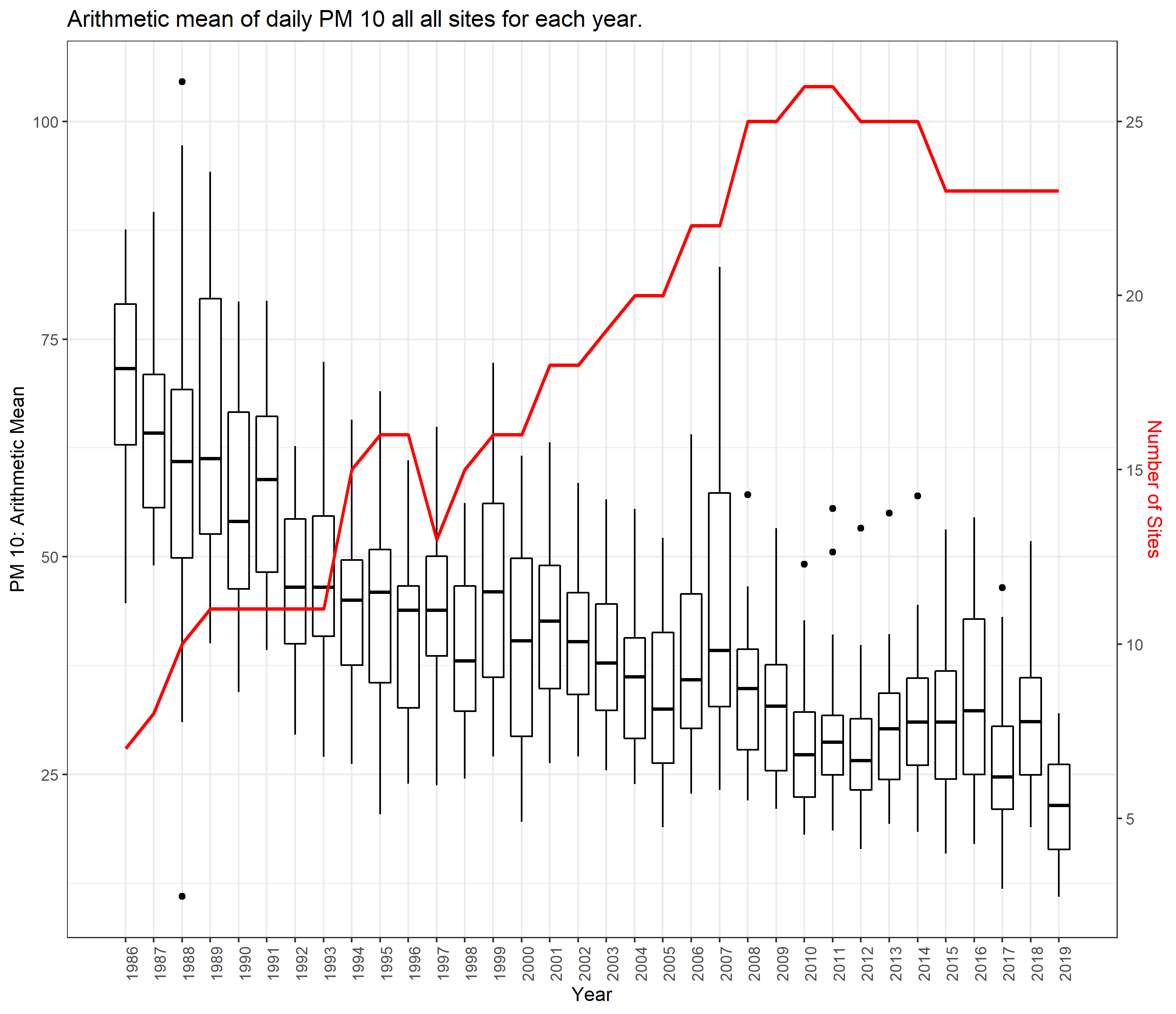}
    \caption{Each box shows the general pattern of mean 
    PM$_{10}$
    observed each year.  The red line shows how the number of active sites recording in the network (and therefore the number of observations feeding into each box) increases to the present day.}
    \label{fig:site_trend-ArthmM_NumSites}
\end{figure} 

From 1986 to 2019 there are 28 unique sites monitoring \ac{PM10} within the \ac{SOCAB}. Figure \ref{fig:site_dotplot} shows when sites are included in the network and when they are removed.   Three sites (Glendora, Upland and Lake Elsinore) stop being recorded in 1997 and then restart in 2008. Why this happens, is unclear, but it only occurs in sites with only continuous \ac{FEM} monitoring (as opposed to scheduled \ac{FEM} sampling).  The timeline coincides with regulatory changes to standards, but we have been unable to learn the reason for the sites' discontinuation and restart.

Figure \ref{fig:site_trend-ArthmM_NumSites} summarizes the yearly mean \ac{PM10} and shows how, over time, the number of sites has increased while the overall concentration in the area has gone down. The trend in \ac{PM10} will be examined later. 

\subsubsection*{Network Trends}
\label{subsubsec:networktrends}
Here are several plots showing traces of each site compared to the rest of the network.  If the network is being biased consistently over time towards a certain goal, we would expect to see some difference between sites kept in the network vs those removed from it.  

\begin{figure}[ht]
    \centering
    \includegraphics[width = \textwidth]{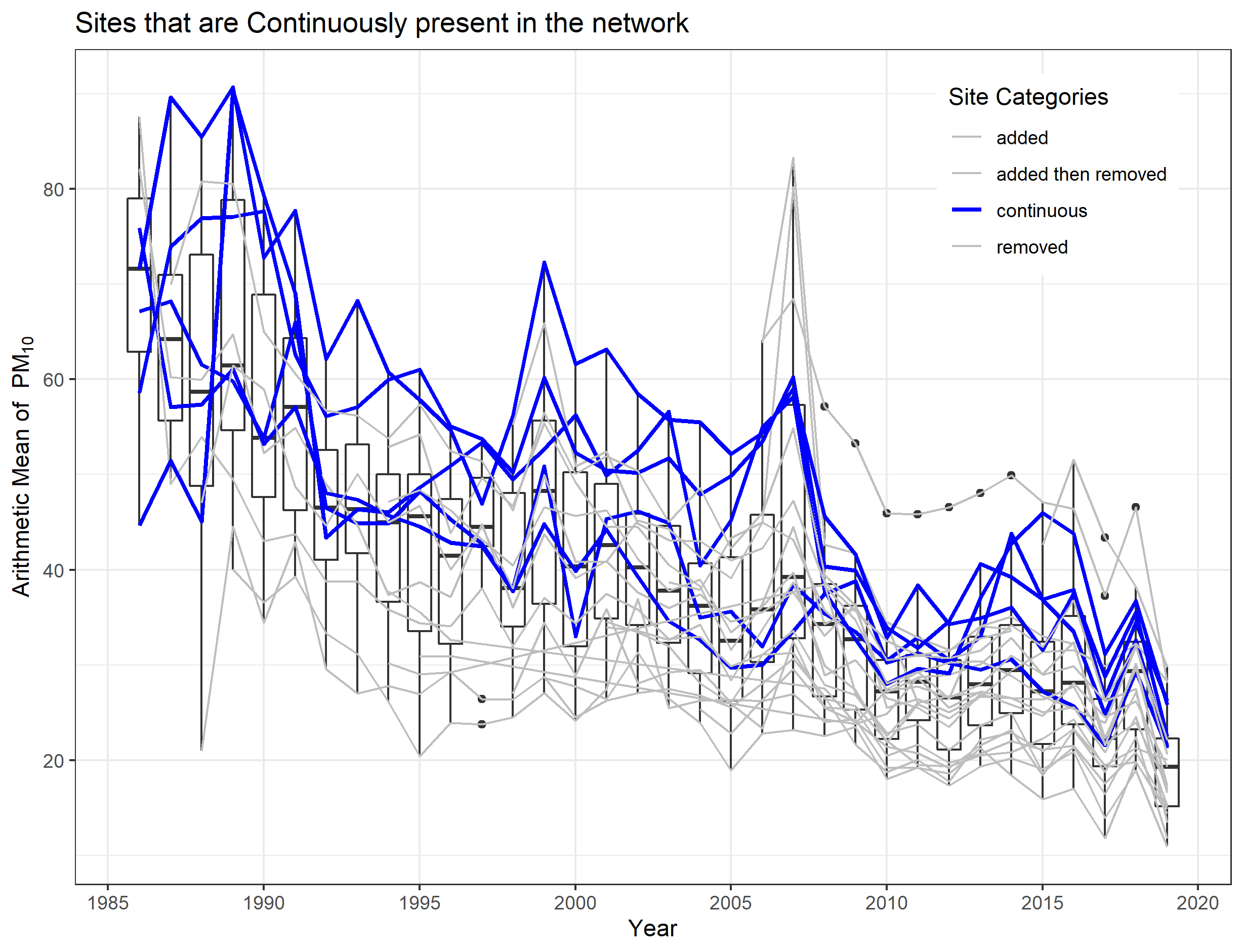}
    \caption{This figure highlights the traces of all the sites that started in the network and have not been removed.  We can see that in later years the lower part of the boxes is not covered by these traces, indicating some possibility of preferential sampling.  In the case when a site had multiple \ac{POC} in a year, the value at the trace is the mean of all \ac{POC} at that site for that year.}
    \label{fig:site_timing_trace-Continuous}
\end{figure}
Figure \ref{fig:site_timing_trace-Continuous} shows the sites that are active from 1986 to the present day, labelled as continuously present.  The continuously present sites tend to be above the mean in more recent years.  If sites maintain their relative position in the overall distribution, this suggests the early years are biased towards higher sites.

\begin{figure}[ht]
    \centering
    \includegraphics[width = \textwidth]{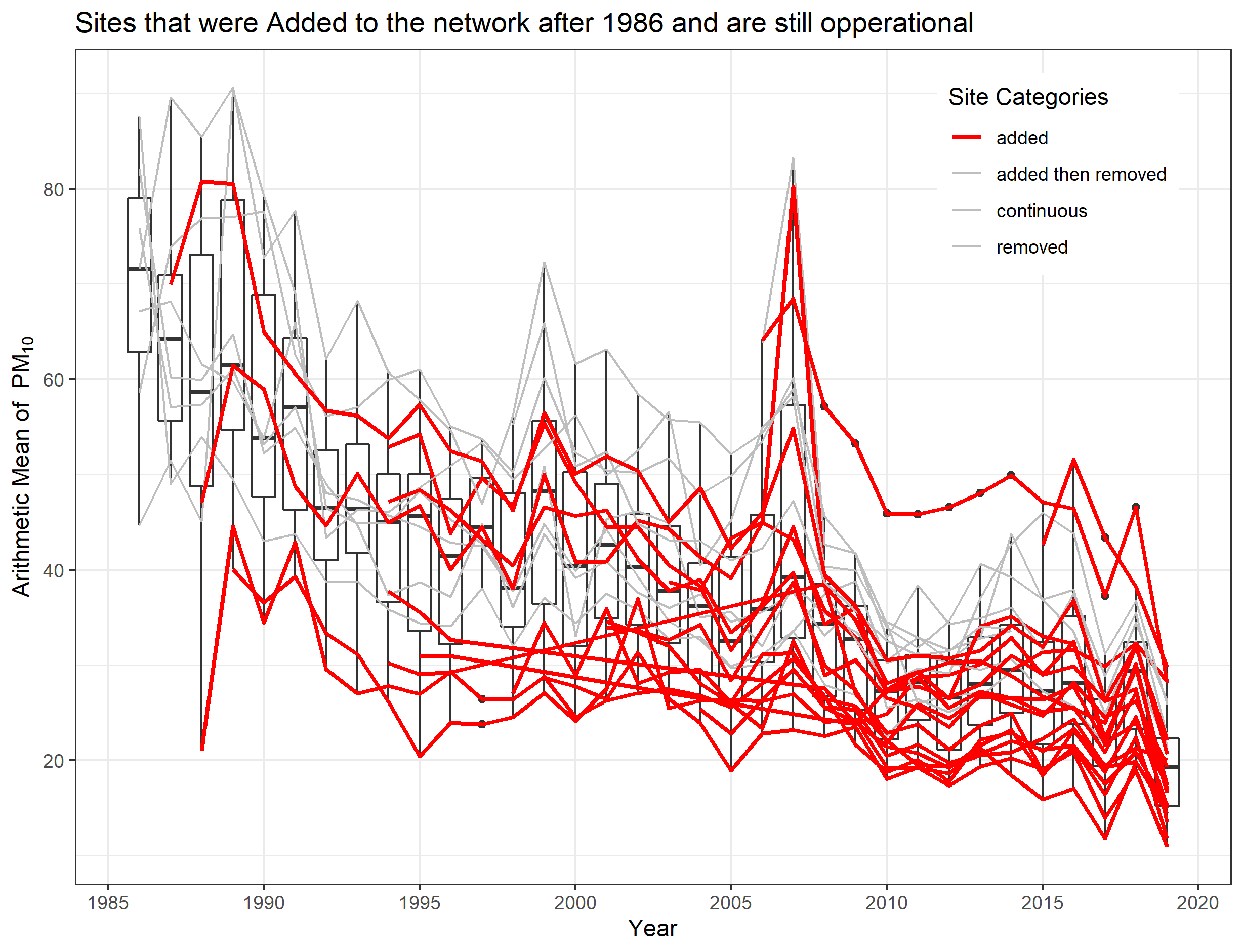}
    \caption{Here are highlights of all the sites that were added after 1986.  It appears that there are more traces in the lower portion of the boxes, which is the required complement of the pattern shown in the previous figure.  Again, in the case when a site had multiple \ac{POC} in a year, the value at the trace is the mean of all \ac{POC} at that site for that year.}
    \label{fig:site_timing_trace-Added}
\end{figure}

Figure     \ref{fig:site_timing_trace-Added} shows how sites added to the network tend to fill out the bottom half of the distribution in later years. This is the inverse of the idea demonstrated in the previous figure (Figure  \ref{fig:site_timing_trace-Continuous}).

\begin{figure}[ht]
    \centering
    \includegraphics[width = \textwidth]{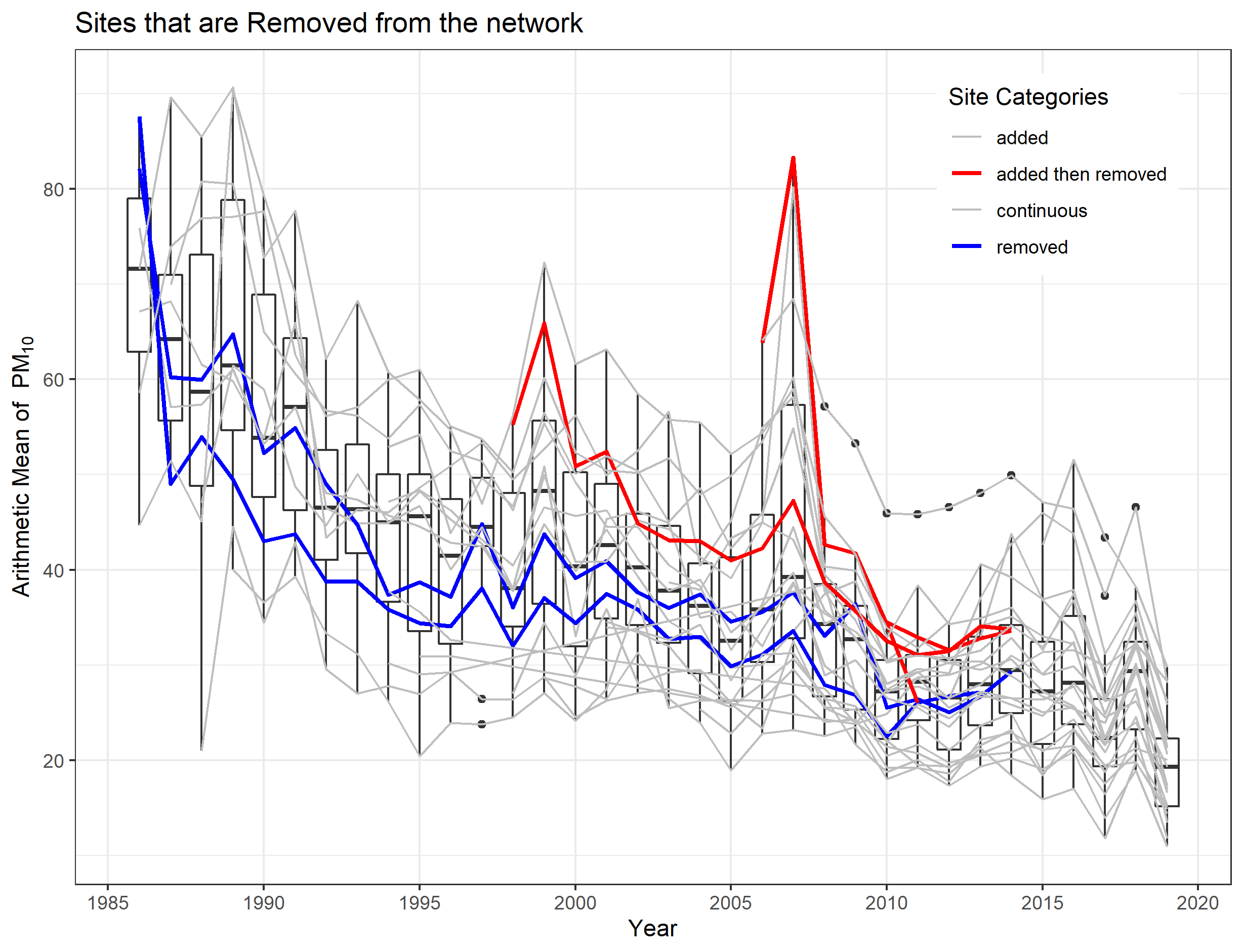}
    \caption{This figure highlights the sites that were dropped from the network.  In Blue are two sites that started in 1986 but were then removed.  In Red are three sites that were added to the network after 1986 and have been since removed.  }
    \label{fig:site_timing_trace-Removed}
\end{figure}
Figure  \ref{fig:site_timing_trace-Removed} highlights the five sites that were dropped from the network.  Two of them were part of the original network in 1986, and tend to fall below the yearly mean.   The other three were added to the network and tend to be above the yearly mean.  This behavior is opposite to that seen in the plots of continuous sites and site that were added and kept.  The sites that start in the network in 1986 and remain throughout tend to be above the mean, but the two that were removed are below the mean.  The sites that were added to the network tend (to a lesser degree) to be below the mean, but those that were added and then removed tend to be above it.

\subsubsection*{Site Location}
\label{subsubsec:sitelocn}
\begin{figure}[ht]
    \centering
    \includegraphics[width = \textwidth]{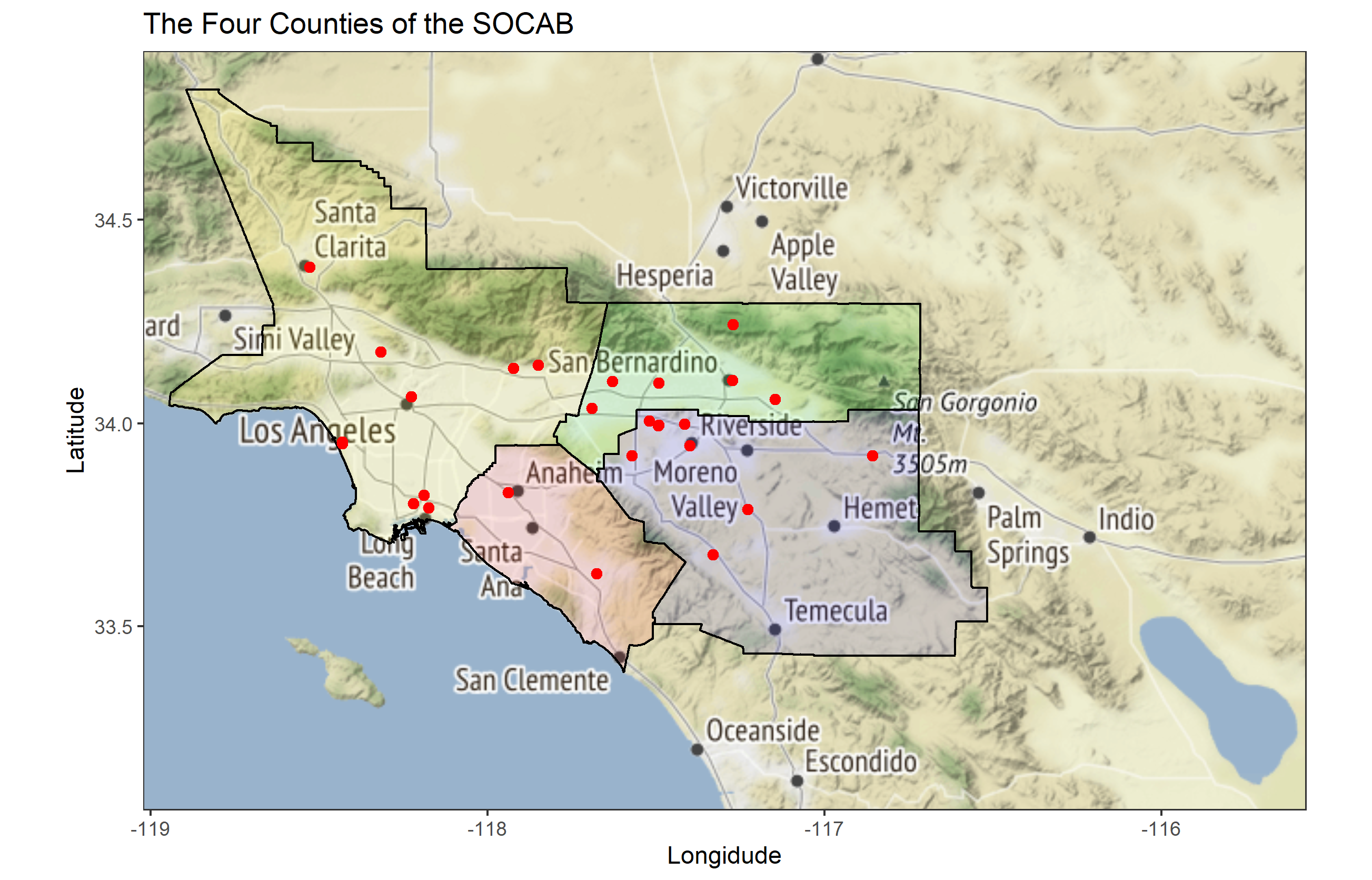}
    \caption{This map shows the \ac{SOCAB} in relation to the Los Angeles region.  It includes 4 counties, Orange County (Pink), Los Angeles County (Yellow), Riverside County (Blue), and San Bernardino County (Green).  Only Orange County is entirely part of the \ac{SOCAB} and the other counties extend into other air basins.  Monitoring sites that are included in this study are red dots.}
    \label{fig:SOCAB_counties}
\end{figure}

Figure  \ref{fig:SOCAB_counties} shows the location of sites in the \ac{SOCAB}.  Note that they are not all present simultaneously.  
It appears that some sites replace others.  For example, Long Beach (North) and Long Beach (Hudson) are very close to each other and one stops while the other starts the next year.  This lack of Independence in site selection was ignored.

\subsubsection*{Site Metadata}
\label{subsubsec:sitemetadata}
The \ac{EPA} and \ac{SCAQMD} record additional descriptive information about each site.  This includes:  

\begin{itemize}
    \item \textbf{Land Use:} Figure  \ref{fig:SOCAB_metadata_Site_Land_use} shows the Land Use, describing whether the site is residential, commercial, industrial, or agricultural.  Most (17) sites are residential, 3 sites are Industrial, 6 sites are Commercial, 1 is Agricultural, and Indio (a site that starts in 1986 and never dropped) has no stated land use.

    \item \textbf{Location Setting:} Figure  \ref{fig:SOCAB_metadata_Site_Status} shows the Location Setting, describing whether the site is Urban (7 sites), Suburban (19 sites), or Rural (2 sites).  Interestingly, the rural and suburban bracket the urban making a sandwich.  

      \item \textbf{Monitoring Objective:} Figure  \ref{fig:SOCAB_metadata_Site_Type} shows the Monitoring Objective, describing what the site is recording.  Options include Extreme Downwind, Highest Concentration, Other, Population Exposure, Unknown, and Upwind Background.   
\end{itemize}

\subsubsection*{Monitoring Purposes}
\label{subsubsec:purposes}
Every site has at least one monitoring purpose.  Some sites have mismatched site categories between the two 5 year reports.  But we have not been able to determine if these are typos (other, more clear-cut typos have been found in the report) or if different monitors at the same location have a different category for some reason.  Here are the sites that mismatch:

\begin{itemize}
    \item San Bernardino (\#060719004) is categorized as High Concentration except for the 2010 continuous monitor which is Population Exposure
    \item LAX Hastings (\#060375005) is categorized as Population exposure and Population Exposure / Background in 2015
    \item Palm Springs (\#060655001) is categorized as High Concentration except for 2010s FEM monitor which is Population Exposure.
\end{itemize}

The site category was only High Concentration and Population Exposure categories (LAX Hastings being the one exception, being both PE and HC in 2015 for the FEM sensors)

\begin{table}[ht]
    \centering
    \begin{tabular}{p{3cm}|p{2cm}||p{3cm}|p{2cm}}
        \multicolumn{2}{c|| }{\small 2010 (Monitoring Purpose)} & \multicolumn{2}{c}{\small 2015 (Monitoring Purpose)}   \\
        \hline  
        Long Description & { \small Two-Letter Code} & Long Description & { \small Two-Letter Code} \\
        \hline
        { \small High Concentration} & HC & { \small Highest Concentration} & HC \\
        {\small Representative Concentration}  & RC & { \small Population Exposure} & PE  \\
        { \small Impact}  & IM & { \small Source Orientated (impact) }& IM \\
        { \small Background} & BK & { \small General Background  }& BK \\
    \end{tabular}
    \caption{Comparison of Terminology describing the category each site is placed into for the 5-year summaries.  Seen in table 2 of the two summaries.}
    \label{tab:site_cat_5yr_summary}
\end{table}

\begin{table}[ht]
    \centering
    \begin{tabular}{p{3cm}|p{2cm}||p{3cm}|p{2cm}}
        \multicolumn{2}{c|| }{2010 (Monitoring Purpose)} & \multicolumn{2}{c}{2015 (Monitoring Purpose)}   \\
        \hline  
        Long Description & Two-Letter Code & Long Description & Two-Letter Code \\
        \hline
        Background Level & BK & & BK \\
        High Concentration & HC & & HC \\
        Pollutant Transport  & TP & & TP \\
        Pollutant Exposure & EX & & EX \\
        Source Impact & SO & & SO \\
        Representative Concentration & RC & & RC \\
        Special Purpose Monitoring   & SPM & -- & -- \\
        Trend Analysis & TR & & TR \\
        Site Comparison & CP & & CP \\
        -- & -- & Real-time Monitoring and Reporting & RM \\
        -- & -- & Collocated & CO \\
    \end{tabular}
    \caption{Comparison of terminology describing the purpose of each site, as described in table 3 of the 2010 and 2015 5-year summaries.  Not all categories exist in both years, when one isn't present a--is placed in the year for which it is absent.}
    \label{tab:monitoring_purpose_5yr_summary}
\end{table}

\subsubsection*{\ac{POC}} \label{seq:POC}
\begin{figure}[ht]
    \centering
    \includegraphics[width = \textwidth]{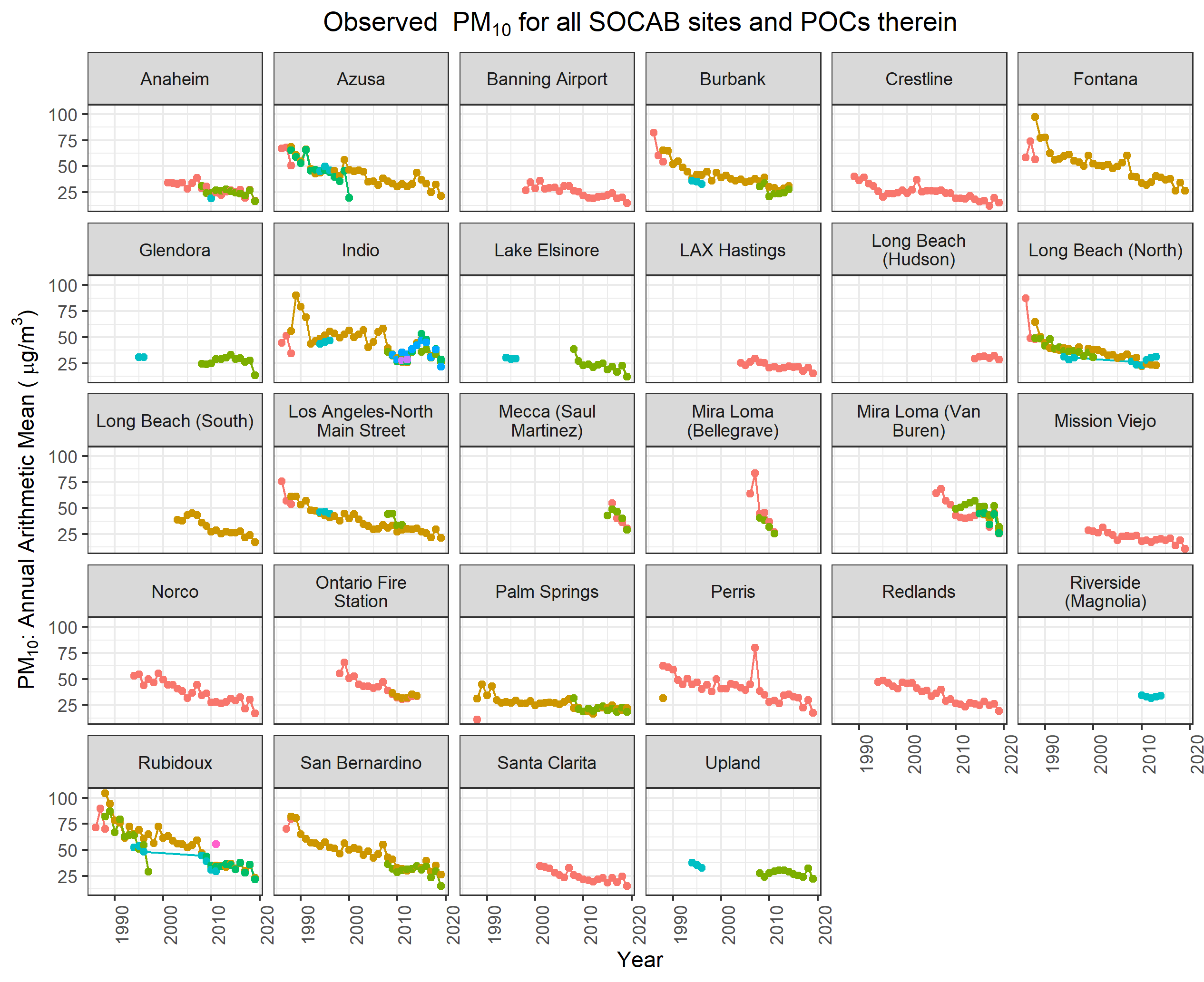}
    \caption{Traces showing every sensor (i.e. \ac{POC}) recorded at each site.  Colors are only to distinguish different \ac{POC} and have no meaning between sites.  Notice Azusa, Burbank, Fontana, Indio, Long Beach, Los Angeles-North, Palm Springs, Perris, Rubidoux, and San Bernardino all have a \ac{POC} that stops being used in 1988 (generally colored red) and is replaced by another \ac{POC} (generally brown) that consistently has a higher concentration of \ac{PM10}.  Rubidoux and Long Beach (North) both have a \ac{POC} (teal) that was discontinued in 1996 and then reestablished in 2007. These are continuous monitoring \ac{FEM} along with Burbank, Glendora, Indio, Lake Elsinore, Los Angeles - North, and Upland which also have a teal sensor discontinued and eventually replaced by a green sensor.}
    \label{fig:POC_all_trace}
\end{figure}

As discussed in Section \ref{subsubsec:DataRows} each site has at least one sensor, but many have more than one.  These could be an \ac{FEM} and a \ac{FRM} monitor, instrument changes, or collocation for trials or calibration.  This redundancy provides an interesting way to estimate the nugget effect, which is sensor uncertainty.  Figure  \ref{fig:POC_all_trace} shows traces for every \ac{POC} at each site.   An example of the instrument changes seems to happen in 1988 when all the sites that had been in operation before 1988 received at least one other \ac{POC} that returned a consistently higher reading of \ac{PM10} and the \ac{POC} that was in operation before 1988 was discontinued.  These sites are Azusa, Burbank, Fontana, Indio, Long Beach (North), Los Angeles-North, Palm Springs, Perris, Rubidoux, and San Bernardino.  

\subsection{Exploratory Data Analysis}
\label{subsec:eda}
Before applying complex INLA modelling to the question of preferential sampling, an exploratory data analysis was carried out.  This analysis provided a sanity check for data acquisition and cleaning, highlight unusual patterns, and provide preliminary suggestions for preferential sampling.

\subsubsection*{Data Transformation}
\label{subsubsec:datatrans}
\cite{ott1990} suggests that particulate counts follow a log-normal distribution due to the physical processes that make the particulates.  Taking the log of the raw counts helps to stabilize the variance.  This was done by \cite{cameletti2011spatio}  in their similar work examining \ac{PM10} concentrations in Italy.

Since log scores are unitless, the raw \ac{PM10} data were normalized to the mean \ac{PM10} in 1986 and the log of that ratio was taken, as described in equation \ref{eq:log_transform}.  In equation \ref{eq:log_transform} $t$ is the year, $s$ is a unique site, $Z$ is the transformed data used for future calculations and modelling, and $PM10_{t,s}$ is the raw data for the year $t$ at site $s$.  Finally, $PM10_{1986,\cdot}$ is the mean of the all observed \ac{PM10} in 1986, 69.65397 $\mu g/m^3$.  This log normalized value of the \ac{PM10} is used for all future analyses.
\begin{equation}
    Z_{t,s} = log(PM10_{t,s}/ PM10_{1986,\cdot})
\end{equation} \label{eq:log_transform}

\subsubsection*{Temporal Effects}
\label{subsubsec:tempeffects}
An initial examination of figure \ref{fig:site_trend-ArthmM_NumSites} shows the concentration of \ac{PM10} decreasing over time, as expected from the known history of particulate matter.  This could be modelled as either a fixed or a random effect.  Both options were examined, and the results are described in this report.  In this preliminary investigation, all spatial correlations
were ignored.

After preliminary modelling, our focus is on describing preferential sampling, not the decrease in \ac{PM10} overtime. Modelling the small perturbations with a Random Walk seemed likely to give a better understanding of what the sites are doing and so a better picture of any possible preferential sampling.  

\subsubsection*{Fixed Temporal effects}
\label{subsubsec:fixedtempeffects}
The first way to model the temporal trend is with a linear model.   While the general trend in means of figure \ref{fig:site_trend-ArthmM_NumSites} appears mostly 1st order to the eye, \cite{shaddick2014case} used a quadratic function to model the decreasing temporal trend in black smoke in the UK. That led us to  investigate both first- and second-order models for the trend over time using equations \ref{eq:linearTime} and \ref{eq:quadraticTime} respectively:
\begin{subequations}
\begin{align}
    Z_{t,\cdot} &= \beta_0 + \beta_1t + \epsilon \label{eq:linearTime} \\ 
    Z_{t,\cdot} &= \beta_0 + \beta_1t + \beta_2t^2 + \epsilon \label{eq:quadraticTime}.
\end{align}
\end{subequations}

\begin{figure}[ht]
    \centering
    \includegraphics[width = \textwidth]{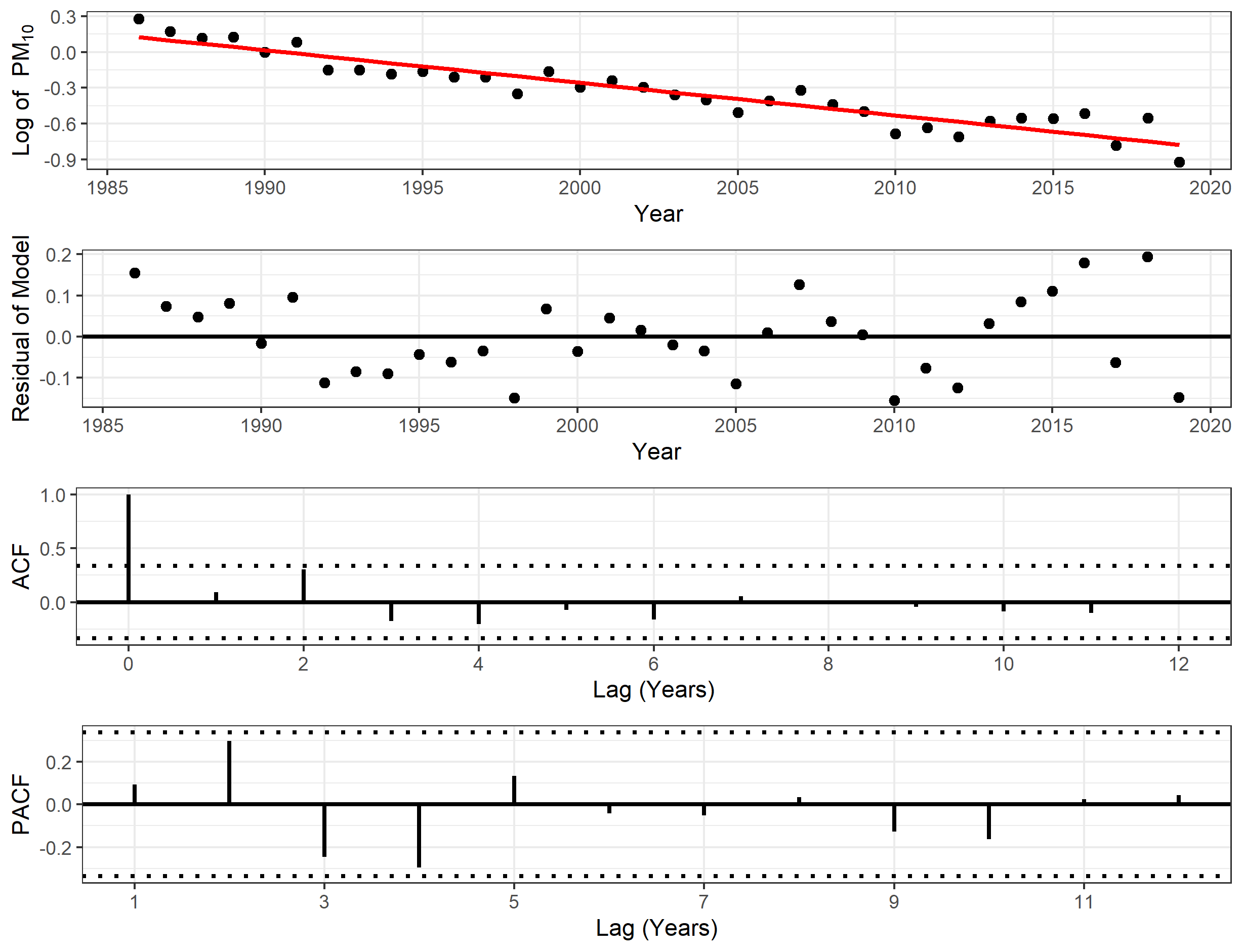}
    \caption{Results of a linear fit to a log transformation of the \ac{PM10} data.  The line is the model $Z_{t,s} \sim Year$, and has a slope of -0.027 and adjusted $R^2$ of 0.88.  The \ac{ACF} and \ac{PACF} suggest that the process remaining is probably white noise.  Dots are the Median of the arithmetic mean of each year.  The dotted line shows a confidence limit of $qnorm((1 + ci)/2)/sqrt(n)$, R's default for ACF and PACF.  This stems from Chatfield's Analysis of Time Series (1980), in which he describes how the variance of the autocorrelation coefficient at lag k, is normally distributed at the limit, and that $Var(rk) \sim 1/N$ (where N is the number of observations).}
    \label{fig:explore_ts_exp_fit}
\end{figure}

\begin{figure}[ht]
    \centering
    \includegraphics[width = \textwidth]{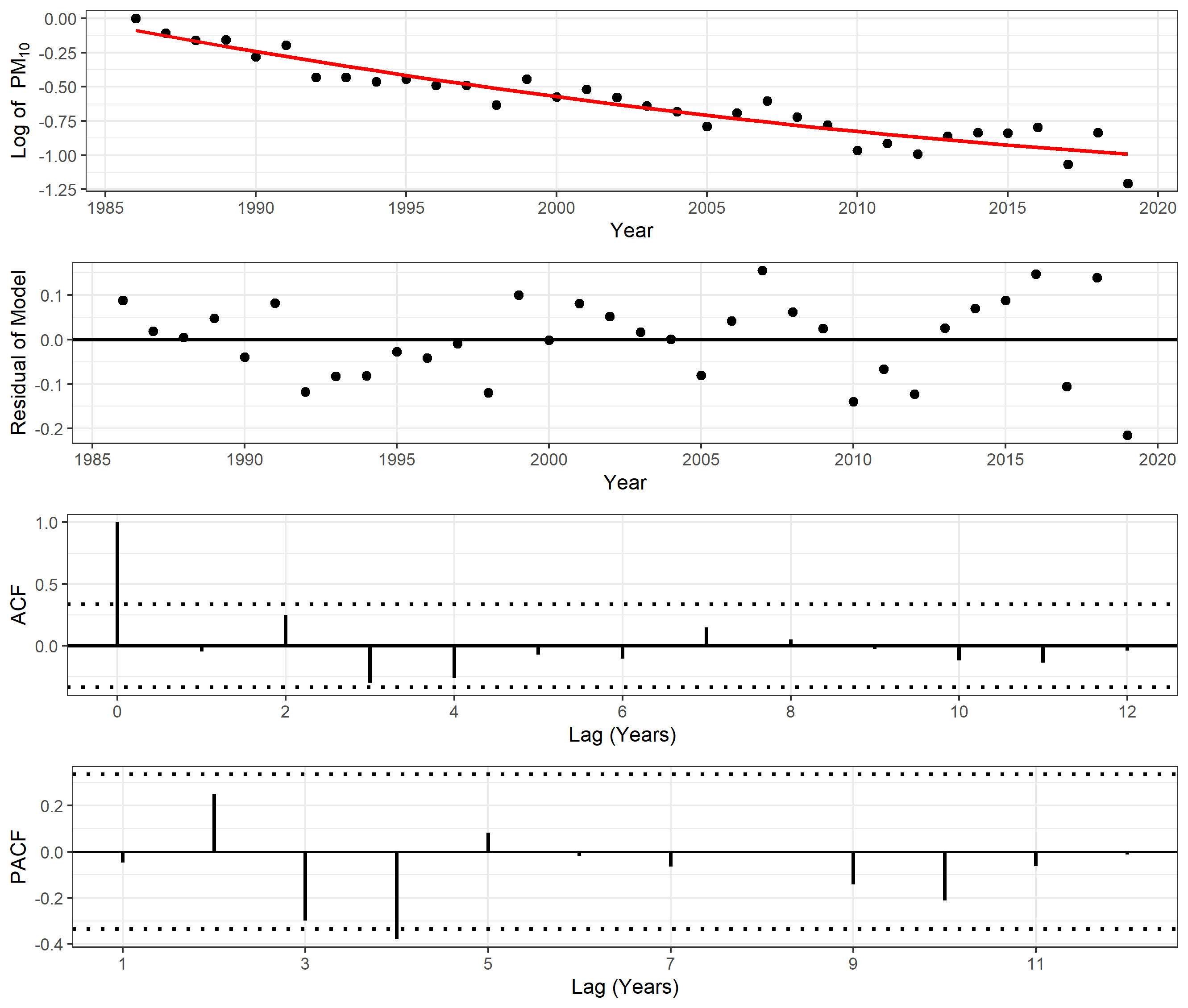}
    \caption{Results of a linear fit to the \ac{PM10} data.  The line is the model $Z_{t,s} \sim \beta_1 Year+ \beta_2 Year^2$ and has coefficients $\beta_1 = -1.56$ and $\beta_2 = 0.00038$ and adjusted $R^2$ of  0.90.  The \ac{ACF} and \ac{PACF} plots of the residuals suggest there might be a MA(1) or AR(1) process.  Dots are the Median of the arithmetic mean of each year. The dotted line shows a confidence limit of $qnorm((1 + ci)/2)/sqrt(n)$, R's default for plots of ACF and PACF. }
    \label{fig:explore_ts_exp_quad_fit}
\end{figure}

Figures  \ref{fig:explore_ts_exp_fit} and \ref{fig:explore_ts_exp_quad_fit} show the results of first and second-order models being fitted to the log normalized \ac{PM10} data.   An \ac{ANOVA} comparing the two models ($anova(model.lm.log, model.lm.log.quad)$) suggests a small but significant improvement of the second order model ($Pr(>F) = 0.0492$) compared to the 1st order. 

Both models do a good job of whitening the overall residual of the annual mean, as seen in the \ac{ACF} and \ac{PACF} plots.   This suggests there might not be an autocorrelation AR(1) process, unlike the model used by \cite{cameletti2011spatio}.

\subsubsection*{Temporal Trend as a Random Walk} \label{subsubsec:RWexploration}
An alternative approach to accounting for a broad-scale trend in time is to use a random variable.  The logic supporting the use of a random variable over a fixed effect lies in our lack of interest in estimating the annual decrease, and a model that can be more adaptive to yearly fluctuations can fit closer and give a better understanding of the field at the time.  Two options are a random walk and a 1-dimensional Mat\'{e}rn interpolation with \ac{INLA}.  Because the random walk is simpler, that was chosen.

\begin{figure}[ht]
    \centering
    \includegraphics[width = \textwidth]{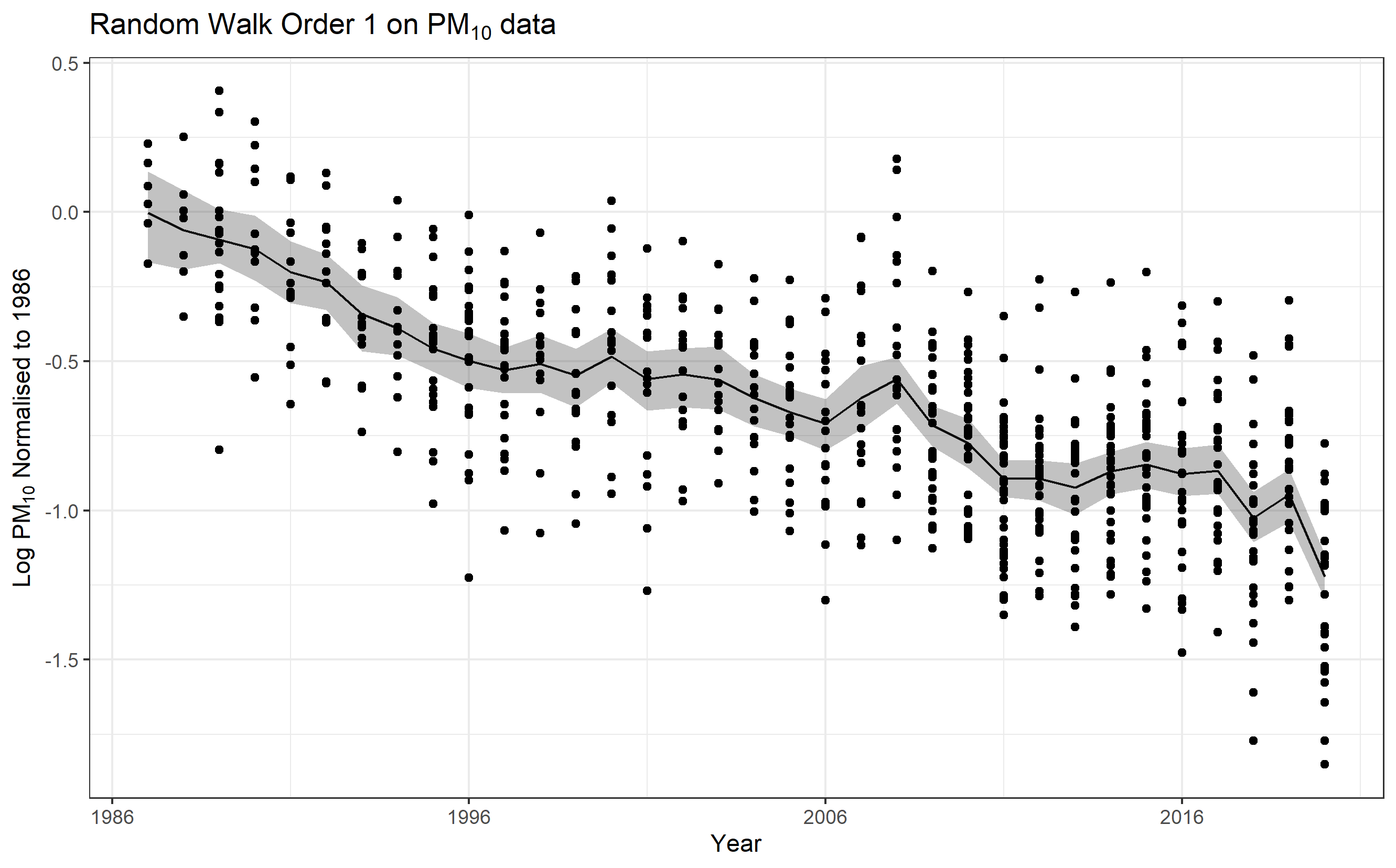}
    \caption{First order Random Walk smoothing sets a prior on the difference between each observed value $f(\kappa_i)$.  Like so: $f(\kappa_{i+1}) - f(\kappa_i) \Tilde{} N(0,\tau)$ }
    \label{fig:Random_Walk1}
\end{figure}

\begin{figure}[ht]
    \centering
    \includegraphics[width = \textwidth]{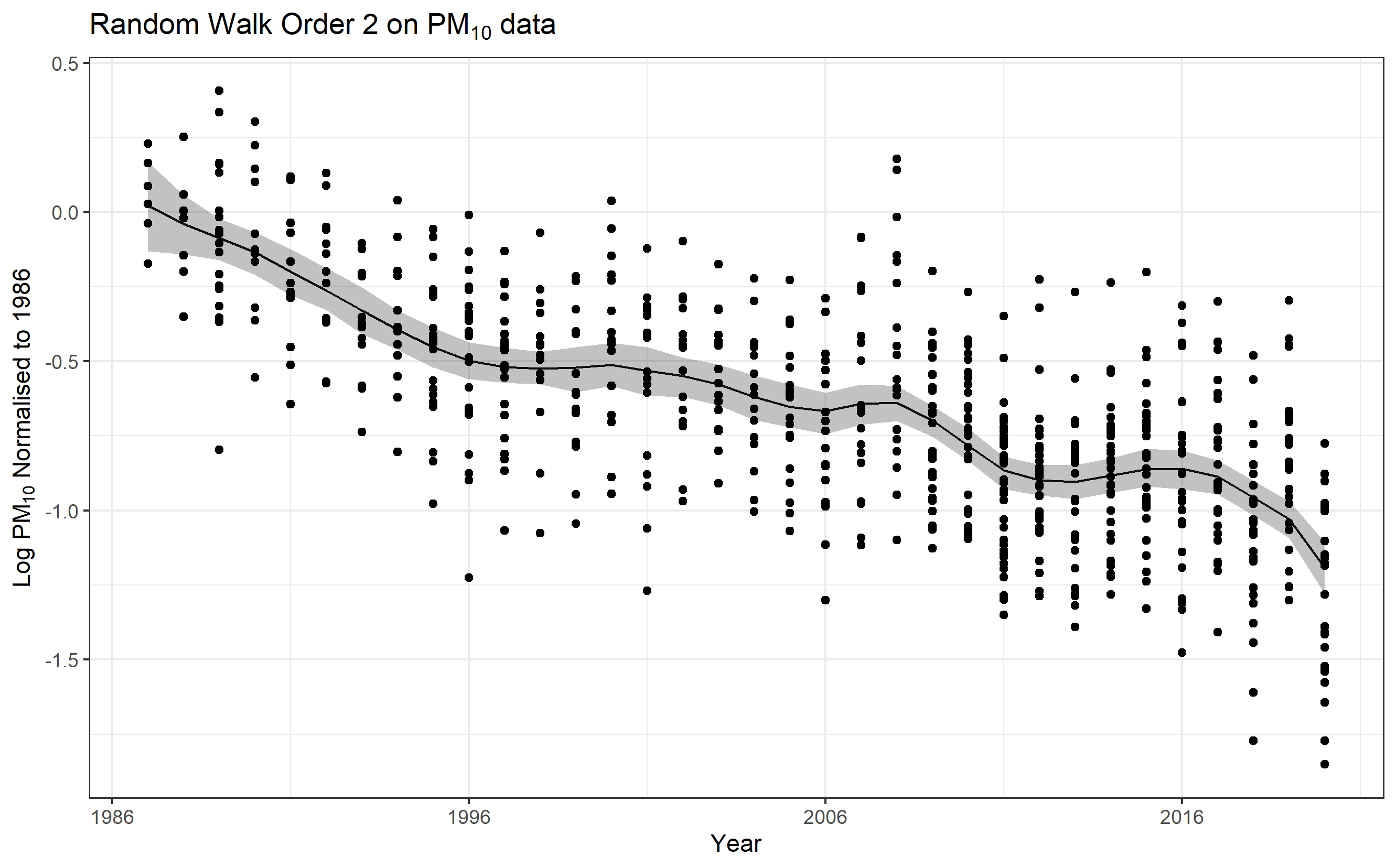}
    \caption{Second order Random Walk smoothing sets a prior on the difference between each observed value $f(\kappa_i)$.  Like so:  $f(\kappa_{i+1}) - 2f(\kappa_i) + f(\kappa_{i-1}) \sim N(0, \tau)$}
    \label{fig:Random_Walk2}
\end{figure}

First-order smoothing, shown in Figure \ref{fig:Random_Walk1}, looks spiky, and second-order smoothing, \ref{fig:Random_Walk2} seems better. Table \ref{tab:RW_parameters} describes the results of the two orders of smoothing, and the DIC criterion suggests that the 1st order smoothing is marginally better than the 2nd order smoothing.

As discussed in the section on priors, section \ref{subsubsec:Priors}, the \ac{PC} prior for the precision of the RW is set as the empirical SD of the data, which is $0.3729848$ when not accounting for the structure of sites and \ac{POC}s.
\begin{lstlisting}[language=R]
    pc_prior <-list(theta =list(prior =``pc.prec'',
                    param =c(data.emp.sd,0.01)))
\end{lstlisting}

\begin{table}[ht]
    \centering
    \begin{tabular}{p{0.25\linewidth}|p{0.30\linewidth}|p{0.30\linewidth}}
        Parameter & RW 1 &  2 \\ \hline
         Model WAIC & 1.187e+02 & 1.285e+02\\
         Model DIC & 1.187e+02 & 1.283e+02 \\
         \hline
         Intercept Mean & 3.407793e-05 & 2.074067e-05 \\
         Intercept SD & 31.62663 & 31.69167 \\
         \hline
         RW trend range, mean: & [-1.2189, -0.00401586] & [-1.19403, 0.0299855] \\
         RW trend sd, range: & [31.6266, 31.6267] & [31.6917, 31.6918] \\
         \hline
         Hyperpar: Precision of Gaussian observation Mean & 14.9708 & 14.38641 \\
         Hyperpar: Precision of Gaussian observation SD & 0.8224243 & 0.5420988 \\
         Hyperpar: Precision of Random Walk Parameter Mean & 106.8693 & 814.72112 \\
         Hyperpar: Precision of Random Walk Parameter SD & 41.5575799 & 961.1867062 \\
         
    \end{tabular}
    \caption{Comparison between exploratory models for a \ac{RW}1 and \ac{RW}2 model for the trend over years.  Neither model has a spatial component, and they treat sites and \ac{POC}s in the same year as \ac{IID} instead of nesting them in any way: y = intercept + RW. }
\label{tab:RW_parameters}
\end{table}

The R package \lstinline{inlabru} to generate a Random walk of order one on the time series using the following code:
\begin{lstlisting}[language = R]
cmp.spline1.PM10 <- log.Arthmt.M ~ Intercept + trend(map = yeari,
                                                model = ``rw1'',
                                                constr = FALSE,
                                                n = n_year,
                                                hyper = pc_prior)
bru.spline1.PM10 <- bru(cmp.spline1.PM10,
                       family = ``gaussian'',
                       data = test)
    
\end{lstlisting}
                       
Similarly, a second-order random walk in R was generated as follows:
\begin{lstlisting}[language = R]
cmp.spline2.PM10 <- log.Arthmt.M ~ Intercept + trend(map = yeari,
                                                model = ``rw2'',
                                                constr = FALSE,
                                                n = n_year,
                                                hyper = pc_prior)
bru.spline2.PM10 <- bru(cmp.spline2.PM10,
                       family = ``gaussian'',
                       data = test)

\end{lstlisting}

\subsubsection{Metadata}
\label{subsubsec:metadata}
Finally, gross patterns in the mean could exist and be described by the metadata available and included in a final model as fixed effects.  Here is a brief discussion of the categorical metadata variables available from the \ac{SCAQMD} and the \ac{EPA} that were examined as possible inclusions.

The three variables, which were examined,  were Land Use (e.g. commercial, residential, industrial, agricultural), Land Density (e.g. urban, rural, suburban), and Site Classification (e.g. background, high concentration, population exposure).

\begin{figure}[ht]
    \centering
    \includegraphics[width = \textwidth]{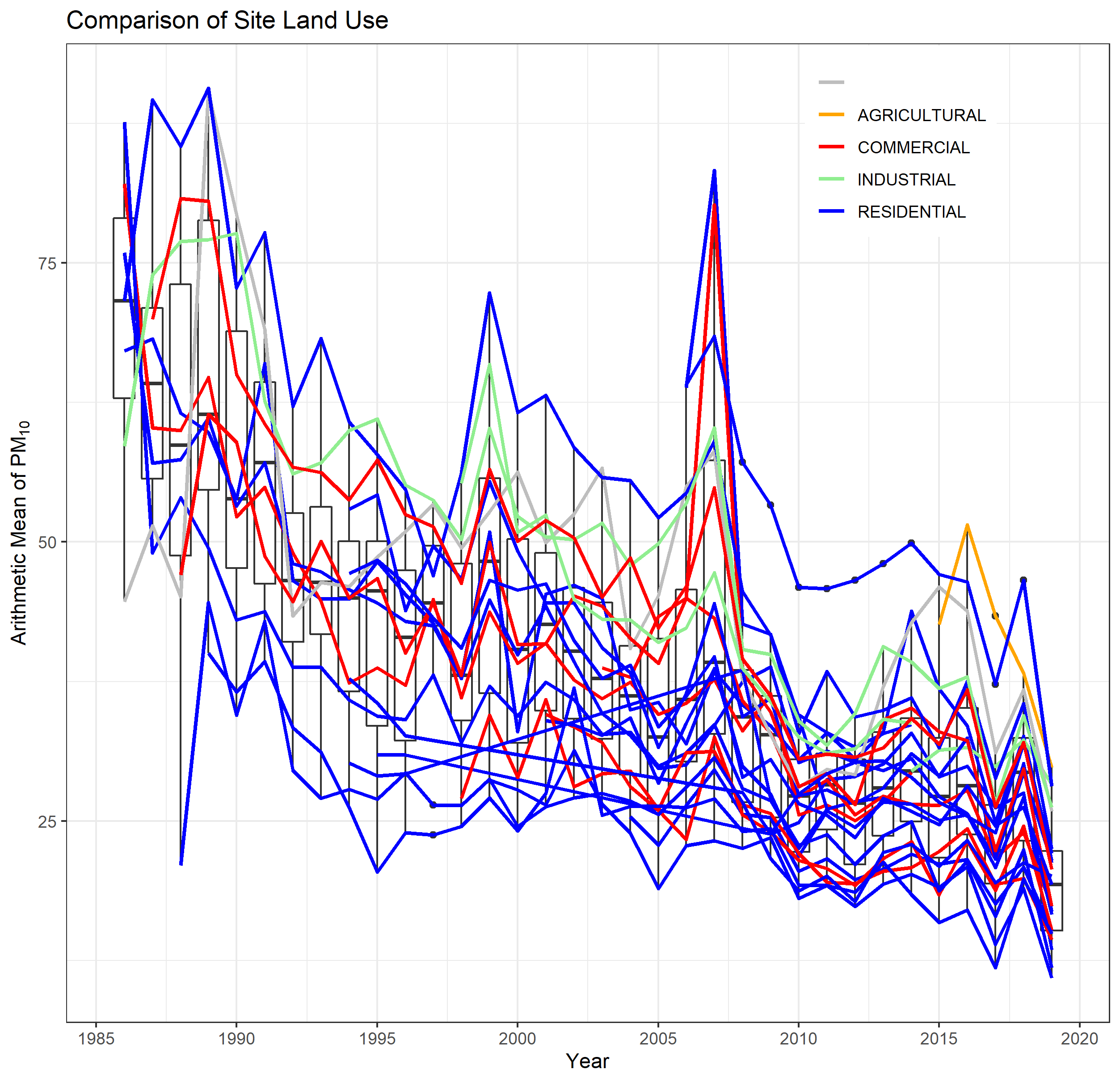}
    \caption{This shows the site traces colored by the type of human activity being carried out in the vicinity of each site as defined by the EPA.  In the case of multiple \ac{POC}s in at one site, the mean of those \ac{POC}s is taken.  One site was given no category and is in Grey.  The vast majority of sites are either Commercial (red, 6 sites total) or Residential (blue, 17 sites total) and these two categories are mixed relatively homogeneously.  Industrial (light green, 3 sites total) and Agricultural (Orange, 1 site) sites stand out as generally being elevated above the IQR for each year's observations, but have a handful of total sites.}
    \label{fig:SOCAB_metadata_Site_Land_use}
\end{figure}

\begin{figure}[ht]
    \centering
    \includegraphics[width = \textwidth]{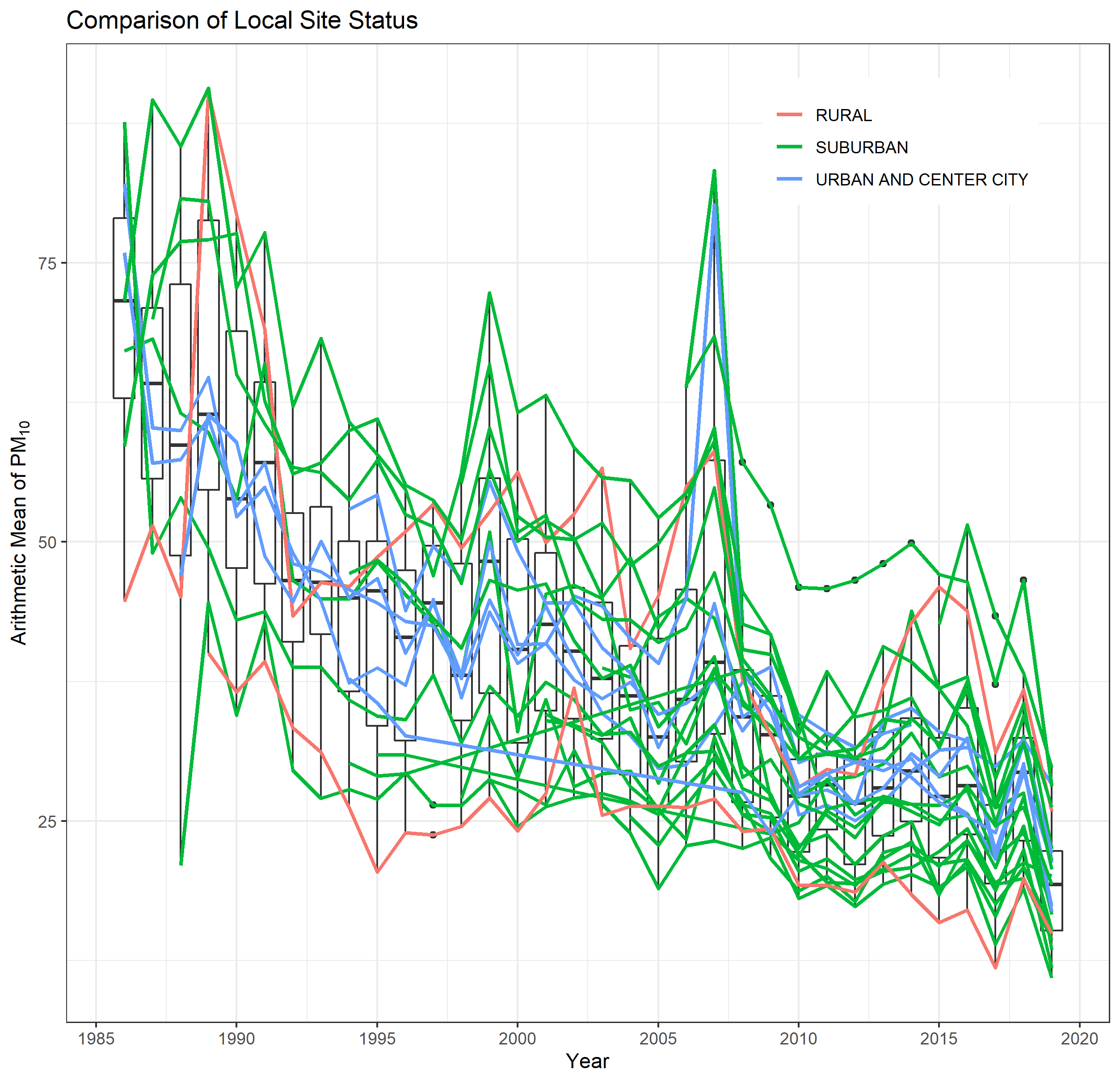}
    \caption{This shows the site traces colored by the density of buildings at each site as defined by the EPA.  In the case of multiple \ac{POC}s in at one site, the mean of those \ac{POC}s is taken.  The three categories of site - Rural (2 sites), Suburban (19 sites), and Urban (7 sites) - are distributed in a way that suggests deliberate choice. Each category seems to be evenly split to have sites above and below the mean.  Urban sites are closely clustered around the overall mean, Suburban generally surround the Urban sites, and the two Rural sites are relatively extreme.  }
    \label{fig:SOCAB_metadata_Site_Status}
\end{figure}

A quick examination of the Land Use and Land Density, (both from the \ac{EPA}'s metadata) in Figures \ref{fig:SOCAB_metadata_Site_Land_use} and \ref{fig:SOCAB_metadata_Site_Status} respectively, show that these categories will have little help in making predictions.

\begin{figure}[ht]
    \centering
    \includegraphics[width = \textwidth]{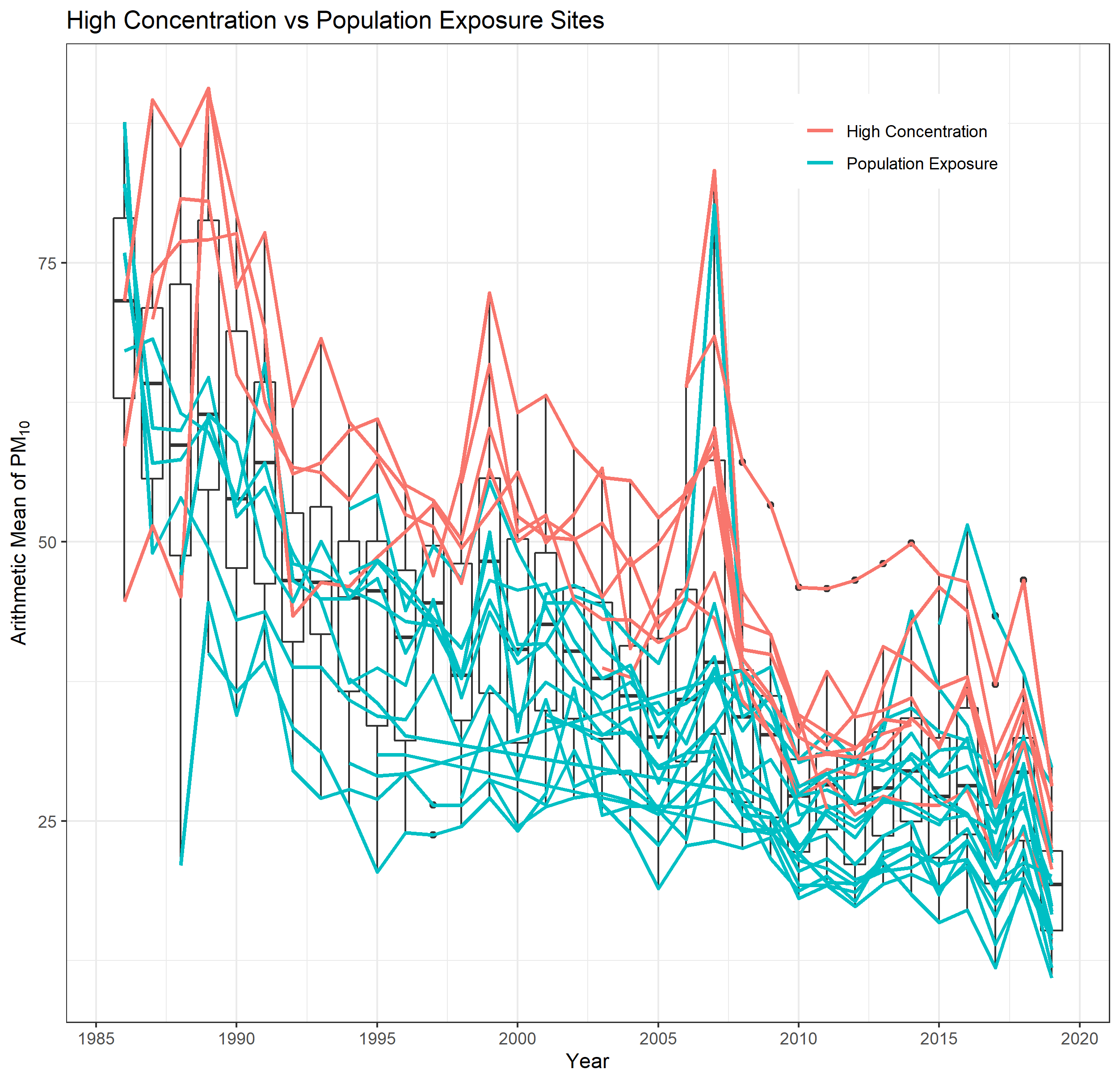}
    \caption{This shows the site traces colored by the Site Type category pulled from the \ac{SCAQMD} 5-year reports.  In the case of multiple \ac{POC}s in at one site, the mean of those \ac{POC}s is taken.  On the rare occasion that a site had a different type in 2010 and 2015 or between \ac{FEM} and \ac{FRM} monitors, the most consistent type was used for that site.  The sites designated High Concentration (9 total) are observing a higher concentration than the sites designated Population Exposure (19 total).  }
    \label{fig:SOCAB_metadata_Site_Type}
\end{figure}

The \ac{SCAQMD} provides the Site Classification.  These categories seem to have different means from each other, see Figure \ref{fig:SOCAB_metadata_Site_Type}, but only in that sites designed to capture high concentration do so.  Since the effort is to describe preferential sampling and not relations to covariates, this tautology seems unhelpful for modelling the \ac{PM10} field.  Future work that tries to model site inclusion or removal might find the Site Classification useful.  

\subsection{Traditional Spatial Modeling}
\label{subsec:tradspatmod}
Initial spatial modelling using Kriging with a Mat\'{e}rn covariance function examined each year's variogram independently of other years. These produced inconsistent results.  Each year's fitted covariance had different parameters.  Some possible reasons for this are:
\begin{itemize}
    \item An insufficient number of sites in a single year leads to instability or non-identifiability in the model.  We think this could be contributing to poor models in the early years.
    \item The sites are too far apart to resolve most of the curve of the covariance function. Since \cite{cameletti2011spatio} found a range of 275 km for \ac{PM10} this seems unlikely to be a issue at the scale of the \ac{SOCAB}.
    \item Biased sampling makes the estimate of the empirical variogram unreliable, with the bias in the semivariance's estimate increasing with $u$ \citep{diggle:07}.  If, as suspected, there is preferential sampling in the \ac{SOCAB} this could be another reason that the semivariograms did not work well.
\end{itemize}

\begin{figure}[ht]
    \centering
    \includegraphics[width = 12cm]{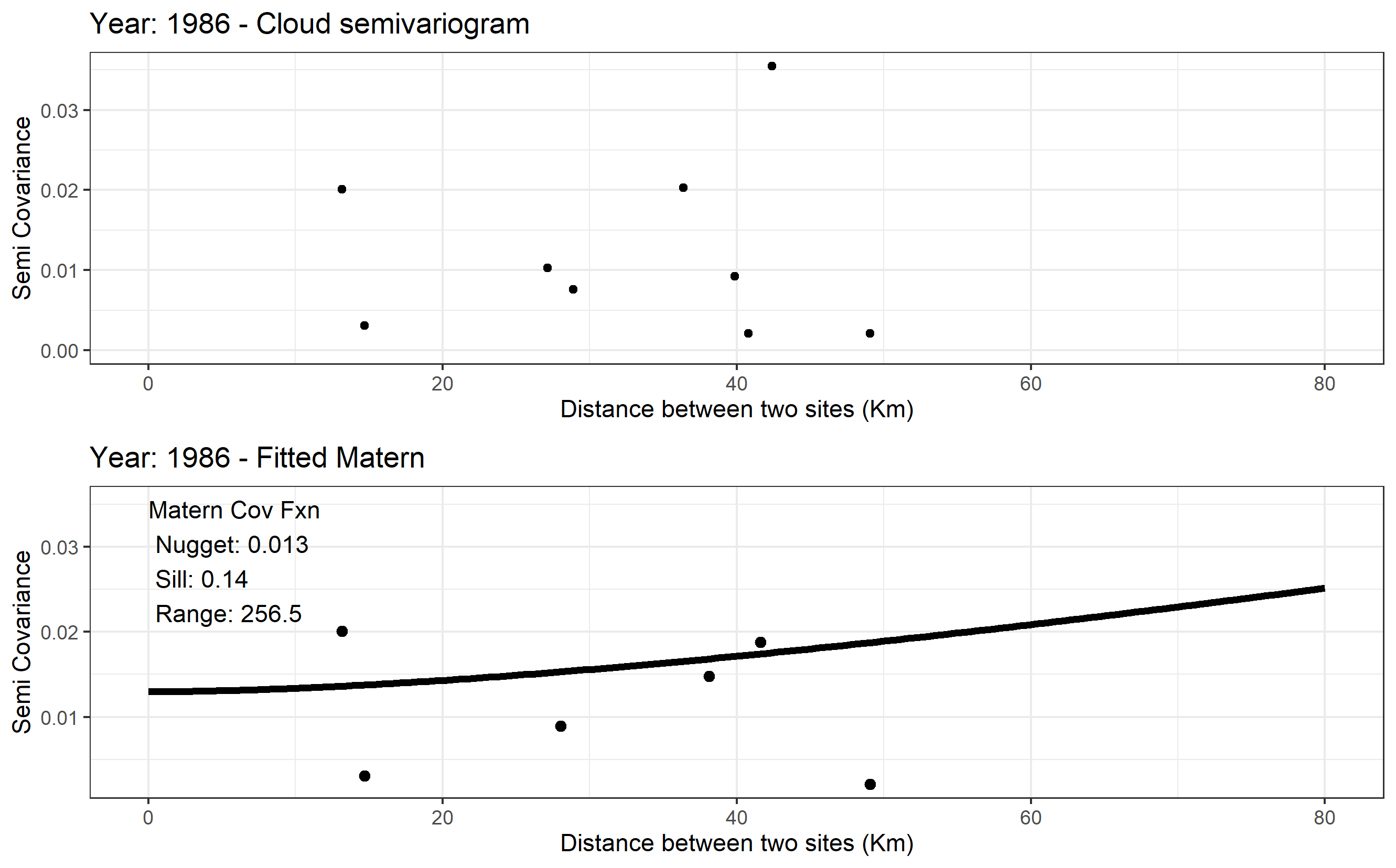}
    \caption{At the start of the network, a lack of sites poses a challenge to obtain a sufficient resolution to resolve a fitted variogram.}
\label{fig:Variogram_1986}
\end{figure}

\begin{figure}[ht]
    \centering
    \includegraphics[width = 12cm]{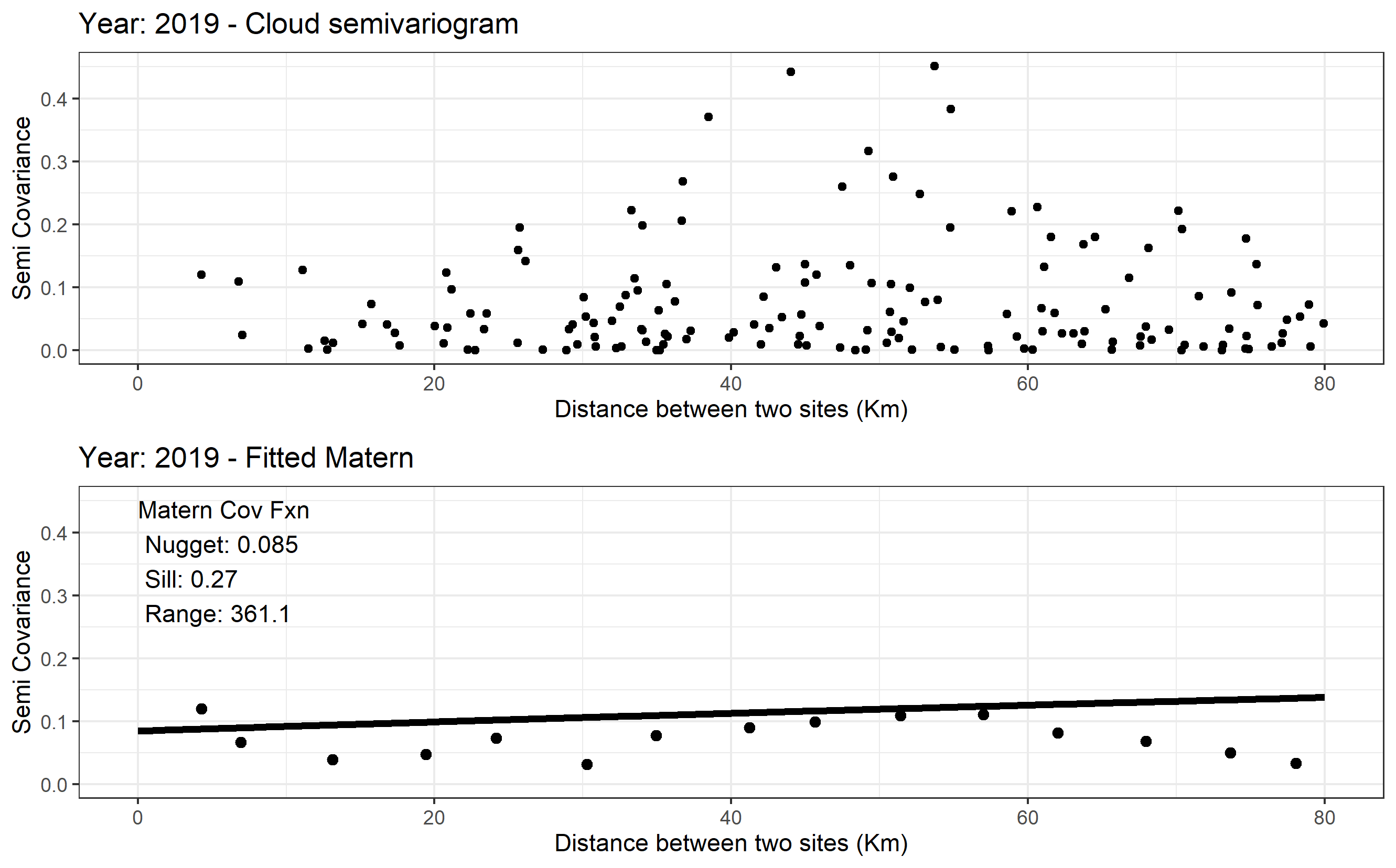}
    \caption{Even in later years, there was no guarantee of a good fit.  Here the variogram has no clear trend early on, preventing the curve from being established.}
    \label{fig:Variogram2019}
\end{figure}

\begin{figure}[ht]
    \centering
    \includegraphics[width = 12cm]{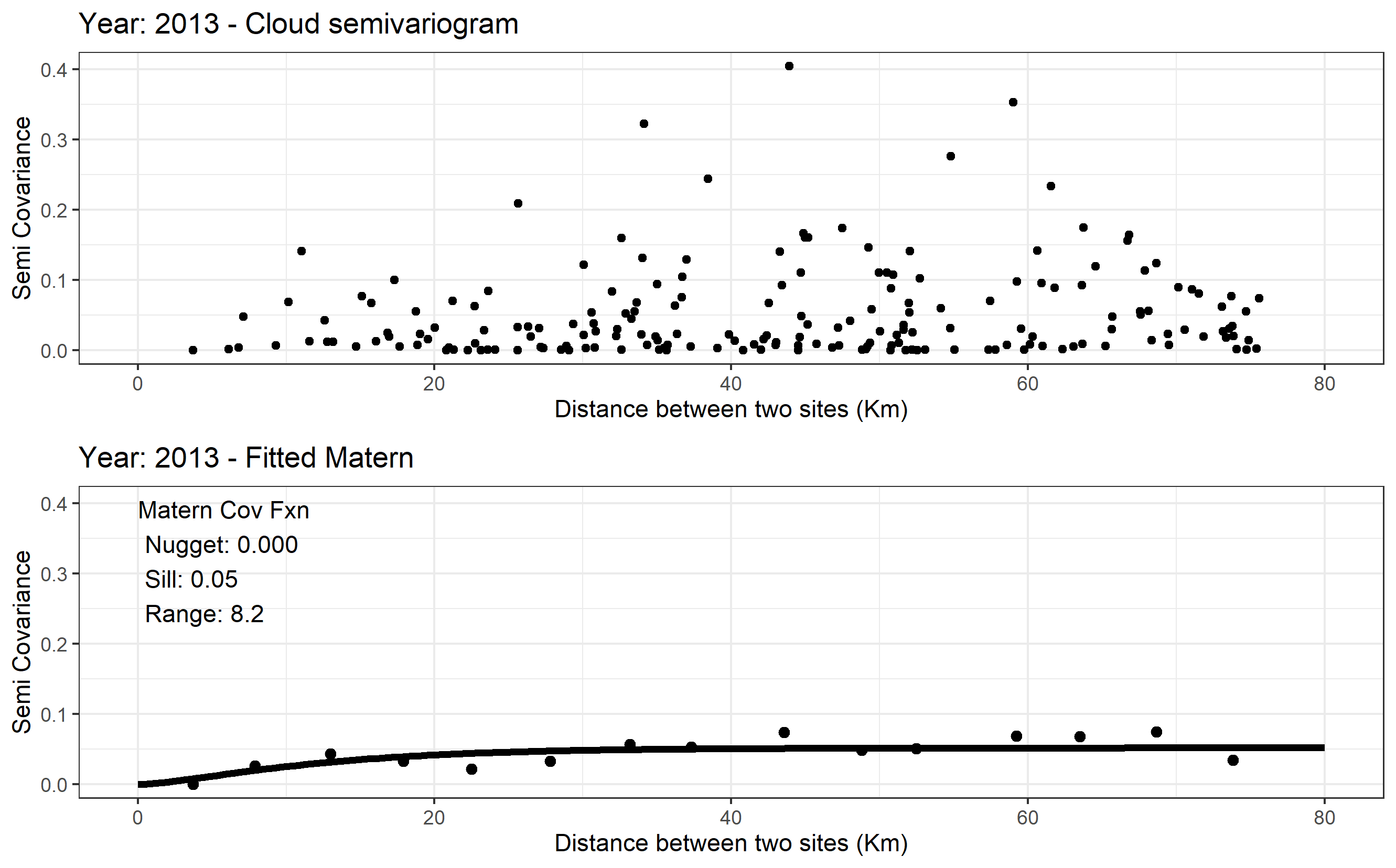}
    \caption{Here is one of the better-fitting years.  Key to the success of these types is the small distance estimates having a lower semivariance than most of the rest of the sites.}
    \label{fig:Variogram2013}
\end{figure}

Variogram plots for each year and their parameter values are not 
included for brevity, 
but here are three examples demonstrating the range of success in modelling the variograms: Figure \ref{fig:Variogram_1986}, Figure \ref{fig:Variogram2013}, and Figure \ref{fig:Variogram2019}.  Figure  \ref{fig:Variogram_1986} does not have enough sites to make a clear variogram.  Figure  \ref{fig:Variogram2013} has plenty of sites, but exhibits a strange behavior with raised semivariance at short range that tails off at longer range.  Finally, Figure  \ref{fig:Variogram2019} exhibits a nice behavior with a rise in semivariance over the closer distances and then a rough flattening as the range increases.  

\section{Modeling with INLA}
\label{subsec:INLAmodelling}

\subsection{Mesh Construction}
\begin{figure}[ht]
    \centering
    \includegraphics[width = \textwidth]{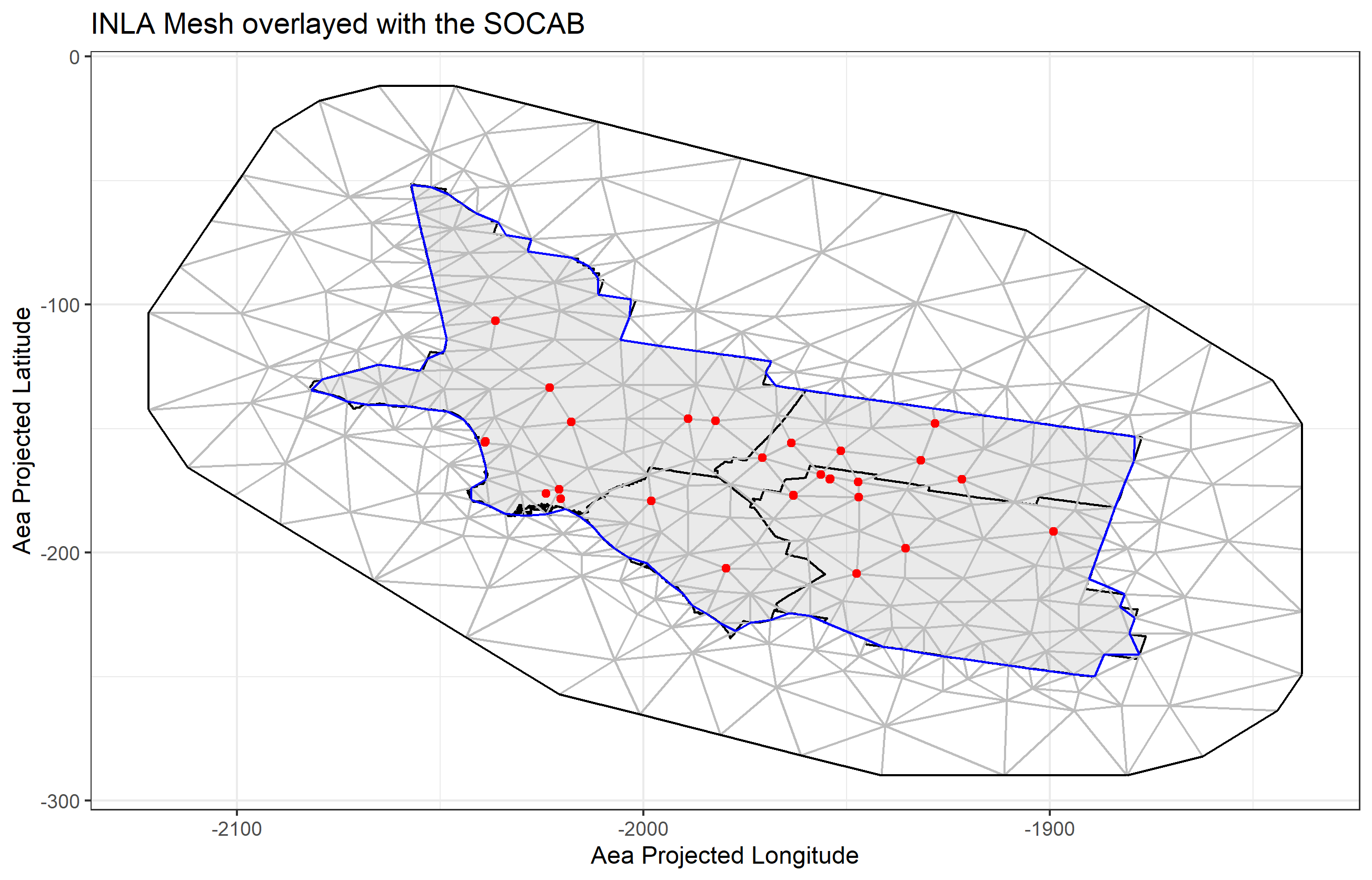}
    \caption{The mesh used for modelling, shown projected in the Albers projection.  The red dots are the location of sites contributing data to the model.  The blue line is the boundary of the \ac{SOCAB} as defined in \ac{INLA}.  Under the blue line is a black line showing the actual legislated boundary.  There are a few sites close to the interior boundary, but no sites near the outer boundary.  By using sites as seed locations for the nodes, the problem of having multiple sites in one triangle was avoided.}
    \label{fig:SOCAB_mesh}
\end{figure}

As discussed in Section \ref{subsec:IntroMesh}, the mesh's design is important for the subsequent modelling.  Preliminary models using coarser meshes suggested the \ac{GRF} has a range of about 100 km.   With that information and following \cite{Righetto2020}, a maximum edge length of 20 km and a cutoff point of 5 km were chosen for the final mesh.  The external maximum edge length was set at 40 km and the width of the offset kept 2-3 edge lengths between the inner boundary and the outer boundary.  The final mesh can be seen in Figure \ref{fig:SOCAB_mesh} and was created using the code below.

\begin{lstlisting}[language = R]
mesh1 <- inla.mesh.2d(loc = PM10.INLA.data.aea.SOCAB@coords, 
                     boundary = SOCAB_union_sp,
                     offset = c(1,40),
                     max.edge = c(20, 40),
                     min.angle = c(21, 21),max.n=c(48000, 16000), 
                     max.n.strict=c(128000, 128000), 
                     cutoff=5
                     ) 
\end{lstlisting} \label{code:inlaMesh}

\subsection{Mat\'{e}rn Parameters}
\label{subsec:maternparas}
The Mat\'{e}rn covariance function has several parameters that must be either estimated or fixed. These are the smoothness, $\kappa$, and the PC priors for the Range and Variance, as discussed in Section \ref{subsubsec:MaternIntro} and Section \ref{subsubsec:Priors}.

\subsubsection*{Smoothness, $\kappa$}
\label{subsubsec:smoothness}
As discussed earlier in Section \ref{subsubsec:MaternIntro} the Mat\'{e}rn's smoothness, $\kappa$, is fixed to make the choice of other parameters clear.  \cite{cameletti2011spatio} used $\kappa = 1$ for their Mat\'{e}rn  function, we chose to use 1 as well.

\subsubsection*{Priors}
\label{subsubsec:priors}
Here are described the choices of \ac{PC} priors for the range, variance, and random walk used to generate the covariance function.

\subsubsection{Range}
\label{subsubsec:range}
The PC prior for the range is based upon the smallest value that is reasonably expected for the range.  This is done with the formula \ref{eq:PC_Matern_Range}.  The documentation for the PC prior suggests setting $p_r$ at 1\%. Table \ref{tab:Empirical_Range_Sill_Quantile} shows the range quantiles from the empirical variograms for each year, providing a guide for what value of $r$ to choose for the PC prior, we used $P(r < 6) = 0.01$.  

\begin{equation} \label{eq:PC_Matern_Range}
    P(r < r_0) = p_{r}.    
\end{equation} 

\begin{table}[ht]
    \centering
    \begin{tabular}{c||c|c}
        Quantile & Range (km) & Partial Sill \\
        \hline
        0\% & 4.34 & 0.00 \\
        1\% & 4.68 & 0.00 \\
        5\% & 6.17 & 0.00 \\
        10\% & 8.42 & 0.00 \\
        15\% & 9.37 & 0.012 \\
        25\% & 10.61 & 0.042 \\
        50\% & 20.55 & 0.066 \\
        75\% & 36.69 & 0.14 \\
        90\% & 349.42 & 1.65 \\
        95\% & 852.55 & 3.82 \\
        99\% & 2247.57 & 32.80 \\
        100\% & 2882.08 & 46.62 \\
        
    \end{tabular}
    \caption{Empirical Quantiles of the range and partial sill of the 34 yearly Variograms using 1986 normalized log \ac{PM10}. }
    \label{tab:Empirical_Range_Sill_Quantile}
\end{table}

\cite{cameletti2011spatio} Found range of 275 km and 1046 km for PM10 in Piedmont valley

The \ac{EPA} describes spatial scale as follows:
\begin{quote}
``Thus, the spatial scale of representativeness is described in terms of the physical dimensions of the air parcel nearest to a monitoring site throughout which actual pollutant concentrations are reasonably similar.''   \end{quote}
In CFR40-58, the \ac{PM10} sensors are defined as having a neighborhood scale up to 4 km.  This implies that it would be physically impossible to resolve a range that is about 4 km or smaller.  With the information about the spatial scale of the sites from CFR40-58 and the combination of the empirical variograms of the \ac{SOCAB} and known range of \ac{PM10} from previous studies, it is reasonable to have $P(\rho < 3) = 0.01$. Implying it is unlikely that the range is smaller than 3 km.

\subsubsection*{Variance}
\label{subsubsec:variance}
The PC prior takes user input on the upper tail quantile and the probability of exceeding it as equation \ref{eq:PC_Matern_Stdev}.  Using the 34 years of empirical variograms, we get the quantiles shown in \ref{tab:Empirical_Range_Sill_Quantile} for the partial sill, and then used $P(\sigma > 35) = 0.01$ as the prior on the partial sill.

\begin{equation}
    P(\sigma > \sigma_0) = p_{\sigma}    
\end{equation} \label{eq:PC_Matern_Stdev}

\subsubsection{Random Walk}
\label{subsubsec:ranwalk}
The \ac{RW} used in the full model was initiated with the same priors as the \ac{RW} performed during the data exploration, see Section \ref{subsubsec:RWexploration}.

\subsection{Choice of Covariance Structure}
\label{subsec:covstructchoice}
The final model was chosen from a range of options by comparing the DIC of models with different structures.  These structures were expanded from the core model of a single Mat\'{e}rn \ac{GRF} through the addition of an RW over time, an AR(1) process over time, and combinations of these.  \cite{cameletti2011spatio} used a Mat\'{e}rn field with an AR(1) process to account for shifts between observations periods.  This modelling was done with the full dataset.

Table \ref{tab:cov_str_DIC} summarizes the model results for these different covariance structures.  The models with smaller DIC are more attractive options for further modelling.  The model chosen is number 4, the combined \ac{GRF} and \ac{AR}(1) process, which is the same structure as that used by \cite{cameletti2011spatio}. 
\begin{table}[ht]
    \centering
    \begin{tabular}{l | c}
        Model Cov Structure & DIC  \\
        \hline
        Mat\'{e}rn & 2.385e+02  \\
        Mat\'{e}rn and RW1 & -8.385e+02 \\
        Mat\'{e}rn and RW2 & -8.349e+02 \\
        Mat\'{e}rn, RW1, and AR1 & -9.173e+02 \\
        Mat\'{e}rn, RW2, and AR1 & -8.969e+02

    \end{tabular}
    \caption{The data used is the log of the normalized data as shown in equation \ref{eq:log_transform}.  The covariance structures are listed in order of increasing complexity. \ac{RW}(1) is the smoothing that describes the transition of the overall mean from one year to the next.  \ac{GRF} is the Mat\'{e}rn function taken as an overall mean for the whole year.  }
    \label{tab:cov_str_DIC}
\end{table}

The equation describing the final model is as follows:

\begin{subequations}
\begin{equation}
    Z(s,t) = \beta_0 + \Delta y_t + y(s,t) + \epsilon(s,t)
\end{equation}
\begin{equation}
    \Delta y_t = y_t - y_{t-1}
\end{equation}
\begin{align}
    y(s,1) &\sim N \left(0, \frac{\sigma^2_w}{(1-a^2)} \right) &, |a| < 1 \\
    y(s,t) &= ay(s,t-1) + w(s,t) &, t>1
\end{align}
\begin{equation}
    cov(w(s,t), w(s,t')) = 
    \begin{cases}
       0 \text{ if } t \neq t' \ , \\
       \sigma^2_w \gamma(u) \text{ otherwise, }  \gamma(u) \sim \text{Mat\'{e}rn.}
    \end{cases}
\end{equation}
\end{subequations}

\subsubsection*{Final Model} \label{seq:finalModel}

Table \ref{tab:model_INLA_full} gives the results of the final model, performed on a randomly selected 90\% of the data, holding the other 10\% for validation, see Section \ref{subsec:validation} for details.  The model has a \ac{RW}1 structure portraying the trend over time, a Mat\'{e}rn covariance function describing the spatial structure and an \ac{AR}(1) function describing how that structure changes over time.

\begin{table}[ht]
    \centering
    \begin{tabular}{l|c|c}
         & Value & SD  \\
         \hline
         WAIC & -7.958e+02   & \\
         DIC & -8.022e+02 & \\
         Intercept & -0.0005082417 & 31.58598  \\
         RW1 & [-1.36122, -0.114299] & [31.5861, 31.5861] \\
         Mat\'{e}rn & [-0.408501, 0.58749] & [0.0812534, 0.349748] \\
         Hyperpar Gaussian Prec. & 73.6006189  &  2.811640470 \\
         Hyperpar RW1 Prec & 60.1411588 & 7.967774153 \\ 
         Hyperpar Mat\'{e}rn Range & 23.8506006 & 6.017936799 \\
         Hyperpar Mat\'{e}rn Stdv & 0.2729190 & 0.023497256 \\
         AR(1) rho & 0.9923009 & 0.001750823 
    \end{tabular}
    \caption{Summary results of model that has RW1 over the years, a Mat\'{e}rn spatial process and an AR(1) term.   The RW1 and Mat\'{e}rn are random variables and so take on a unique value at each site. }
    \label{tab:model_INLA_full}
\end{table}

The following code is how the model was constructed in R with \lstinline{INLABru}.
\begin{lstlisting}
cmp.Matern.RW1.AR1.PM10.subsample = log.Arthmt.M ~ 
  Intercept + trend(map = year, model = ``rw1'', 
  constr = FALSE, n = n_year, hyper = rw_pc_prior) +
    myspde(map = coordinates, group = year,\ngroup = n.year,
    model = inla.spde2.pcMatern(mesh1, alpha = Matern_alpha,
 prior.range = Matern_pc_prior_range,prior.sigma = 
 Matern_pc_prior_sigma),mesh = mesh1,control.group=list(model=``ar1''))
bru.Matern.RW1.AR1.PM10.subsample = bru(cmp.Matern.RW1.AR1.PM10.subsample,             
  family = ``gaussian'', data= PM10.INLA.data.aea.subsample.SOCAB)
\end{lstlisting}

\subsection{Prediction and Validation} \label{subsec:validation}
The \ac{PM10} surface for each discrete year was interpolated using the model described in Section \ref{seq:finalModel}.  This surface was used for subsequent preferential sampling testing as well as model validation.  The prediction was performed on a grid of pixels covering the \ac{SOCAB} with each pixel approximately 2 km square as defined by the Lambert projection.  After predicting the surface, the results were used to examine whether the model did a ``good job'' of describing the known observations.

\subsubsection*{Withholding 10\% of Data}
\label{subsubsec:withholding}
\begin{figure}[ht]
    \centering
    \includegraphics[width = \textwidth]{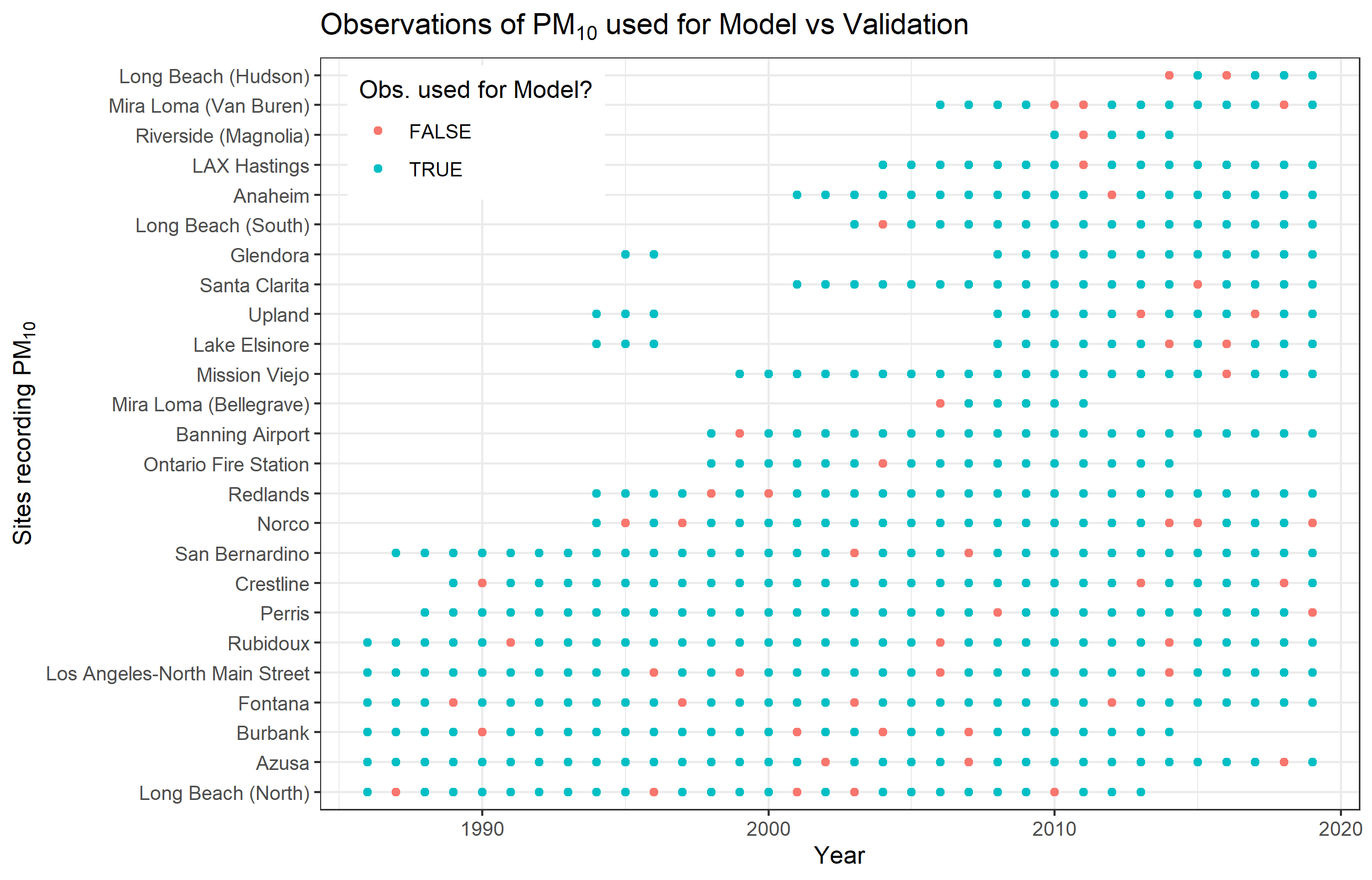}
    \caption{The red dots are observations that were kept out of the model for use in future validation.  Eighty (ten percent) of the 822 total observations were held back, chosen at random}
    \label{fig:validate_dotplot}
\end{figure}

\begin{figure}[ht]
    \centering
    \includegraphics[width = \textwidth]{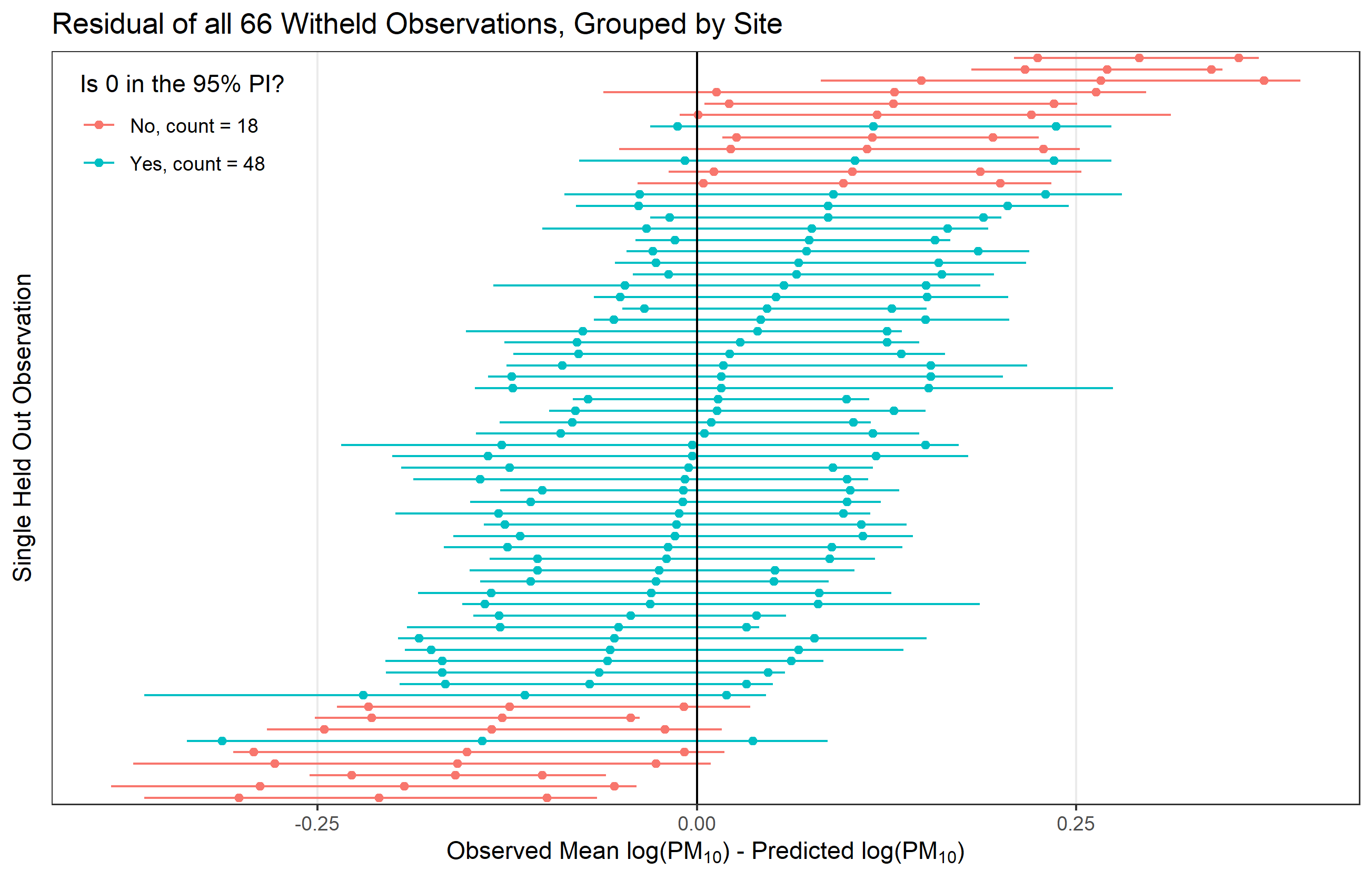}
    \caption{This shows the posterior distribution of the withheld observations, centered at 0 by subtracting them from the observed observations.  For each site, the central point is the posterior mean, the left and right points are the posterior 0.025 and 0.975 percentiles respectively, and the line goes from the smallest to the largest value in the 100 samples from the posterior.  It is concerning that 35\% of the validation points don't have the observed value within the 95\% prediction interval, which suggests that some part of the model could be improved.}
    \label{fig:validate_delta_mean}
\end{figure}

\begin{figure}[ht]
    \centering
    \includegraphics[width = \textwidth]{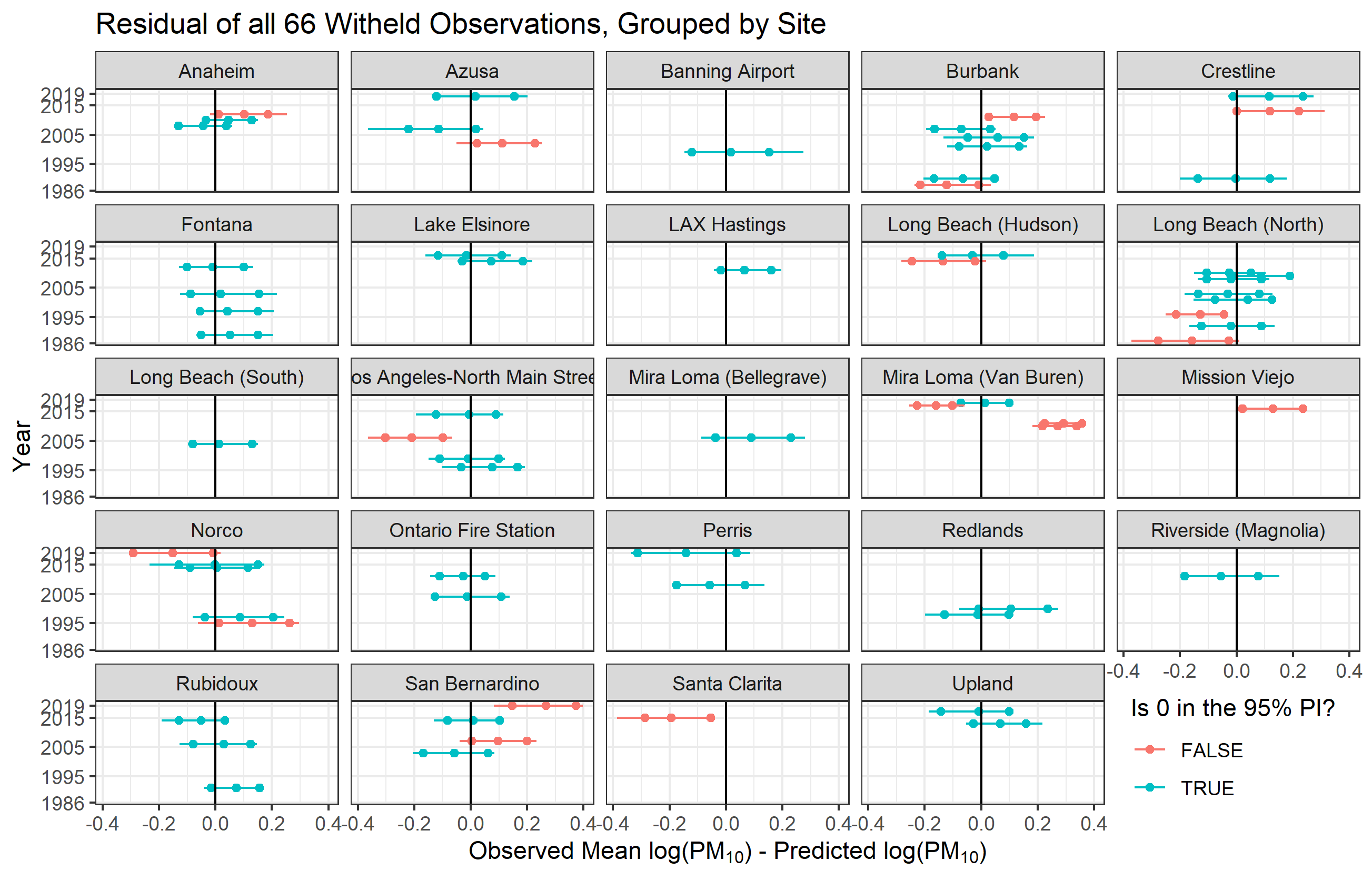}
    \caption{Here the posterior distributions shown in the previous figure are split up by site and by year in an attempt to see any patterns suggesting where the model could be improved.  It looks like sites have more heterogeneity in the distribution of out-of-bounds predictions than years, so perhaps the spatial covariance function needs tweaking.}
    \label{fig:validate_delta_site}
\end{figure}

.
Model validation was done while holding out a randomly selected 10\% of the data.  The sites and years that were withheld can be seen in Figure \ref{fig:validate_dotplot}.  The model was then used to predict that 10\% and the results compared to the actual values.  The difference between the prediction and the actual observed value can be seen in Figures \ref{fig:validate_delta_mean} and \ref{fig:validate_delta_site}.   This is a basic but easy-to-implement method that is not computationally intensive.

It is concerning that Figure \ref{fig:validate_delta_mean} shows such a high percentage of sites whose prediction interval does not contain the actual observed value.  Theoretically, only 5\% of the validations should not contain the actual value within the 95\% prediction interval.  The variance could be underestimated, or the model could be too smooth.

Figure \ref{fig:validate_delta_site} suggests that there isn't any obvious prediction problem at individual sites, so maybe it is the variance that is too small.

\section{Preferential Sampling}
\label{sec:prefsamp}

As discussed in Section \ref{subsec:PreferentialSampling}, \ac{PS} results in sampling sites that have a stochastic dependence upon the latent field of interest and results in a biased estimation of the true field.  This chapter presents evidence  \ac{PS} found in the course of this project.

Because the test for preferential sampling is done by comparing the location of all sites to simulated sites sampled on the pollutant field, a field calculated from all the sites.  To that end, after the validation modelling that held back 10\% of the data, the same model using all the data was run .

\subsection{Governmental Acknowledgement of Preferential Sampling} \label{subsec:govPrefSamp}
The first and perhaps most clear-cut evidence for \ac{PS} are direct statements from the agencies responsible for site selection, the \ac{SCAQMD} and \ac{EPA}.  In their five-year reports, the \ac{SCAQMD} describes some of how the monitoring locations are distributed.  On page 63 of the 2010 \ac{SCAQMD} 5-year report and page 83 of the 2015 version is the following statement:
\begin{quote}
  ``Real time monitors, for the most part, are clustered in the high concentration areas...''  ``Real time \ac{PM10} monitors also support ongoing health studies in the region.''    
\end{quote} \cite{CASCAQMD:2010}, \cite{CASCAQMD:2015}, \cite{AQMNP:2019}
Site clustering is one way that \ac{PS} is described, and its presence is used by \cite{watson2020} as an indicator for the presence of \ac{PS}. 

\begin{quote}
    ``Though the current PM 10 network is relatively stable, monitoring agencies may  continue divesting of some of the PM10 monitoring stations where concentration levels are low relative to the NAAQS.''
\end{quote} \cite{EPA:IntegratedReview}
This divesting of low-concentration sites from the network is similar behavior to that found in the UK for black smoke \citep{zidek2010monitoring}.

\subsection{Four Site Categories}\label{sec:4sitecategories}
In Figures \ref{fig:site_timing_trace-Added}, \ref{fig:site_timing_trace-Continuous}, and \ref{fig:site_timing_trace-Removed} We demonstrated the first sign of preferential sampling in the data, by examining the traces of sites grouped by when they entered and left the network.  Sites were split by a 2-by-2 table based on whether a site was A) present at the start of the network or B) still monitoring at the end of the network (see table \ref{tab:2X2_site_category}).

\begin{table}[ht]
    \centering
    \begin{tabular}{| c c | c c |}
        \hline
         \multirow{2}{*}{} & {} & \multicolumn{2}{c}{Present in 1986} \\
                           & {} & Yes & No \\
        \hline 
        \multirow{2}{*}{Present in 2019} & Yes & Continuous (many) & Added (many) \\
                                         & No & Removed (2 sites) & Added then removed (3 sites)   \\
         \hline
    \end{tabular}
    \caption{Naming conventions for sites categorized according to the two-way table made by whether the site is A) Present in the network in 1986, and B) Present in the network in 2019.}
    \label{tab:2X2_site_category}
\end{table}

Calling back to those three figures, the ``Continuous'' and ``Added then Removed'' sites have very similar overall means, an increase compared to the overall mean of all sites.  In contrast, the ``Removed'' and ``Added'' sites have similar means that are lower compared to the overall mean.
This pattern could be a result of preferential sampling early (starting biased high, then corrected by adding low pollution sites later) or late (ending biased low, the mean dragged down by the sites added later). Alternatively, a change in the pollution field's distribution could explain this; if the tail shifts and drags the mean over time.  However, the box plots of each year do not seem to support that explanation,
as they stay roughly symmetrical throughout.

Table \ref{tab:model_INLA_4site_Retention} shows the result of including those four categories as fixed effects in the model

\begin{table}[ht]
    \centering
    \begin{tabular}{l|c|c}
         & Value & SD  \\
         \hline
         WAIC & -9.532e2 & \\
         DIC & -9.645e2 & \\
         Intercept & -0.0004295981  & 31.59055710 \\
         Continuous & 0.0094430558  & 0.12665671   \\
         Removed & -0.2024973929   & 0.10773814   \\
         Temporary & -0.2746064961   & 0.04417141\\
         RW2 & [-1.38555, -0.106361] & [31.5907, 31.5908] \\
         Mat\'{e}rn & [-0.394883, 0.70015] & [0.107131, 0.427265] \\
         Hyperpar Gaussian Prec. & 82.1063483 & 5.363589751  \\
         Hyperpar RW2 Prec. & 30.2078731   &  8.969724584 \\
         Hyperpar Mat\'{e}rn Range & 28.0537004  & 11.325932812  \\
         Hyperpar Mat\'{e}rn Stdv & 0.2980578  & 0.043042302   \\
         Hyperpar AR(1) rho & 0.9948262   & 0.001907155  
    \end{tabular}
    \caption{Including fixed effects for the 4 site retention category }
    \label{tab:model_INLA_4site_Retention}
\end{table}


\subsection{\texttt{PStestR}: A Preferential Sampling Package}
\label{subsec:prefsamppkg}

As described in Section \ref{subsubsec:WatsonPrefSample}, \citet{watson2020} proposed a theory to detect preferential sampling.  They also provide an R package called \texttt{PStestR} to implement the test, which is used below.

\subsubsection*{Implementation in \texttt{PStestR}}
\label{subsubsec:implementation}
Having obtained a predicted pollutant surface and knowing the location of sites, the package calculates the mean of the \gls{k} nearest neighbors at each site and correlates that with the estimated concentration of the pollutant. The same result is produced for many Monte Carlo samples of possible sites over the whole network area.
The package described in \cite{watson2020} can be used under two paradigms. 
\begin{enumerate}
    \item The number of nearest neighbors to use is known.
    \item The number of nearest neighbors is uncertain, and testing a range of options is part of the research question.
\end{enumerate}
In the first case, a single test is performed comparing the known sites to the distribution of the Monty Carlo samples.  In the second case, a multiple comparison test is implemented with a comparison for each value of \gls{k}, the number of nearest neighbors used.  \texttt{PStestR} then provides the following outputs:
\begin{itemize}
    \item Test Rho:  The calculated Spearman's Rho for the network during each year.
    \item Empirical P-Value:  Compares the Test Rho for the actual network to the distribution from the Monte Carlo simulation.
\end{itemize}

\subsubsection*{Tuning Parameters}
\label{subsubsec:tuneparameters}
\texttt{PStestR} has several parameters that can be adjusted to affect the simulation and test.  Here we describe those parameters and what we chose to use.

\begin{itemize}
    \item Number of Nearest Neighbors, \gls{k}:
As described earlier, the number of nearest neighbors included in the test can be tuned to improve the power of the overall test at the cost of precision.  Looking at the points showing site locations in fig \ref{fig:site_dotplot}, the largest cluster seems to be about 3.  So we set $k = 3$.  

\item Number of Monte Carlo Samples:
Increasing the number of Monte Carlo samples improves the posterior's precision at the cost of computational time.  An M of 1000 is large enough to be reasonable while still being manageable by the  computer on which this analysis was performed.
\item Year:
Each year is not independent of the others, but each test on a year will assume that it is.  To avoid multiple comparisons, it is necessary to choose one year to test.  We choose 2019 because it will show the current state of the network.

\end{itemize}

\subsection{Results of Test}
\label{subsec:testresults}

\subsubsection*{Result of Test for 2019}
\label{subsubsec:test2019}

The network in 2019 has a correlation of -0.822 and an Empirical P-Value of 0.00300. 
This correlation is very close to -1 and the P-Value implies that the observed network of sites would be very unlikely to be chosen in a sampling regime that is not stochastically dependent upon the pollutant field.  

\begin{figure}
    \centering
    \includegraphics[width = \textwidth]{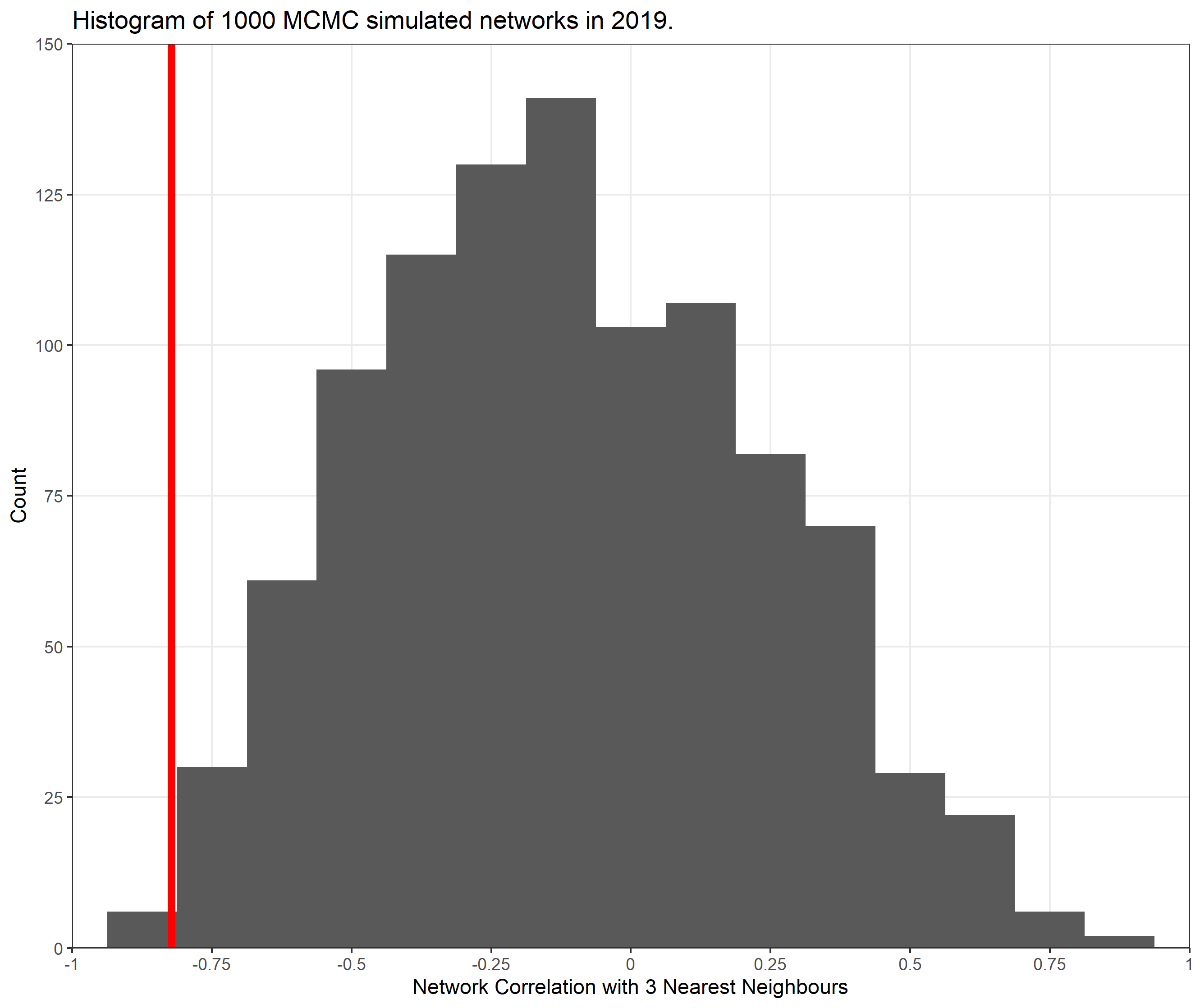}
    \caption{Histogram of sampled network Correlation of 1000 MCMC samples showing their empirical distribution.  The red line shows where the observed correlation lies in relation to the samples.}
    \label{fig:MCMC_hist}
\end{figure}

Figure \ref{fig:MCMC_hist} shows the distribution of the test Rho for each of the MCMC samples and the position on that distribution of the test Rho for the actual network during 2019.

\subsubsection*{Time Series of Test Score} \label{subsubsec:TestScoreTimeSeries}
Despite knowing there is a lack of independence between years, we chose to assess the data as a time series.  What follows is a more qualitative exploration rather than a quantitative result.

Before 1994 the network did not have enough sites to produce a result from the \texttt{PStestR} algorithm.  However, except for 1997, from 1994 to the present there are enough sites for a correlation score.  

\begin{figure}
    \centering
    \includegraphics[width = 
   \textwidth]{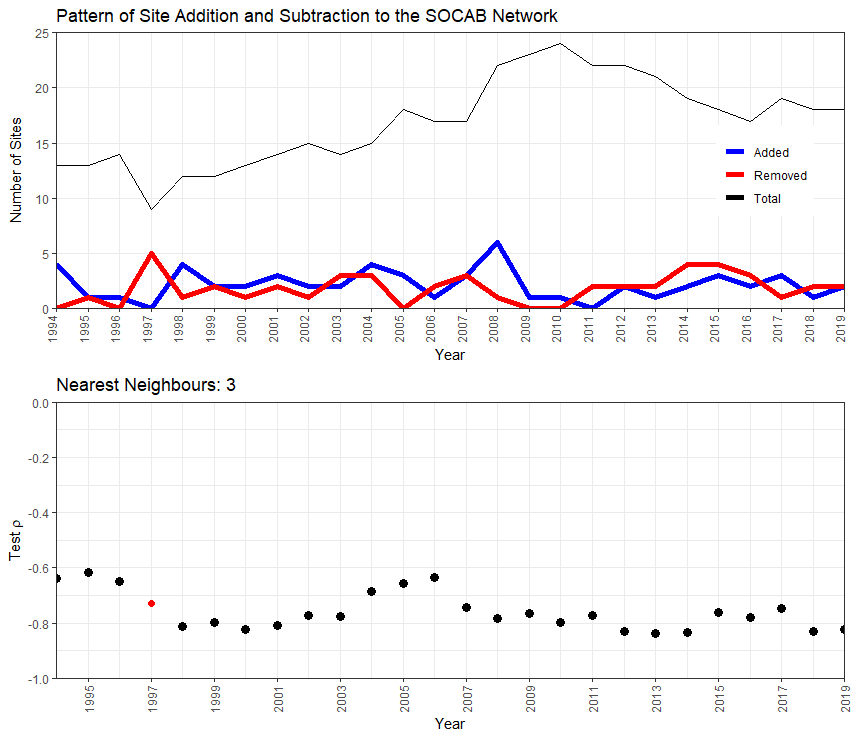}
    \caption{Top:  Counts of sites in the network for each year.  Total sites are in black, the number of sites that were removed compared to the previous year is in red, and several sites that were added compared to the previous year are in blue. 
    Bottom: Test $\rho$ (Spearman's Rank Correlation) for the three nearest neighbors.  A negative correlation implies a bias towards high-concentration monitoring. 1997 had too few sites to calculate a score and is interpolated as the mean of 1996 and 1998 scores (red dot).}
    \label{fig:PS_site_counts_test_rho}
\end{figure}

Figure \ref{fig:PS_site_counts_test_rho} shows that each year has a negative correlation score, implying preferential sampling for locations with a higher concentration of \ac{PM10}.  In addition, the Test Rho's time series in the bottom half of Figure \ref{fig:PS_site_counts_test_rho} shows a decrease in time, which would imply an increase in preferential sampling from earlier to later dates.  On the other hand, the ``trend'' could easily result from a stationary time series. 

\begin{figure}
    \centering
    \includegraphics[width = \textwidth]{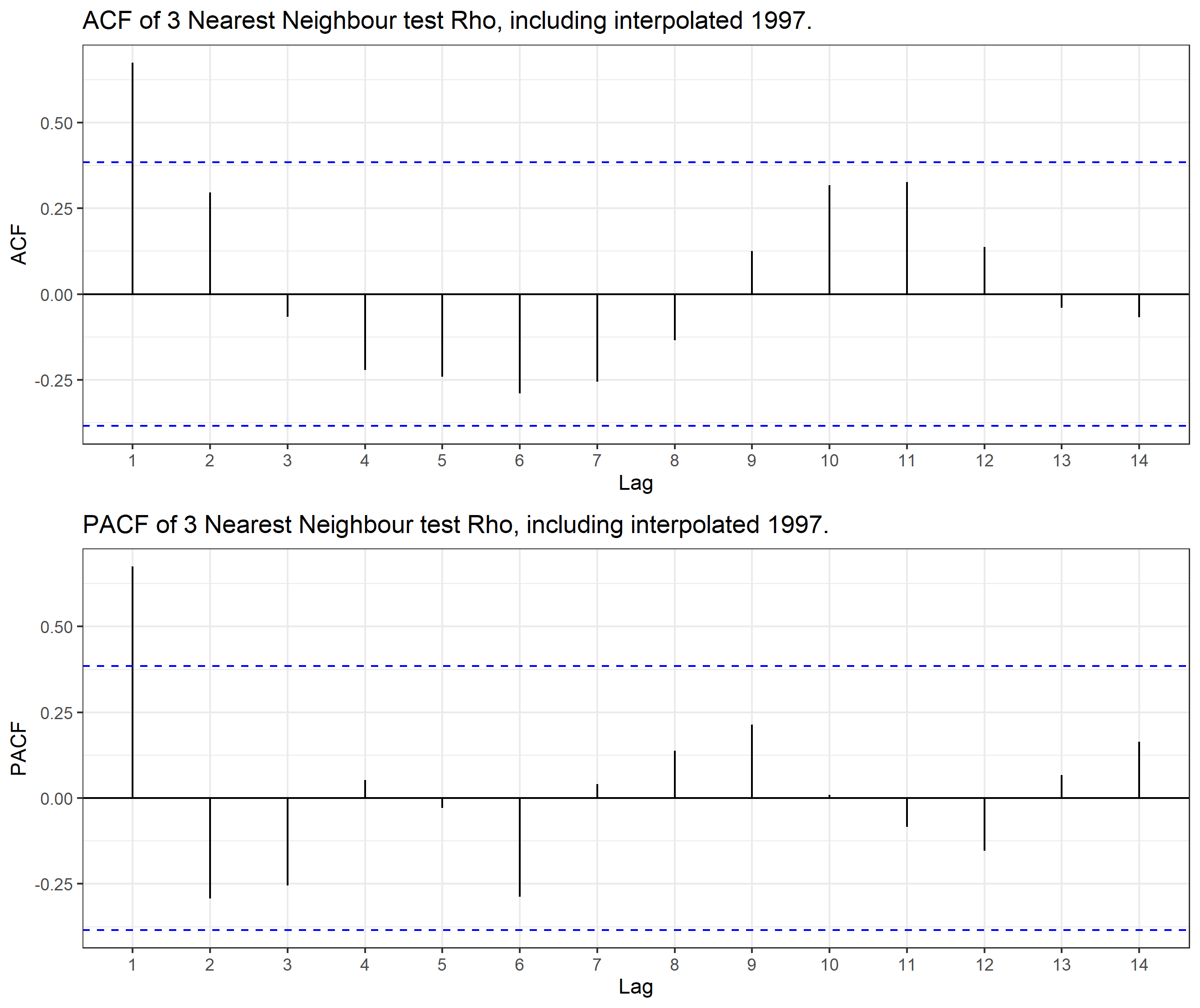}
    \caption{An examination of the autocorrelation of the nearest neighbors.  With both quickly decaying, it seems unlikely that there is nonstationarity in the time series.   Note that since sites carry over from year to year, each year's correlation is not an independent observation.  This uses the interpolated value for 1997 as part of the time series.  With the spike in ACF and PACE at lag 1, the smooth decay in the ACF, and the sharp drop-off in the PACF, an AR(1) process seems a plausible model choice.}
    \label{fig:test_rho_acf_pacf}
\end{figure}

To examine whether the scores can be explained by a stationary time series, ACF and PACF plots of the Test Rho (Figure \ref{fig:test_rho_acf_pacf}) were produced.  There are two ways of handling the missing value for 1997 while calculating the ACF and PACF:  (1) interpolate the missing year; (2) cut out the time series before the missing year.  We chose option (1) and so interpolated the missing year's Test Rho by using the mean of 1996 and 1998.  Figure \ref{fig:test_rho_acf_pacf} suggests the Test Rho time series is stationary, with a possible AR(1) process.  

The final issue we investigated is whether a  relationship between the Test Rho and the addition or removal of sides from the network exists, ie the dots and the red and blue lines in Figure \ref{fig:PS_site_counts_test_rho}.  This could be another indicator of administrative choices resulting in preferential sampling over time as the network evolves.  Spearman's correlation between the time series test scores and the two-time series of network site changes was used.   One correlation with the interpolated point included was done, and one with that year removed from the site counts.

\begin{table}[ht]
    \centering
    \begin{tabular}{c|c|c}

         &  Interpolated 1997 & 1997 Absent\\
        \hline
        Number of Added Sites & 0.02921195 & 0.09615902 \\
        Number of Removed Sites & -0.1937513 & -0.2670485
    \end{tabular}
    \caption{Spearman's Rho between the network's Test Rho and the number of sites added or removed from the network compared to the previous year.}
    \label{tab:testRho_siteChange}
\end{table}


Table \ref{tab:testRho_siteChange} shows the results of the correlation.  Adding sites seems to have very little correlation with the network's Test Rho over the years of monitoring.  Site removal has a stronger correlation, but still not much.

\section{Conclusions}\label{sec:conclusions}

\subsection{Presence of Preferential Sampling}
\label{subsec:presenceprefsamp}

This report found evidence of Preferential Sampling in the \ac{SOCAB}.  Government reports provided textual support, while statistical support came from both the pattern of network adjustment and the \texttt{PStestR} package.  This \ac{PS} results in observed pollution higher than the regional mean.

\subsection{Complications Not Considered}
\label{subsec:notconsidered}
Throughout this work, several ways of adding complexity were sidelined.  Also, having finished, several extensions or new angles of inquiry have occurred to me.  These include considerations when modelling the field, ways to interpret preferential sampling, and how to extend the work to applicability.

\subsubsection*{Modelling the field}
\label{subsubsec:modellingfield}
As discussed in Section \ref{subsubsec:DataRows}, many sites have multiple instruments recording \ac{PM10} at the same time.  These could provide an understanding of the nugget effect by being replicated measurements.  However, this would require modifications to the \ac{INLA} model to account for the unusual presence of information on the nugget.  Therefore, the instruments were combined into a mean to simplify modelling.

Section \ref{subsubsec:SpatialDomain} described how the study area to \ac{SOCAB} was constrained.  However, the decisions are made by the \ac{SCAQMD} which has jurisdiction over a wider area. If the study domain could be extended to the jurisdictional boundary instead of the geographic airshed boundary, a fuller understanding of the preferential sampling process would be achieved. However, this requires modelling the discontinuity in the \ac{PM10} surface.

The exploratory examination of the preferential sampling time series in Section \ref{subsubsec:TestScoreTimeSeries} did not account for the lack of independence between years.  It is not known what modifications would have to be made to the \texttt{PStestR} algorithm to account for this, but it is another area that could be examined.
 
\subsubsection*{Understanding Preferential Sampling}
\label{subsubsec:understandingPF}

The MCMC samples of hypothetical networks generated by \texttt{PStestR} placed sites with a uniform distribution over the SOCAB's area.  However, sites have numerous constraints on their actual real-world location.  Implementing these constraints would require finding documentation of what considerations are made for site selection, and then using a GIS tool with data layers describing those considerations.  Another way to examine the network for preferential sampling is whether a given site complies with \ac{EPA} standards.  If decisions are being made based upon a site's compliance or lack thereof, including it into the \ac{PS} model could help understand the decisions.

\subsubsection*{Future Applicability}
\label{subsubsec:futureapps}
Much of the motivation for detecting preferential sampling is from its potential impact on studies using the observed data as an unbiased sample of population exposure.  Having found evidence for Preferential Sampling, a reasonable next step could be to determine how this bias has affected various studies.  A sensitivity study testing the impact of \ac{PS} on health studies might help.

\subsection{Workflow}
\label{subsec:workflow}
We now summarize the flow of the work we learned was necessary to reach the findings reported in this report. We hope that this blueprint might be of value in future work aimed at the same objective but in a different geographical domain.  Of note is an R package developed partway through this work that uses the EPA API called \texttt{raqdm}.  This package makes data acquisition for future work more streamlined.

\begin{itemize}
    \item Obtain \ac{EPA} data, e.g. through the R package \texttt{raqdm}.
    \item Obtain a spatial shape file for the area of interest, e.g. from an online GIS service.
    \item Match the projection of the spatial shape file to the coordinate system of the data and then filter the data for the pollutant of interest, exclusion criteria, and area of interest.
    \item Perform preliminary data exploration for spatial and temporal trends, presence of anisotropy, and utility of metadata for modelling. 
    \item Create a mesh, using the area of interest as a boundary, and preliminary range as edge lengths.
    \item Model the data with splines for spatial and temporal trends, Mat\'{e}tern function for spatial covariance and probably an AR(1) for temporal covariance.  
    \item After validating the model, use it to predict a surface.
    \item Run a preferential sampling test using \texttt{PStestR} and the predicted surface.
\end{itemize}


\bibliographystyle{plainnat}
\bibliography{prefsamp.Jan03_2017,PreferentialSampling_2020118_JZ,personal_work,AJ_Gov_reports_docs,mybib}

\begin{thebibliography}{25}
\providecommand{\natexlab}[1]{#1}
\providecommand{\url}[1]{\texttt{#1}}
\expandafter\ifx\csname urlstyle\endcsname\relax
  \providecommand{\doi}[1]{doi: #1}\else
  \providecommand{\doi}{doi: \begingroup \urlstyle{rm}\Url}\fi

\bibitem[Bermudez and Fine(2010)]{CASCAQMD:2010}
Rene~M Bermudez and Philip~M. Fine.
\newblock \emph{SOUTH COAST AIR QUALITY MANAGEMENT DISTRICT 5 YEAR NETWORK
  ASSESSMENT}.
\newblock 2010.

\bibitem[Bermudez et~al.(2015)Bermudez, Pakbin, and Low]{CASCAQMD:2015}
Rene~M Bermudez, Payam Pakbin, and Jason~C Low.
\newblock \emph{SOUTH COAST AIR QUALITY MANAGEMENT DISTRICT 5 YEAR NETWORK
  ASSESSMENT}.
\newblock 2015.

\bibitem[Cameletti et~al.(2011)Cameletti, Lindgren, Simpson, and
  Rue]{cameletti2011spatio}
M.~Cameletti, F.~Lindgren, D.~Simpson, and H.~Rue.
\newblock Spatio-temporal modeling of particulate matter concentration through
  the spde approach.
\newblock \emph{AStA Advances in Statistical Analysis}, pages 1--23, 2011.

\bibitem[Cressie and Wikle(2011)]{cressie2011statistics}
N.~Cressie and C.K. Wikle.
\newblock \emph{Statistics for spatio-temporal data}, volume 465.
\newblock Wiley, 2011.

\bibitem[Diggle and Ribeiro(2007)]{diggle:07}
P.J. Diggle and P.J. Ribeiro.
\newblock \emph{{Model Based Geostatistics}}.
\newblock Springer, 2007.

\bibitem[Diggle and Ribeiro~Jr(2010)]{diggle10}
P.J. Diggle and P.J. Ribeiro~Jr.
\newblock \emph{Model based geostatistics}.
\newblock Springer Verlag, 2010.

\bibitem[Diggle et~al.(2010)Diggle, Menezes, and Su]{diggle2010geostatistical}
P.J. Diggle, R.~Menezes, and T.~Su.
\newblock Geostatistical inference under preferential sampling.
\newblock \emph{Journal of the Royal Statistical Society: Series C (Applied
  Statistics)}, 59\penalty0 (2):\penalty0 191--232, 2010.

\bibitem[EPA(2016)]{EPA:IntegratedReview}
EPA.
\newblock Integrated review plan for the national ambient air quality standards
  for particulate matter, 2016.

\bibitem[EPA(2021)]{CFR:Title40-58}
EPA.
\newblock Title 40: Projection of the environment, 2021.
\newblock URL \url{https://www.law.cornell.edu/cfr/text/40/part-58}.

\bibitem[Fuglstad et~al.(2017)Fuglstad, Simpson, Lindgren, and
  Rue]{fuglstad2017constructing}
Geir-Arne Fuglstad, Daniel Simpson, Finn Lindgren, and H{\aa}vard Rue.
\newblock Constructing priors that penalize the complexity of gaussian random
  fields.
\newblock \emph{Journal of the American Statistical Association}, \penalty0
  (just-accepted), 2017.

\bibitem[G{\'o}mez-Rubio(2020)]{gomezGitBook}
Virgilio G{\'o}mez-Rubio.
\newblock \emph{Bayesian inference with INLA}.
\newblock Chapman and Hall/CRC Press, 2020.
\newblock URL
  \url{https://becarioprecario.bitbucket.io/inla-gitbook/index.html}.

\bibitem[Isaaks and Srivastava(1988)]{isaaks1988spatial}
EH~Isaaks and R~Mohan Srivastava.
\newblock Spatial continuity measures for probabilistic and deterministic
  geostatistics.
\newblock \emph{Mathematical geology}, 20\penalty0 (4):\penalty0 313--341,
  1988.

\bibitem[Miyasato et~al.(2019)Miyasato, Low, and Bermudez]{AQMNP:2019}
Matt Miyasato, Jason Low, and Rene~M Bermudez.
\newblock \emph{ANNUAL AIR QUALITY MONITORING NETWORK PLAN}.
\newblock 2019.

\bibitem[Ott(1990)]{ott1990}
W~Ott.
\newblock A {P}hysical {E}xplanation of the {L}ognormality of {P}ollutant
  {C}oncentrations.
\newblock \emph{Journal of the {A}ir {W}aste {M}anagement {A}ssociation},
  40:\penalty0 1378--1383, 1990.

\bibitem[Righetto et~al.(2020)Righetto, Faes, Vandendijck, and
  Jr.]{Righetto2020}
Ana~Julia Righetto, Christel Faes, Yannick Vandendijck, and Paulo
  Justiniano~Ribeiro Jr.
\newblock On the choice of the mesh for the analysis of geostatistical data
  using r-inla.
\newblock \emph{Communications in Statistics - Theory and Methods}, 49\penalty0
  (1):\penalty0 203--220, 2020.
\newblock \doi{10.1080/03610926.2018.1536209}.
\newblock URL \url{ttps://doi.org/10.1080/03610926.2018.1536209}.

\bibitem[Schlather et~al.(2004)Schlather, Ribeiro, and
  Diggle]{schlather2004detecting}
Martin Schlather, Paulo~J Ribeiro, and Peter~J Diggle.
\newblock Detecting dependence between marks and locations of marked point
  processes.
\newblock \emph{Journal of the Royal Statistical Society: Series B (Statistical
  Methodology)}, 66\penalty0 (1):\penalty0 79--93, 2004.

\bibitem[Shaddick and Zidek(2012)]{shaddick2012preferential}
G.~Shaddick and J.~V. Zidek.
\newblock Preferential sampling in long term monitoring of air pollution: a
  case study.
\newblock Technical report, Technical Report 267, Department of Statistics,
  University of British Columbia, 2012.

\bibitem[Shaddick and Zidek(2014)]{shaddick2014case}
Gavin Shaddick and James~V Zidek.
\newblock A case study in preferential sampling: Long term monitoring of air
  pollution in the uk.
\newblock \emph{Spatial Statistics}, 9:\penalty0 51--65, 2014.

\bibitem[Simpson et~al.(2017)Simpson, Rue, Riebler, Martins, S{\o}rbye,
  et~al.]{simpson2017penalising}
Daniel Simpson, H{\aa}vard Rue, Andrea Riebler, Thiago~G Martins, Sigrunn~H
  S{\o}rbye, et~al.
\newblock Penalising model component complexity: A principled, practical
  approach to constructing priors.
\newblock \emph{Statistical science}, 32\penalty0 (1):\penalty0 1--28, 2017.

\bibitem[Snyder(1987)]{USGS:MapProjections}
John~P. Snyder.
\newblock \emph{Map Projections - A Working Manual}.
\newblock Washington, 1987.

\bibitem[Watson(2021)]{watson2020}
Joe Watson.
\newblock A perceptron for detecting the preferential sampling of locations and
  times chosen to monitor a spatio-temporal process.
\newblock \emph{Spatial Statistics}, 43:\penalty0 100500, 2021.

\bibitem[Watson et~al.(2019)Watson, Zidek, Shaddick, et~al.]{watson2019}
Joe Watson, James~V Zidek, Gavin Shaddick, et~al.
\newblock A general theory for preferential sampling in environmental networks.
\newblock \emph{Annals of Applied Statistics}, 13\penalty0 (4):\penalty0
  2662--2700, 2019.

\bibitem[Wong et~al.(2004)Wong, Yuan, and Perlin]{wong2004comparison}
David~W Wong, Lester Yuan, and Susan~A Perlin.
\newblock Comparison of spatial interpolation methods for the estimation of air
  quality data.
\newblock \emph{Journal of Exposure Science \& Environmental Epidemiology},
  14\penalty0 (5):\penalty0 404--415, 2004.

\bibitem[Zidek et~al.(2012)Zidek, Le, and Liu]{Zidek:2012}
James Zidek, Nhu Le, and Zhong Liu.
\newblock Combining data and simulated data for space--time fields: application
  to ozone.
\newblock \emph{Environmental and Ecological Statistics}, 19\penalty0
  (1):\penalty0 37--56, 2012.
\newblock ISSN 1352-8505.
\newblock \doi{10.1007/s10651-011-0172-1}.

\bibitem[Zidek and Zimmerman(2010)]{zidek2010monitoring}
James~V Zidek and Dale~L Zimmerman.
\newblock Monitoring network design.
\newblock \emph{Handbook of Spatial Statistics}, pages 131--148, 2010.

\end{thebibliography}
\end{document}